\documentclass[fleqn,usenatbib]{mnras}

\usepackage[T1]{fontenc}
\usepackage{ae,aecompl}

\usepackage{graphicx}	% Including figure files
\usepackage{amsmath}	% Advanced maths commands
\usepackage{pdflscape}
\usepackage{multirow}
\usepackage{float}
\usepackage{rotating}
\usepackage{lipsum}
\usepackage{threeparttable, tablefootnote}
\usepackage{array}
\title[VKL: exploring lens reconstruction regularization]{The Very Knotty Lenser: exploring the role of regularization in source and potential reconstructions using Gaussian Process Regression}

\author[Vernardos \& Koopmans]{
	\parbox{\textwidth}{
		G. Vernardos$^{1,2}$\thanks{E-mail: georgios.vernardos@epfl.ch}
		and L. V. E. Koopmans$^{2}$\\
	}
\\
	\vspace{0.4cm}\
	\parbox{\textwidth}{
		$^{1}$Institute of Physics, Laboratory of Astrophysics, Ecole Polytechnique F\'{e}d\'{e}rale de Lausanne (EPFL), Observatoire de Sauverny, 1290 Versoix, Switzerland \\
		$^{2}$Kapteyn Astronomical Institute, University of Groningen, PO Box 800, NL-9700AV Groningen, the Netherlands \\
	}
}

\date{Accepted XXX. Received YYY; in original form ZZZ}

\begin{document}
\label{firstpage}
\pagerange{\pageref{firstpage}--\pageref{lastpage}}
\maketitle

\begin{abstract}
Reconstructing lens potentials and lensed sources can easily become an underconstrained problem, even when the degrees of freedom are low, due to degeneracies, particularly when potential perturbations superimposed on a smooth lens are included.
Regularization has traditionally been used to constrain the solutions where the data failed to do so, e.g. in unlensed parts of the source.
In this exploratory work, we go beyond the usual choices of regularization and adopt observationally motivated priors for the source brightness.
We also perform a similar comparison when reconstructing lens potential perturbations, which are assumed to be stationary, i.e. permeate the entire field of view.
We find that physically motivated priors lead to lower residuals, avoid overfitting, and are decisively preferred within a Bayesian quantitative framework in all the examples considered.
For the perturbations, choosing the wrong regularization can have a detrimental effect that even high-quality data cannot correct for, while using a purely smooth lens model can absorb them to a very high degree and lead to biased solutions.
Finally, our new implementation of the semi-linear inversion technique provides the first quantitative framework for measuring degeneracies between the source and the potential perturbations.
\end{abstract}


%%%%%%%%%%%%%%%%%%%%%%%%%%%%%%%%%%%%%%%%%%%%%%%%%%%%%%%%%%%%%%%%%%%%%%%%%%%%%%%%%%%%%%%%%%%%%%%%%%%%%%%%%%%%%%%%%%%%%%%%%%%%%%%%%%%%%%%%%%%%%%%%%%%%%
\section{Introduction}
The standard cosmological model, comprising the still unknown dark energy and dark matter, has been successful in describing the large scale structure of the Universe and its properties \citep[$>$1 Mpc, e.g.][]{Komatsu2011,PlanckCollaboration2018vi}. % "Planck 2016 results. XIII.
The dark matter component in particular, plays an important role throughout cosmic evolution by participating in the collapse of baryons via gravitational instability to form galaxies \citep{White1978}.
Verifying the validity of the current Cold Dark Matter paradigm down to sub-galactic scales, and what this implies for the microscopic properties of the dark matter particle, is masked by the onset of highly non-linear physical mechanisms attributed to baryons, e.g. stellar winds, supernovae, feedback from Active Galactic Nuclei, etc, that appear in such high density environments \citep{Vogelsberger2014,Schaye2015}. %Vogelsberger: Nature

The tension between dark matter theory and observations on galactic and sub-galactic scales \citep[$<$1 Mpc,][]{Bullock2017} has several manifestations, e.g. the ``missing satellites'' \citep{Moore1999,Klypin1999}, the ``cusp-core'' \citep{Moore1994,Oh2015}, the ``too-big-to-fail'' \citep{BoylanKolchin2011}, and the ``bulge-halo conspiracy'' \citep{Dutton2014} problems.
Regardless of the role of baryons and their gravitational interactions with dark matter in each case, aspects of which constitute independent major research fields \citep[e.g. the efficiency of star formation,][]{McKee2007}, measuring the overall shape and smoothness of the mass density in galaxies is critical.
In the local Universe, this can be achieved by understanding the statistics \citep[e.g.][]{Papastergis2015}, instrumental effects \citep{Kim2018}, and dynamics \citep[e.g.][]{Helmi2012} of dwarf galaxies and stellar streams \citep[e.g.][]{Carlberg2012,Erkal2016}.

As soon as one leaves the neighbourhood of the Milky Way, the only way to achieve such measurements is via gravitational lensing - the deflection of light from a distant source by the intervening mass of a galaxy.
In this way, the overall shape of the total mass distribution has been measured for massive elliptical galaxies \citep{Koopmans2006,Koopmans2009,Gavazzi2007,Auger2010,Barnabe2011,Sonnenfeld2013,Suyu2014,Oldham2018} and massive substructures down to the order of $10^8$ M$_{\odot}$ have been detected out to cosmological distances \citep{Vegetti2010,Vegetti2012,Fadely2012,MacLeod2013,Nierenberg2014,Li2016,Hezaveh2016b,Birrer2017}.
Strong lensing analysis has also been combined with other techniques, e.g. stellar kinematics \citep{Barnabe2011,Yildirim2020}, stellar kinematics and weak lensing \citep{Sonnenfeld2018}, stellar population analysis \citep{Barnabe2013,Smith2015,Spiniello2015}, and quasar microlensing \citep{Oguri2014}, in order to disentangle the baryonic and dark mass components.

The gravitational imaging technique \citep{Koopmans2005,Vegetti2009a} is a powerful method to study the non-smoothness of the lensing mass distribution, analyzing perturbations of lensing features, such as arcs and Einstein rings\footnote{This can also be achieved by analyzing flux ratios from lensed quasars \citep[][]{Dalal2002}, however, this requires carefully planned spectroscopic observations, taking into account the possible effect of microlensing \citep[e.g.][]{Nierenberg2014}.}.
Based on the semi-linear inversion method of \citet{Warren2003}, which can reconstruct the light distribution of the lensed source on a grid once the lensing potential is given, \citet{Koopmans2005} provided an extension that simultaneously obtains a grid-based reconstruction of potential perturbations to an overall smooth (parametric) lens potential: in the presence of substructure, dark or luminous, the smooth modelling residuals are remodelled in terms of lens potential perturbations using the smooth potential and its corresponding source as a starting point.
\citet{Vegetti2009a} improved this technique in a number of ways, expanding the work of \citet{Suyu2006} by casting the problem in a Bayesian framework that includes the potential perturbations and using an adaptive grid for the source.
With careful control over the regularization level of the solutions, the presence of substructure in a lens can be uncovered by accumulating small potential corrections within an iterative scheme.
The detection is then justified by comparing the Bayesian evidence to the best purely smooth lensing model \citep{Vegetti2010}.

The regularization scheme plays a critical role in such a strong lensing Bayesian analysis approach, as it enables the matrix inversions to find a unique solution \citep{MacKay1992,MacKay2003}.
Focusing only on the reconstruction of the source, there are several pixel-based methods\footnote{The possibility of using basis sets to reconstruct the source has been explored by \citet{Birrer2015,Joseph2019,Galan2021} and the use of deep neural networks was investigated by \citet{Morningstar2019}. Both methods do not explicitly require regularization, but rely on the number of independent basis vectors and a descriptive training set respectively, to model higher order statistics of the source.} that employ a brightness, gradient, or curvature based regularization scheme, or a combination thereof \citep{Dye2005,Suyu2006,Vegetti2009a,Tagore2014,Nightingale2015,Yildirim2020}, i.e they assume that each of these source properties is drawn from a normal distribution, whose variance is determined by the regularization parameter that itself can be optimized for, and whose correlation properties are set by a corresponding covariance matrix.
However, a poor choice of the regularization parameter in each case is known to cause problems with over- and under- fitting of the data in some cases, which in turn might affect the mass model parameters \citep{Nightingale2015}.
\citet{Suyu2006} solve exactly for the value of the regularization parameter that maximizes the evidence.
To allow for more flexibility, \citet{Nightingale2018} have introduced a non-constant (adaptive) regularization scheme, whose principle is to vary the strength of the regularization (width of the normal distribution) across the source, based on its surface brightness profile.
Some form of regularization is necessary to be able to solve the equations, however all of these methods are equivalent to setting priors for the different source properties that are not necessarily astrophysically motivated.

Upon combining the source reconstruction with potential perturbations, which enter the equations in a very similar way to the source and require their own regularization scheme, an additional non-linear dependence of the perturbations on the source is introduced \citep{Koopmans2005}.
Again, the regularization of the two fields, the source and the perturbations, plays an important role in reaching a unique solution.
\citet{Vegetti2009a} follow a line-search optimization, starting with finding the best smooth lens-mass model and then proceeding with calculating potential corrections based on the corresponding source (see also equation \ref{eq:dpsi_residuals} here).
In their iterative scheme, the source and potential perturbations are solved for at each step and then updated: the new surface brightness derivatives are calculated across the source and the perturbations are added to the overall smooth potential in the form of corrections.
The regularization parameter of the perturbations is carefully controlled, initially set to very high values (very smooth fields) and later reduced to allow for more structure.
This is similar to a Gauss-Newton optimization scheme that is known to be sensitive to the step size; any spurious structure appearing in the solutions would be added to the overall lensing potential with the risk of irrecoverably drifting away from the true solution.
Although this is a powerful approach, it is limited by two caveats: some manual fine-tuning is needed in setting up the algorithm to converge to a meaningful solution, and there is no obvious means to quantify degeneracies between the reconstructed source and the potential perturbations.
The latter is inherent to the technique and has not been studied in depth before \citep[see][for an example]{Chatterjee2019}.

In this paper, we more rigorously investigate the importance of new forms of regularization, introducing more realistic priors on the source surface brightness distribution that are more flexible in capturing higher order statistical properties, and a statistical approach to finding the best regularization parameters via sampling.
The latter is based on the theory of Gaussian Process Regression \citep{Rasmussen2006} and is quite powerful as it provides a way to quantify degeneracies between the source and perturbation fields.
In addition, this sampling approach is better suited to describe extended perturbations, which are not necessarily restricted to compact and well-localized perturbers that might be more accurately detected by an iterative and additive scheme \citep[as in][]{Vegetti2009a}.
The outcome is a statistical approach to generic perturbations of a smooth lensing potential, which can be directly linked to the underlying statistical properties of baryonic and dark matter (e.g. via the power spectrum), or to higher order structure in the lens potential, such as the presence of a galactic disc \citep[as was recently found by][]{Hsueh2017}.

The structure of the paper is as follows.
In Section \ref{sec:method} we set up the theoretical framework, provide the Bayesian evidence equation extending the work of \citet{Suyu2006} and \citet{Vegetti2009a}, and demonstrate the use of this approach under various regularization schemes.
Section \ref{sec:results} presents a set of selected applications of the method on mock lens systems, which are discussed further in Section \ref{sec:discussion}.
Our conclusions are summarized in Section \ref{sec:conclusions}.

%%%%%%%%%%%%%%%%%%%%%%%%%%%%%%%%%%%%%%%%%%%%%%%%%%%%%%%%%%%%%%%%%%%%%%%%%%%%%%%%%%%%%%%%%%%%%%%%%%%%%%%%%%%%%%%%%%%%%%%%%%%%%%%%%%%%%%%%%%%%%%%%%%%%%%
\section{Method}
\label{sec:method}
The Bayesian formalism applied to grid-based strong lensing analyses was introduced by \citet{Suyu2006} and \citet{Vegetti2009a}.
Here, we use the same framework and repeat some of the steps, while we point out the differences, particularly with respect to the regularization and our sampling approach.
In addition, an explicit equation describing the Bayesian evidence is derived, which has not appeared in the literature so far (\citealt{Suyu2006} give such an expression but including only the source).

First, we formulate the problem in terms of a lensing operator depending on a parametrized smooth lens potential and a source brightness distribution defined on a grid, and then we introduce potential perturbations.
Solving the resulting equations directly is an ill-posed problem.
We therefore need to look for solutions minimizing some form of penalty function that includes regularization.
This leads to a new set of linear equations with respect to the source and the potential perturbations that has an exact solution.
The problem is then re-cast using a Bayesian formalism and the expression of the evidence is derived.
The general treatment is independent of any assumption on the particular type of regularization, however, several physically motivated schemes are examined in more detail.
Finally, we present a sampling approach to determine the probability distribution of all non-linear parameters of the problem.

\subsection{The lensing operator and the source grid}
\label{sec:grids}
The problem at hand is finding how the brightness of the lensed images relates to the background source brightness via gravitational lensing, and can be cast in the following way \citep[similarly to][]{Warren2003,Koopmans2005,Vegetti2009a}:
\begin{equation}
\label{eq:cast_source}
\boldsymbol{d} = BL(\psi) \boldsymbol{s} + \boldsymbol{n}
\end{equation}
where $\boldsymbol{d}$ and $\boldsymbol{n}$ are the vectors of brightness measurements (the ``data'') and the associated noise (the ``noise'') of the image pixels, $\boldsymbol{s}$ is the vector of the source brightness (the ``source''), $B$ is the blurring operator that is linked to the point spread function (PSF), and $L$ is the lensing operator that depends on the lensing potential $\psi$.
The data and noise vectors correspond to a rectangular M$\times$N grid of N$_{\rm d}$ pixels in total on the image plane, which delineates the part of the pixel array of the optical detector covering the lensed images.
The blurring operator (N$_{\rm d} \times $N$_{\rm d}$) is assumed constant\footnote{The PSF can in fact vary for each pixel based on the spectral energy distribution of the source for that specific pixel, or due to atmospheric effects if we deal with ground-based observations.} and mimics the effect of the PSF; it acts on (blurs) the resulting image plane pixels with a fixed weighting scheme, after the source has been lensed.
Assuming that the source can also be described by a pixelated grid of N$_{\rm s}$ pixels and arbitrary form on the (unobserved) source plane, then the lensing operator (N$_{\rm d} \times$ N$_{\rm s}$) couples each data pixel position to the source grid via the lens equation \citep{Vegetti2009a}.
This can introduce multiplicity because different image pixels can be associated with the same source location, thus creating multiple images.
Equation (\ref{eq:cast_source}) is a linear transformation between the image and source planes that depends on the gradient of the lensing potential $\psi$.
We note that the lensing potential is typically a non-linear function of the lens plane coordinates, $\boldsymbol{x}$, and some parameters, $\boldsymbol{\eta}$, that can vary in complexity.

Once the positions of the data pixels are traced back to the source plane, they are matched to pixels on the source grid via an interpolation scheme that guarantees the conservation of surface brightness \citep[see fig. 1 in][]{Koopmans2005}.
The source grid can have any arbitrary structure, e.g. fixed or free-floating regular grids, irregular, adaptive, etc.
On a regular grid, bi-linear interpolation is sufficient, while higher order schemes could also be used (e.g. bi-cubic, natural neighbour, etc).
An irregular grid has a unique Delaunay triangulation and its corresponding dual Voronoi tesselation, whose cells can both be considered as source ``pixels'' \citep{Gallier2011}.
Data pixels that are cast back onto the source plane land inside a Delaunay triangle and their value is interpolated linearly between the triangle's vertices (the centers of the irregular Voronoi source grid ``pixels'').
Hence, the brightness values inside any such triangle lie on a tilted plane defined by the values at the triangle vertices.
Barycentric coordinates are used to perform these triangular interpolations, which is equivalent to the procedure described in \citet[][figs. 1 and 2]{Vegetti2009a}.

An irregular source grid can also be constructed randomly \citep[e.g.][]{Nightingale2015} or by a recipe designed to facilitate the source reconstruction.
An example is a so-called adaptive grid that is reconstructed every time the lens potential $\psi(\boldsymbol{\eta})$ changes.
Here, we create such adaptive grids by casting back one out of every $n\times n$ block of the data pixels, with $1 \leq n < 6$ (fixed throughout the reconstruction).
Alternative gridding techniques are known to affect the ``discreteness-noise'' in the computed Bayesian evidence and $\chi^2$ terms \citep{Tagore2014,Nightingale2015}.
However, exploring different grids is out of the current paper's scope and left for future improvements to our method.
For very large values of $n$ the resulting grid will be too coarse to successfully describe a detailed lensed image brightness distribution.
For $n=1$, there is no need for any interpolation as all the data pixels have been used to create the source grid ($\mathrm{N_s = N_d}$).
However, in this case the system of equations to solve is under-constrained and heavily relies on the regularization (i.e. assumed prior on the source surface brightness).

Applying this procedure for any given lens potential $\psi(\boldsymbol{\eta})$ results in a set of points on the source plane representing the positions of the source brightness values $\boldsymbol{s}$ and a N$_{\rm d} \times$ N$_{\rm s}$ operator $L$, whose rows contain the interpolation weights on the source grid for each data pixel.
The procedure is repeated each time the lens potential $\psi$ changes \citep{Vegetti2009a}.

\subsection{Lens potential corrections}
Often, an elliptical power law mass model is assumed for the lens \citep{Kassiola1993,Barkana1998}.
However, such smooth lens potential models may well be too simplified to capture more detailed structure of real lenses.
Deviations from smoothness could be the result of dark matter substructure or higher order moments in the mass distribution of the lens galaxy itself, originating from its morphology \citep[e.g.][find a non-negligible disc component]{Hsueh2017} or evolution history (e.g. mergers).
If such deviations exist in an observed system, they will manifest themselves as residuals, $\delta\boldsymbol{d}$, left behind after modelling the lens with a smooth potential:
\begin{equation}
\label{eq:smooth_residuals}
\delta\boldsymbol{d} = M \boldsymbol{s}_{\rm p} - \boldsymbol{d},
\end{equation}
where $M \equiv M(\boldsymbol{\eta}) = B L(\boldsymbol{\eta})$, and $\boldsymbol{s}_{\rm p}$ is the solution for the source after inverting the smooth model as described in Section \ref{sec:smooth_inversion}.
Such residuals will persist regardless of the smooth potential used to describe the lens, although they may be absorbed to some degree into the source surface brightness or by modifying the values of the parameters $\boldsymbol{\eta}$.
%In the following, the matrix $M$ is always assumed to be a function of $\boldsymbol{\eta}$ unless otherwise stated.

If the residuals from the smooth modelling are not noise-like, then the inclusion of a new lens potential component may be warranted in order for $\delta\boldsymbol{d} \rightarrow 0$ (or, more precisely, $\delta\boldsymbol{d}$ reaching the properties of the noise).
The most general treatment of such a component is assuming a potential perturbations field, $\delta\boldsymbol{\psi}$, which to first order can de described by \citep{Koopmans2005}:
\begin{equation}
\label{eq:dpsi_residuals}
\delta\boldsymbol{d} = -B D_{\rm s}(\boldsymbol{s}_{\rm p}) D_{\rm \delta\psi} \delta \boldsymbol{\psi},
\end{equation}
where $D_{\rm s}(\boldsymbol{s}_{\rm p})$ is a matrix containing the gradient of the previously known source $\boldsymbol{s}_{\rm p}$, and $D_{\rm \delta\psi}$ is the gradient operator of the potential perturbations that yield $\delta\boldsymbol{\alpha}$, the perturbative deflection angle vector \citep[see appendix A of][for a derivation of this equation]{Koopmans2005}.
This equation describes how potential perturbations induce additional deflections, causing the positions in the image plane to become associated with a different position in the source plane, and hence with a different source brightness.
These deflections are assumed to be small enough for the source to be well approximated by a first order Taylor expansion around the original unperturbed locations.
In this way, the residual image plane brightness of the smooth model can be associated with the gradient of the source brightness and some small potential perturbation field.

Equations (\ref{eq:cast_source}), (\ref{eq:smooth_residuals}), and (\ref{eq:dpsi_residuals}) can be combined to reformulate the lensing problem as \citep{Koopmans2005,Vegetti2009a}:
\begin{equation}
\label{eq:source_dpsi_combined}
\boldsymbol{d} = M_{\rm r} \boldsymbol{r} + \boldsymbol{n},
\end{equation}
where $M_{\rm r}$ is the block matrix:
\begin{equation}
\label{eq:combined_M}
M_{\rm r} \equiv M_{\rm r}(\psi_{\rm p},\boldsymbol{s}_{\rm p}) = B [L(\psi_{\rm p}) | -D_{\rm s}(\boldsymbol{s}_{\rm p}) D_{\rm \delta\psi} ],
\end{equation}
and:
\begin{equation}
\label{eq:r}
\boldsymbol{r} \equiv 
\begin{pmatrix}
\boldsymbol{s} \\
\delta\boldsymbol{\psi} \\
\end{pmatrix}.
\end{equation}
The similarity with equation (\ref{eq:cast_source}) is striking, however, there is one important difference: some prior knowledge of the source brightness is necessary to construct the matrix $D_{\rm s}(\boldsymbol{s}_{\rm p})$.
The lens potential $\psi_{\rm p}$ can either depend on $\boldsymbol{\eta}$, as is the case in equation (\ref{eq:cast_source}), or it can include accumulated corrections $\delta\boldsymbol{\psi}_{\rm p}$ derived at previous stages - similarly to a Gauss-Newton scheme where a small update to the previous solution is calculated via a linear extrapolation.

The $\delta\boldsymbol{\psi}$ field can be approximated by N$_{\rm \delta\psi}$ pixels on the image plane, which we here assume to be on a fixed regular P$\times$Q grid (as opposed to, for example, being adaptive).
The $D_{\rm s}(\boldsymbol{s}_{\rm p})$ matrix entries are calculated at the locations of the deflected data pixels on the source grid.
Similarly, the $D_{\rm \delta\psi}$ operator determines the derivatives of $\delta\boldsymbol{\psi}$ at the locations of the data pixels on the image plane.
The product $D_{\rm s}(\boldsymbol{s}_{\rm p})D_{\rm \delta\psi} \delta\boldsymbol{\psi}$ is a N$_{\rm d}\times$N$_{\rm \delta\psi}$ matrix, whose rows contain the terms:
\begin{equation}
\label{eq:expanded_dsdpsi}
[D_{\rm s}(\boldsymbol{s}_{\rm p})D_{\rm \delta\psi} \delta\boldsymbol{\psi}]_{\rm i} = \frac{\partial s_{\rm p} (\boldsymbol{y}_{\rm i})}{\partial y_{\rm 1}} \frac{\partial \delta\psi(\boldsymbol{x}_{\rm i})}{\partial x_{\rm 1}} + \frac{\partial s_{\rm p} (\boldsymbol{y}_{\rm i})}{\partial y_{\rm 2}} \frac{\partial \delta\psi(\boldsymbol{x}_{\rm i})}{\partial x_{\rm 2}},
\end{equation}
where $\boldsymbol{x}$ are the data pixel coordinates on the image plane and $\boldsymbol{y}$ their corresponding source plane positions.
If the data and perturbation grids coincide this matrix is diagonal, but usually the $\delta\boldsymbol{\psi}$ grid has a lower resolution such that each matrix row will contain the terms and corresponding weights resulting from a bilinear (in our case) interpolation on the $\delta\boldsymbol{\psi}$ grid (i.e. $\delta\psi(\boldsymbol{x}_{\rm i}) = \sum_{\rm j=1}^{4} w_{\rm i,j} \delta\psi_{\rm i,j}$, where the j-th index goes over the four vertices of the $\delta\boldsymbol{\psi}$ pixel encompassing the i-th data pixel).

\subsection{Model inversion}
\label{sec:smooth_inversion}
The observed data result from the physical and instrumental processes of lensing and blurring, described as operators acting on a gridded source (their order is important), the finite detector pixel size, and the inclusion of noise with some properties (e.g. statistical Poisson noise of photon counts, correlated noise introduced at data reduction, cosmic rays, etc).
Even in the absence of noise, inverting equation (\ref{eq:cast_source}) for the source is generally an ill-posed problem that does not have a unique or exact solution.
One way to proceed is by searching for a source solution that minimizes a regularized penalty function.
First, we define the penalty function, which is a likelihood function under the assumption of Gaussian errors in the data, excluding the perturbations $\delta\boldsymbol{\psi}$, to be the sum of a generalized $\chi^2$ and a regularization term:
\begin{equation}
\label{eq:penalty_source}
\begin{split}
G(\boldsymbol{s}) \equiv  \, & G(\boldsymbol{s}|\boldsymbol{d},\boldsymbol{\eta},\boldsymbol{g}_{\rm s},\lambda_{\rm s}) \\
	                 =    \, & \frac{1}{2}(M \boldsymbol{s} - \boldsymbol{d})^T C^{-1}_{\rm d} (M \boldsymbol{s} - \boldsymbol{d}) +  \frac{1}{2}\lambda_{\rm s}\boldsymbol{s}^T C^{-1}_{\rm s} (\boldsymbol{g}_{\rm s}) \boldsymbol{s},
\end{split}
\end{equation}
where $M$ is the operator used in equation (\ref{eq:smooth_residuals}), and $C_{\rm d}$ and $C_{\rm s}$ are the covariance matrices of the data and source, which, in the case of the source, may in general be a function of another set of non-linear regularization parameters, $\boldsymbol{g}_{\rm s}$ - we take out the regularization parameter $\lambda_{\rm s}$ to separate the effect of the source brightness from the shape of its correlations and make its effect more explicit.
The parameter $\lambda_{\rm s}$ sets the level of contribution to the overall penalty of the regularization term with respect to the value of $\chi^2$.
In the following, the covariance matrix $C_{\rm s}$ is always assumed to be a function of $\boldsymbol{g}_{\rm s}$, while specific regularization schemes are discussed in Section \ref{sec:reg}.

The source property used for regularization (gradient, curvature, etc) is assumed to be distributed normally, i.e. a quadratic form in equation (\ref{eq:penalty_source}), similarly to the $\chi^2$ term, guaranteeing that the source for which $\nabla_{\rm s} G = 0$ minimizes the penalty function \citep{Suyu2006}.
Using this condition, after some basic algebraic manipulations, we get:
\begin{equation}
\label{eq:min_source}
(M^T C^{-1}_{\rm d} M + \lambda_{\rm s} C^{-1}_{\rm s}) \boldsymbol{s} = M^T C^{-1}_{\rm d} \boldsymbol{d},
\end{equation}
where the matrix $M^T C^{-1}_{\rm d} M + \lambda_s C^{-1}_{\rm s}$ is now positive-definite and can be inverted using standard techniques.
The source that minimizes the penalty function is found in this way for each set of fixed $\boldsymbol{\eta}$, $\boldsymbol{g}_{\rm s}$, and $\lambda_{\rm s}$.
This solution implicitly assumes that the Gaussian random field describing the source has
a zero mean.
Although this is not formally correct because of the finite dimensions of the
source grid, this offset is in general easily absorbed by the shape of the covariance matrix, and as our tests later will show, this assumption holds to very good
approximation.

Often, masking the data is required to isolate and model only the lensed image features.
This can be achieved by an operator $S$, acting on the image plane and excluding all the pixels outside the mask from the modelling, which is simply a diagonal matrix with values of 1 or 0 for included and excluded pixels, respectively.
In equations (\ref{eq:penalty_source}) and (\ref{eq:min_source}), this can be incorporated into a ``masked'' covariance matrix $S^T C^{-1}_{\rm d} S$, all rest being the same.
In the remaining treatment, $C^{-1}_{\rm d}$ and $S^T C^{-1}_{\rm d} S$ can be used interchangeably.

Reformulating the problem to include the potential perturbations is straightforward due to the similarity of equations (\ref{eq:cast_source}) and (\ref{eq:source_dpsi_combined}).
As before, in general equation (\ref{eq:source_dpsi_combined}) cannot be directly inverted and we have to proceed by minimizing some penalty function.
Here we define such a function similarly to equation (\ref{eq:penalty_source}), including an additional regularization term for the potential perturbations in the same way as for the source:
\begin{equation}
\label{eq:penalty_source_dpsi}
\begin{split}
G(\boldsymbol{r}) \equiv & \, G(\boldsymbol{r}|\boldsymbol{d},\boldsymbol{s}_{\rm p},\psi_{\rm p},\boldsymbol{g}_{\rm s},\lambda_{\rm s},\boldsymbol{g}_{\rm \delta\psi},\lambda_{\rm \delta\psi}) \\
                     =   & \, \frac{1}{2}(M_{\rm r} \boldsymbol{r} - \boldsymbol{d})^T C^{-1}_{\rm d} (M_{\rm r} \boldsymbol{r} - \boldsymbol{d}) + \frac{1}{2}\boldsymbol{r}^T R \, \boldsymbol{r},
\end{split}
\end{equation}
where:
\begin{equation}
\label{eq:combined_reg}
R =
\begin{pmatrix}
\lambda_{\rm s} C^{-1}_{\rm s} & 0 \\
0 & \lambda_{\rm \delta\psi} C^{-1}_{\rm \delta\psi} \\
\end{pmatrix}.
\end{equation}
We underline again the important difference with equation (\ref{eq:penalty_source}), which is the dependence on a previously known source, $\boldsymbol{s}_{\rm p}$ (through $M_{\rm r}$).
This equation has the same form as equation (\ref{eq:penalty_source}), and the condition $\nabla_{\rm r} G = 0$ leads to:
\begin{equation}
\label{eq:min_r}
(M_{\rm r}^T C^{-1}_d M_{\rm r} + R) \boldsymbol{r} = M_{\rm r}^T C^{-1}_{d} \boldsymbol{d},
\end{equation}
which can be solved for $\boldsymbol{r}$ by inverting the positive-definite matrix on the left hand side.

\subsection{Bayesian framework}
The number of free parameters involved in the lens potential and source reconstruction may vary between different models.
For example, one may choose different parametric models for the smooth mass distribution, with or without additional perturbations, and regularization schemes (see Section \ref{sec:reg}).
As in \citet{Suyu2006} and \citet{Vegetti2009a}, we use a Bayesian approach to quantitatively justify the inclusion of extra free parameters and compare models to find the one most consistent with the data - assuming all quantities are Gaussian processes.
By recasting the problem in Bayesian terms, the evidence term necessary to compare models can be computed.
In addition, the solutions for the source and the perturbations obtained in the previous section, which minimize the penalty function, coincide with the most probable solutions that maximize the posterior probability.
A similar treatment is followed in \cite{Suyu2006} and \cite{Vegetti2009a}, however, here we explicitly derive the expression for the evidence.

Bayes' theorem states that the posterior probability density of the source and potential perturbations given the data, lensing operator, and some form of prior (regularization) described by parameters $\boldsymbol{g}$ and $\lambda$ is:
\begin{equation}
\label{eq:bayes}
\begin{split}
P(\boldsymbol{r}) \equiv & \, P(\boldsymbol{r}|\boldsymbol{d},\boldsymbol{\eta},\boldsymbol{g}_{\rm s},\boldsymbol{g}_{\rm \delta\psi},\lambda_{\rm s},\lambda_{\rm \delta\psi}) \\
 = & \,
\frac{
	P(\boldsymbol{d}|\boldsymbol{r},\boldsymbol{\eta}) \;
	P(\boldsymbol{s}|\boldsymbol{g}_{\rm s},\lambda_{\rm s}) \;
	P(\delta\boldsymbol{\psi}|\boldsymbol{g}_{\rm \delta\psi},\lambda_{\rm \delta\psi})
}{
	\cal{E}(\boldsymbol{d}|\boldsymbol{\eta},\boldsymbol{g}_{\rm s},\boldsymbol{g}_{\rm \delta\psi},\lambda_{\rm s},\lambda_{\rm \delta\psi})
},
\end{split}
\end{equation}
where the numerator terms are the likelihood, source prior, and potential perturbations prior respectively, and the denominator is the evidence.
Assuming the likelihood and priors are normal distributions and associating them with the previously introduced $\chi^2$ and regularization terms, their individual probability densities can be written as:
\begin{align}
\label{eq:prob_densities}
P(\boldsymbol{d}|\boldsymbol{r},\boldsymbol{\eta})  = \, &\frac{1}{Z_{\rm d}}\exp[-\frac{1}{2}(M_{\rm r} \boldsymbol{r} - \boldsymbol{d})^T C^{-1}_{\rm d} (M_{\rm r} \boldsymbol{r} - \boldsymbol{d})], \nonumber \\
P(\boldsymbol{s}|\boldsymbol{g}_{\rm s},\lambda_{\rm s}) 			= \, & \frac{1}{Z_{\rm s}}\exp[-\frac{1}{2}\lambda_{\rm s} \boldsymbol{s}^T C^{-1}_{\rm s} \boldsymbol{s}], \nonumber \\
P(\delta\boldsymbol{\psi}|\boldsymbol{g}_{\rm \delta\psi},\lambda_{\rm \delta\psi}) = \, &  \frac{1}{Z_{\rm \delta\psi}}\exp[-\frac{1}{2}\lambda_{\rm \delta\psi} \delta\boldsymbol{\psi}^T C^{-1}_{\rm \delta\psi} \delta\boldsymbol{\psi}],
\end{align}
where the normalization factors are given by:
\begin{align}
\label{eq:prob_density_norms}
Z_{\rm d} = \,			& (2\pi)^{N_{\rm d} / 2} (\mathrm{det} C_{\rm d} )^{1/2}, \nonumber \\
Z_{\rm s}(\boldsymbol{g}_{\rm s},\lambda_{\rm s}) = \, 			& (\frac{2\pi}{\lambda_{\rm s}})^{N_{\rm s} / 2} (\mathrm{det} C_{\rm s} )^{1/2}, \nonumber \\
Z_{\rm \delta\psi}(\boldsymbol{g}_{\rm \delta\psi},\lambda_{\rm \delta\psi}) = \, & (\frac{2\pi}{\lambda_{\rm \delta\psi}})^{N_{\rm \delta\psi} / 2} (\mathrm{det} C_{\rm \delta\psi} )^{1/2}.
\end{align}
The above set of equations assumes that we already have a decent solution for the source, $\boldsymbol{s}_{\rm p}$, in order to derive $M$ (see equation \ref{eq:combined_M}), which could come, for example, by solving the smooth version of the problem \citep[see][]{Koopmans2005}.
The most probable solution - the one that maximizes the posterior probability - can be derived by requiring $\nabla_{\rm r} P = 0$ in equation (\ref{eq:bayes}), and it can be calculated independently of the evidence term (a constant factor in this case).
This is the solution that also minimizes the penalty function in equation (\ref{eq:penalty_source_dpsi}), which has already been given in equation (\ref{eq:min_r}).

The posterior in equation (\ref{eq:bayes}) is the product of equations (\ref{eq:prob_densities}), hence it is itself a normal distribution and can be written as:
\begin{equation}
\label{eq:post_gauss}
P(\boldsymbol{r}) = \frac{1}{Z_{\rm G}} \exp[-G(\boldsymbol{r})],
\end{equation}
where $G(\boldsymbol{r})$ is given in equation (\ref{eq:penalty_source_dpsi}).
Taking a Taylor expansion of $G$ around the most probable solution $\boldsymbol{r}_{\rm MP}$, which satisfies $\nabla_{\rm r} G = 0$ (equation \ref{eq:min_r}), we get:
\begin{equation}
\label{eq:taylor_g}
G(\boldsymbol{r}) = G(\boldsymbol{r}_{\rm MP}) + \frac{1}{2} (\boldsymbol{r} - \boldsymbol{r}_{\rm MP})^T H \, (\boldsymbol{r} - \boldsymbol{r}_{\rm MP}),
\end{equation}
where $H$ is the Hessian of $G$:
\begin{equation}
\label{eq:hessian}
H \equiv \nabla_{\rm r}^{2} G =  M_{\rm r}^T C^{-1}_{\rm d} \, M_{\rm r} + R.
\end{equation}
Equation (\ref{eq:taylor_g}) is in fact exact - assuming we already know $\boldsymbol{s}_{\rm p}$ - because all terms in equation (\ref{eq:penalty_source_dpsi}) are quadratic in $\boldsymbol{r}$.
Equation (\ref{eq:post_gauss}) can now be rewritten as:
\begin{equation}
\label{eq:post_gauss_rewritten}
P(\boldsymbol{r}) = \frac{1}{Z_{\rm G}} \exp[-G(\boldsymbol{r}_{\rm MP}) -\frac{1}{2}(\boldsymbol{r} - \boldsymbol{r}_{\rm MP})^T H \, (\boldsymbol{r} - \boldsymbol{r}_{\rm MP})],
\end{equation}
where:
\begin{equation}
\label{eq:post_gauss_norm}
\begin{split}
Z_{\rm G} \equiv & \, Z_{\rm G}(\boldsymbol{\eta},\boldsymbol{g}_{\rm s},\lambda_{\rm s},\boldsymbol{g}_{\rm \delta\psi},\lambda_{\rm \delta\psi}) \\
			   = & \, e^{-G(\boldsymbol{r}_{\rm MP})} (2\pi)^{(\mathrm{N}_{\rm s}+\mathrm{N}_{\rm \delta\psi})/2} (\mathrm{det} H)^{-1/2}.
\end{split}
\end{equation}
Combining equations (\ref{eq:penalty_source_dpsi}), (\ref{eq:prob_densities}), (\ref{eq:taylor_g}), and (\ref{eq:post_gauss_rewritten}) the evidence term from equation (\ref{eq:bayes}) can be computed for the most probable solution $\boldsymbol{r}_{\rm MP}$:
\begin{equation}
\label{eq:evidence_first}
\mathcal{E}(\boldsymbol{d}|\boldsymbol{\eta},\boldsymbol{g}_{\rm s},\boldsymbol{g}_{\rm \delta\psi},\lambda_{\rm s},\lambda_{\rm \delta\psi}) = \frac{Z_{\rm G}(\boldsymbol{\eta},\boldsymbol{g}_{\rm s},\lambda_{\rm s},\boldsymbol{g}_{\rm \delta\psi},\lambda_{\rm \delta\psi})}
{Z_{\rm d} Z_{\rm s}(\boldsymbol{g}_{\rm s},\lambda_{\rm s}) Z_{\rm \delta\psi}(\boldsymbol{g}_{\rm \delta\psi},\lambda_{\rm \delta\psi})}.
\end{equation}
Substituting the normalization factors from equations (\ref{eq:prob_density_norms}) and (\ref{eq:post_gauss_norm}), and taking the logarithm of the evidence we get:
\begin{align}
\label{eq:evidence}
\log \mathcal{E} =& \, - \frac{\mathrm{N}_{\rm d}}{2} \log (2\pi) + \frac{\mathrm{N}_{\rm s}}{2} \log (\lambda_{\rm s}) + \frac{\mathrm{N}_{\rm \delta\psi}}{2} \log (\lambda_{\rm \delta\psi}) \nonumber \\
		& \, - \frac{1}{2} \log (\det C_{\rm d}) - \frac{1}{2} \log (\det C_{\rm s}) - \frac{1}{2} \log (\det C_{\rm \delta\psi}) \nonumber \\
		& \, - G(\boldsymbol{r}_{\rm MP}) - \frac{1}{2} \log (\det H).
\end{align}
Computing and comparing this value for models with different sets of parameters $\boldsymbol{\eta}$, $\boldsymbol{g}_{\rm s}$, and $\boldsymbol{g}_{\rm \delta\psi}$ allows one to rank the different mass models and regularization schemes, finding the combination most consistent with the data \citep{MacKay2003}.

\subsection{Regularization schemes}
\label{sec:reg}
Adding regularization terms to the penalty function (equation \ref{eq:penalty_source_dpsi}), or equivalently using priors in the posterior probability density (equation \ref{eq:bayes}), is necessary in order to find a solution for the source and the potential perturbations by inverting the matrices in equations (\ref{eq:min_source}) and (\ref{eq:min_r}).
Quadratic terms (Gaussian priors), such as the ones used here, as opposed to other forms of regularization\footnote{\citet{Wayth2005a} used maximum entropy regularization that has the benefit to prevent negative values for the source at the cost of not having quadratic penalty functions anymore. The solution minimizing the penalty function has to found numerically at a higher computational cost.}, have the advantage of leading to linear equations that have exact and efficient to calculate analytic solutions (equations \ref{eq:min_source} and \ref{eq:min_r}), and put the problem in the context of Gaussian Process Regression \citep{Rasmussen2006}.

The effect of the regularization on the source and perturbation fields is captured in the detailed structure of the generic covariance matrices $C_{\rm s}$ and $C_{\rm \delta\psi}$, while the overall contribution to the penalty function (posterior probability) is moderated by the $\lambda_{\rm s}$ and $\lambda_{\rm \delta\psi}$ parameters.
Here we examine different physically motivated forms of the covariance matrices $C_{\rm s}$ and $C_{\rm \delta\psi}$, and because the treatment is the same for both source and perturbations, we simply use $C$ and $\lambda$ in the following.

The usual forms of regularization impose some sort of ``smoothness'' condition on the solution \citep[see][]{Press1992}.
Choices in the literature are based on source derivatives of some order \citep[e.g.][]{Dye2005,Suyu2006,Vegetti2009a,Tagore2014,Nightingale2015,Yildirim2020}.
For example, a zero-th order derivative of the source \citep[the usual Tikhonov regularization, or ridge regression,][]{Tikhonov1963} means that $C$ is the identity matrix and brightness values are derived from a normal distribution centered on zero with standard deviation $\lambda^{-1/2}$.
Similarly, the gradient and curvature regularizations constrain the corresponding source derivatives, imposing a varying degree of smoothness to the solution.
However, although such schemes are useful to find a solution to the problem, they are not physically motivated (there is no reason for the gradient or curvature of a galaxy's brightness profile to follow a normal distribution centered at zero or any other value), may introduce spurious properties to the solutions, and cause degeneracies between the source and the lens potential.
In other words, the assumed covariance matrix resulting from
these choices imposes a correlation function (or power spectrum) on the source or
potential perturbations that might not reflect reality.

A more realistic and general approach can involve covariance matrices $C$ that do not correspond to any particular derivative and impose a physically motivated structure, via its covariance, directly on the solutions.
%In fact, the source and the perturbations can be modelled as Gaussian random processes with their properties defined by the covariance function \citep[e.g.][]{Gelman2013}.
Here we examine two forms of such covariance kernels:
\begin{align}
\label{eq:covariance_exp}
C(\boldsymbol{y}_{\rm i},\boldsymbol{y}_{\rm j},l) =& \exp \left( -\frac{d_{\rm i,j}}{l} \right), 			& \mathrm{(exponential)} \\
\label{eq:covariance_exp_squared}
C(\boldsymbol{y}_{\rm i},\boldsymbol{y}_{\rm j},l) =& \exp \left( -\frac{d_{\rm i,j}^2}{2 \, l^2} \right), 	& \mathrm{(Gaussian)}
\end{align}
where $\boldsymbol{y}$ are the source pixel coordinates, $d_{\rm i,j}$ the Euclidean distance between pixels i and j, and $l$ is the characteristic correlation length of the kernels \citep{Rasmussen2006}.
These two choices (also known as Ornstein-Uhlenbeck and squared exponential kernels respectively) constitute two opposite extremes of the more general Mat\'{e}rn kernel \citep[e.g.][]{Mertens2017}, and have a single free parameter, $l$ (which belongs to the $\boldsymbol{g}$ set of non-linear regularization parameters), which gives more freedom for additional structure in the covariance matrices $C$ beyond the fixed-form derivative-based regularization.
Also, these covariance kernels appear in better agreement with various sources, as it will be shown by two examples later on.
The variance level (i.e. the diagonal of the covariance
matrix) is set by $\lambda$, and hence we assume here the diagonal values of of $C$ are
by definition equal to unity.

The potential perturbations given in equation (\ref{eq:min_r}) provide a measure of sub-galactic scale mass density fluctuations.
The covariance matrix $C_{\rm \delta\psi}$ is equivalent to the correlation function (or two-point correlation function), which is related to the power spectrum via the Fourier transform.
Hence, measuring the covariance of $\delta\boldsymbol{\psi}$ can probe the sub-galactic matter power spectrum.
Although there have been theoretical and applied studies on this connection  \citep[][]{Hezaveh2016a,DiazRivero2018,Chatterjee2018,Bayer2018}, this work is the first consistent approach using the gravitational imaging technique.
The derived power spectrum/covariance of $\delta\boldsymbol{\psi}$ can then be associated to higher order moments in the lens mass distribution \citep[e.g.][]{Hsueh2017,Gilman2018} or dark matter substructure \citep[e.g.][]{Hezaveh2016a}.
For the latter, a more realistic approach to disentangle the effect of baryons would be to compare the observed sub-galactic scale perturbations to predictions from hydrodynamical simulations \citep[e.g.][]{Vogelsberger2014,Schaye2015}.

\subsection{Optimization}
\label{sec:optimization}
There are three main components in our approach to modelling gravitational lenses: the parametrized smooth lens potential, $\psi$, the grid-based potential perturbations, $\delta\boldsymbol{\psi}$, and the grid-based source brightness, $\boldsymbol{s}$.
The task is to find the linear solutions and non-linear parameter values that are the most consistent with the data, i.e. maximizing the evidence.
The linear part of the problem provides an exact solution for $\boldsymbol{s}$ and $\delta\boldsymbol{\psi}$ - assuming that we already know $\boldsymbol{s}_{\rm p}$ - that minimizes the penalty function and maximizes the posterior (equation \ref{eq:min_r}), for fixed non-linear parameters.
Here we describe our treatment of the non-linear parameters, namely, the smooth potential parameters $\boldsymbol{\eta}$, and the regularization parameters $\boldsymbol{g}$ and $\lambda$ for the source and the potential perturbations.

Firstly, we emphasize that the lens potential is dominated by the smooth model and any resulting perturbations are required to be small in order for equation (\ref{eq:dpsi_residuals}) to be valid.
This is also motivated by decent agreement between data and smooth models \citep[e.g.][]{Koopmans2009,Auger2010,Suyu2014,Oldham2018}, as well as evidence for lens perturbations \citep[e.g.][]{Vegetti2012,MacLeod2013,Nierenberg2014,Hezaveh2016b,Birrer2017}.
Solving simultaneously for $\boldsymbol{\eta}$ and $\delta\boldsymbol{\psi}$, however, is degenerate\footnote{The $\delta\boldsymbol{\psi}$ can in principle mimic almost any potential
$\psi(\boldsymbol{\eta})$ and hence only the sum of the total potential is relevant. In practice however, $\psi(\boldsymbol{\eta})$ is set by general
processes of galaxy formation and phase-space mixing and is expected to be smooth, while $\delta\boldsymbol{\psi}$ describes any remaining structure such as sub-halos, streams,	etc.} and very inefficient; if $\boldsymbol{\eta}$ is far from the truth then $\delta\boldsymbol{\psi}$ will try to make up for the correct sum of the smooth potential and the perturbations, leading away from a realistic solution.
Hence, as a first step we optimize for the parameters $\boldsymbol{\eta}$ assuming $\delta\boldsymbol{\psi} = 0$, while simultaneously solving the linear equations for the source (equation \ref{eq:min_source}).
The parameter space of $\boldsymbol{\eta}$ is explored using a nested-sampling approach \citep{Skilling2004}, which provides several benefits: it computes the evidence term in equation (\ref{eq:evidence}) with the $\delta\boldsymbol{\psi}=0$ assumption, finds the most probable parameters, provides confidence intervals, and explores a large part of the parameter space with a limited chance of getting stuck in local extrema.
There is the additional option to start a Monte Carlo Markov Chain exploration of the parameter space near the most probable solution to obtain smoother posterior probability distributions.

Once the smooth model is determined, the parameters $\boldsymbol{\eta}$ are kept fixed to their maximum a posteriori values and solving for $\boldsymbol{r}$ (the potential perturbations and the source) is conducted.
The varying non-linear parameters are now the regularization parameters $\boldsymbol{g}$ and $\lambda$, together describing the source and perturbation covariance matrices.
Although it is possible to solve approximately for the $\lambda$ parameters, at least in the case with $\delta\boldsymbol{\psi}=0$ \citep{Koopmans2005,Suyu2006}, here we incorporate them in the full non-linear treatment.
This allows one to infer confidence intervals and, most importantly, degeneracies between the source and the potential solutions.

The perturbations investigated here are assumed to be small and originate from an extended field of mass density fluctuations permeating the lens, as opposed to specific and localized massive substructures, such as dark sub-halos.
For such prominent and confined perturbers, an iterative approach\footnote{At the end of each iteration, the lens potential is updated by adding the newly determined $\delta\boldsymbol{\psi}$ and the $D_{\rm s}(\boldsymbol{s_{\rm p}})$ matrix in equation (\ref{eq:combined_M}) is recalculated based on the derivatives of the newly determined source.} would indeed be expected to perform better in locating and measuring the mass of putative massive substructures, carefully controlling the regularization parameters in the process \citep[e.g.][]{Suyu2006,Vegetti2009a,Nightingale2015,Hezaveh2016b}.
In this work, however, we do not assume any restriction on the regularization parameters and solve for $\boldsymbol{r}$ for each set of sampled non-linear parameters without updating the lens potential and the source.
This approach is sufficient to capture the statistical properties of the perturbations field, provided that its amplitude is small.
Mixing the two approaches, i.e. sampling the regularization parameters and then iterating up to a given number of steps for each combination, could be another possibility, especially when the extended field of perturbations also includes prominent mass concentrations such as sub-halos, but this is out of the scope of this paper.

\begin{figure*}
	\includegraphics[width=\textwidth]{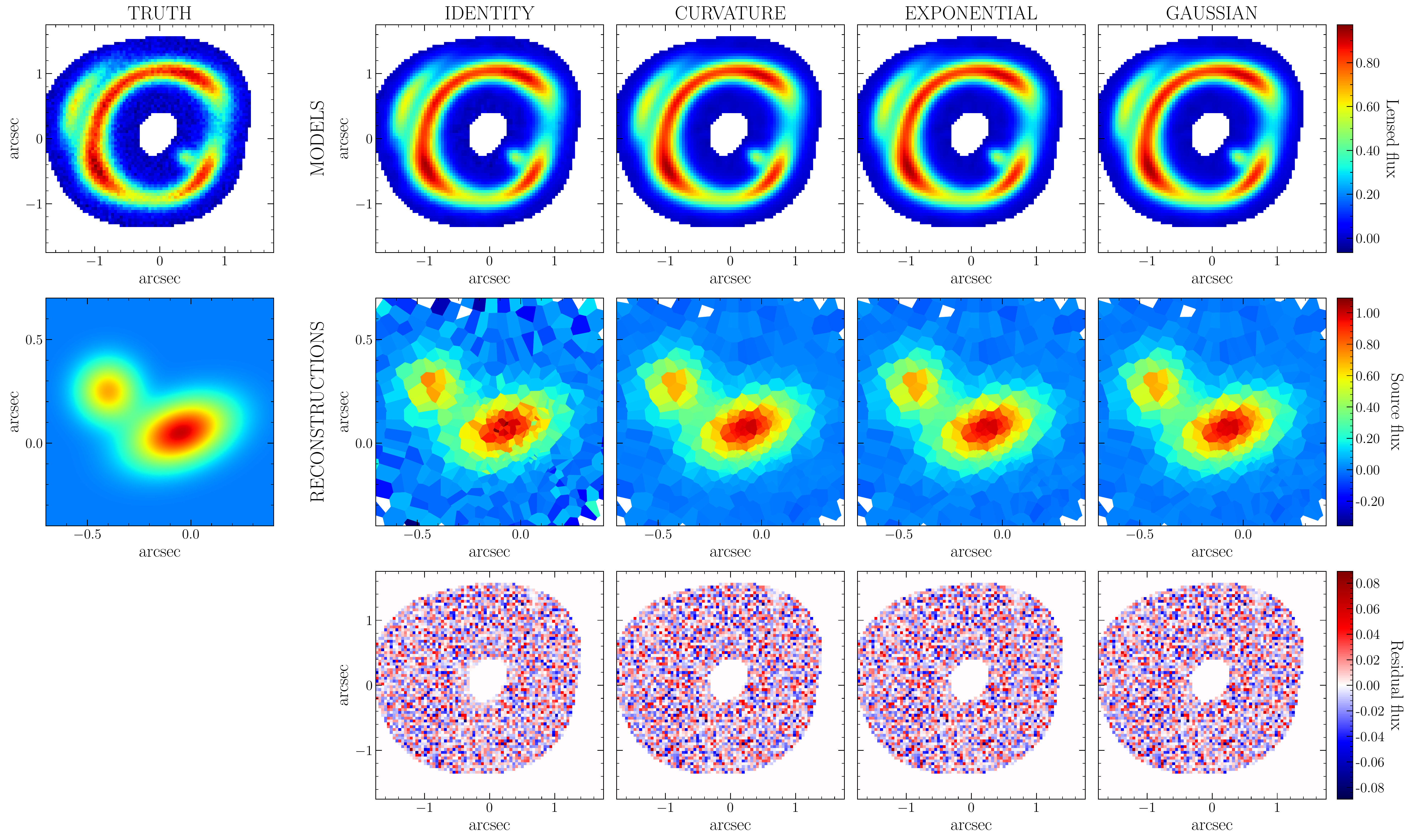}
	\caption{Lensed images (top), source (middle), and residuals (bottom) for the mock data (truth) and the reconstructions with different source regularization kernels: identity, curvature, exponential, and Gaussian. The source brightness, shown as Voronoi cells of an adaptive grid (see Section \ref{sec:grids}), has been reconstructed using $n=3$. The corresponding parameters for the lens potential and the source regularization are shown in Table \ref{tab:table_map}.}
	\label{fig:results_smooth_images}
\end{figure*}

%%%%%%%%%%%%%%%%%%%%%%%%%%%%%%%%%%%%%%%%%%%%%%%%%%%%%%%%%%%%%%%%%%%%%%%%%%%%%%%%%%%%%%%%%%%%%%%%%%%%%%%%%%%%%%%%%%%%%%%%%%%%%%%%%%%%%%%%%%%%%%%%%%%%%%
\section{Results}
\label{sec:results}
In order to demonstrate the capabilities of our method, we examine modelling aspects of mock lenses combining smooth and complex lens potentials and source light profiles.
In all cases, we use a point spread function (PSF) simulated for the Hubble Space Telescope (HST) using the \textsc{tiny-tim}\footnote{\url{http://www.stsci.edu/hst/observatory/focus/TinyTim}} software \citep{Krist2010}, which is assumed to be the same in creating and modelling the mock data, uniform Gaussian random noise with a signal-to-noise ratio of $\approx$40 at peak brightness, and a mask to exclude regions of the image without lensing features (also the central part of the image that may hold residuals after removing the lens galaxy light, which we do not include or model).
We generated the mocks using the MOLET\footnote{\url{https://github.com/gvernard/molet}} software package \citep{Vernardos2021}.

The smooth parametric model used for the lens potential is a Singular Isothermal Ellipsoid \citep[SIE,][]{Kassiola1993,Kormann1994}.
We follow the notation of \citet{Schneider2006}, with convergence given by:
\begin{equation}
\label{eq:kappa_sie}
\kappa(\omega) = \frac{b}{2 \omega},
\end{equation}
where $\omega = \sqrt{q^2 x^2 + y^2}$, $q$ is the minor to major axis ratio, and $b$ (in arcsec) describes the overall potential strength\footnote{We set $b=\sqrt{q} \, \theta_{\mathrm{Ein}}$, where the Einstein radius, $\theta_{\mathrm{Ein}}$, is defined as the radius within which the integral of equation (\ref{eq:kappa_sie}) becomes equal to unity.}.
This relation holds in the reference system whose x-axis is aligned with the ellipsoid's major axis, rotated by the position angle, $\theta$, and whose origin coincides with the lens center ($x_0,y_0$).
External shear with magnitude $\gamma$ and direction $\phi$ is included, leading to 7 free parameters in total, hereafter denoted as $\boldsymbol{\eta}$.
All angles are measured east-of-north, in order to remain consistent with the standard celestial definition.

\renewcommand{\arraystretch}{1.2}
\begin{table*}
	\caption{Values of the lens potential ($\boldsymbol{\eta}$) and source regularization parameters ($\lambda_{\rm s},\boldsymbol{g}_{\rm s}$) that maximize the posterior probability density, i.e. Maximum A Posteriori (MAP) values, and corresponding terms from equation (\ref{eq:evidence}). The smooth source (top part of the table) is described in Section \ref{sec:model_smooth} and the lensed images corresponding to the parameters listed here are shown in Fig. \ref{fig:results_smooth_images}. Similarly, NGC3982 and NGC2623 (middle and bottom parts) are described in Section \ref{sec:smooth_complex} and shown in Figs. \ref{fig:results_spiral_images}, and \ref{fig:results_merger_images}.}
	\label{tab:table_map}
	\begin{threeparttable}
		\begin{tabular}{rrr@{\hskip 0.8cm}rrrr@{\hskip 0.8cm}rrrr@{\hskip 0.8cm}rrrr}
			& name & units & Truth & Identity & Curvature & Exponential & Gaussian \\ 
\hline
\multirow{16}{*}{\begin{sideways}Smooth source\end{sideways}}&$b$ & arcsec & 0.9 & $   0.894$ & $   0.897$ & $   0.898$ & $   0.897$ \\ 
&$q$ & - & 0.8 & $   0.790$ & $   0.795$ & $   0.796$ & $   0.795$ \\ 
&$\theta$ & $^\circ$ & -135 & $-136.208$ & $-135.031$ & $-135.176$ & $-135.135$ \\ 
&$x_0$ & arcsec & 0 & $  -0.022$ & $  -0.021$ & $  -0.021$ & $  -0.021$ \\ 
&$y_0$ & arcsec & 0 & $   0.022$ & $   0.022$ & $   0.022$ & $   0.022$ \\ 
&$\gamma$ & - & 0.03 & $   0.028$ & $   0.029$ & $   0.029$ & $   0.029$ \\ 
&$\phi$ & $^\circ$ & -40 & $ -37.278$ & $ -39.436$ & $ -39.352$ & $ -39.340$ \\ 
\cline{2-8}
&$\lambda_{\rm s}$ & - & - & $   7.945$ & $   0.121$ & $  29.379$ & $  86.581$ \\ 
&$l_{\rm s}$ & arcsec & - & - & - & $   0.675$ & $   0.128$ \\ 
\cline{2-8}
&\multicolumn{3}{r}{$-\frac{\mathrm{N}_{\rm d}}{2} \log(2\pi)$\tnote{$\dagger$}} & -3536.08 & -3536.08 & -3536.08 & -3536.08  \\ 
&\multicolumn{3}{r}{$\frac{\mathrm{N}_{\rm s}}{2} \log(\lambda_{\rm s})$} &   755.44 &  -769.81 &  1232.11 &  1626.06  \\ 
&\multicolumn{3}{r}{$-\frac{1}{2} \log(\det C_{\rm d})$\tnote{$\dagger$}} & 24020.04 & 24020.04 & 24020.04 & 24020.04  \\ 
&\multicolumn{3}{r}{$-\frac{1}{2} \log(\det C_{\rm s})$} &     0 &  3175.55 &   907.62 &   583.66  \\ 
&\multicolumn{3}{r}{$-\frac{1}{2}\chi^2$} & -1648.13 & -1797.81 & -1765.54 & -1765.10  \\ 
&\multicolumn{3}{r}{$-\frac{1}{2}\lambda_{\rm s} \boldsymbol{s}^T C_{\rm s}^{-1} \boldsymbol{s}$} &  -257.99 &  -109.26 &  -133.33 &  -127.33  \\ 
&\multicolumn{3}{r}{$-\frac{1}{2} \log(\det H)$} & -1916.13 & -2757.57 & -2525.20 & -2556.19  \\ 
&\multicolumn{3}{r}{$\log P$} & 17417.16 & 18225.07 & 18199.64 & 18245.07  \\ 
\hline
\hline
\multirow{16}{*}{\begin{sideways}NGC3982\end{sideways}}&$b$ & arcsec & 0.9 & $   0.896$ & $   0.891$ & $   0.895$ & $   0.895$ \\ 
&$q$ & - & 0.8 & $   0.793$ & $   0.785$ & $   0.791$ & $   0.791$ \\ 
&$\theta$ & $^\circ$ & -135 & $-134.743$ & $-133.491$ & $-133.837$ & $-134.473$ \\ 
&$x_0$ & arcsec & 0 & $  -0.023$ & $  -0.021$ & $  -0.022$ & $  -0.022$ \\ 
&$y_0$ & arcsec & 0 & $   0.023$ & $   0.025$ & $   0.024$ & $   0.024$ \\ 
&$\gamma$ & - & 0.03 & $   0.029$ & $   0.027$ & $   0.028$ & $   0.029$ \\ 
&$\phi$ & $^\circ$ & -40 & $ -40.203$ & $ -41.950$ & $ -41.619$ & $ -40.533$ \\ 
\cline{2-8}
&$\lambda_{\rm s}$ & - & - & $  16.172$ & $   0.126$ & $  69.969$ & $  56.937$ \\ 
&$l_{\rm s}$ & arcsec & - & - & - & $   0.385$ & $   0.194$ \\ 
\cline{2-8}
&\multicolumn{3}{r}{$-\frac{\mathrm{N}_{\rm d}}{2} \log(2\pi)$\tnote{$\dagger$}} & -4307.06 & -4307.06 & -4307.06 & -4307.06  \\ 
&\multicolumn{3}{r}{$\frac{\mathrm{N}_{\rm s}}{2} \log(\lambda_{\rm s})$} &  1014.51 &  -755.05 &  1548.42 &  1473.29  \\ 
&\multicolumn{3}{r}{$-\frac{1}{2} \log(\det C_{\rm d})$\tnote{$\dagger$}} & 25672.50 & 25672.50 & 25672.50 & 25672.50  \\ 
&\multicolumn{3}{r}{$-\frac{1}{2} \log(\det C_{\rm s})$} &     0 &  3181.83 &   711.78 &   684.29  \\ 
&\multicolumn{3}{r}{$-\frac{1}{2}\chi^2$} & -2083.00 & -2383.13 & -2271.33 & -2207.01  \\ 
&\multicolumn{3}{r}{$-\frac{1}{2}\lambda_{\rm s} \boldsymbol{s}^T C_{\rm s}^{-1} \boldsymbol{s}$} &  -291.20 &  -151.58 &  -172.64 &  -194.16  \\ 
&\multicolumn{3}{r}{$-\frac{1}{2} \log(\det H)$} & -2335.42 & -2929.98 & -2766.41 & -2685.90  \\ 
&\multicolumn{3}{r}{$\log P$} & 17670.32 & 18327.53 & 18415.24 & 18435.95  \\ 
\hline
\hline
\multirow{16}{*}{\begin{sideways}NGC2623\end{sideways}}&$b$ & arcsec & 0.9 & $   0.900$ & $   0.903$ & $   0.901$ & $   0.901$ \\ 
&$q$ & - & 0.8 & $   0.802$ & $   0.808$ & $   0.804$ & $   0.803$ \\ 
&$\theta$ & $^\circ$ & -135 & $-135.151$ & $-132.769$ & $-134.165$ & $-134.464$ \\ 
&$x_0$ & arcsec & 0 & $  -0.020$ & $  -0.018$ & $  -0.020$ & $  -0.020$ \\ 
&$y_0$ & arcsec & 0 & $   0.020$ & $   0.019$ & $   0.021$ & $   0.021$ \\ 
&$\gamma$ & - & 0.03 & $   0.033$ & $   0.033$ & $   0.033$ & $   0.033$ \\ 
&$\phi$ & $^\circ$ & -40 & $ -39.932$ & $ -44.236$ & $ -41.881$ & $ -41.180$ \\ 
\cline{2-8}
&$\lambda_{\rm s}$ & - & - & $  25.623$ & $   0.016$ & $  64.835$ & $  37.979$ \\ 
&$l_{\rm s}$ & arcsec & - & - & - & $   0.078$ & $   0.065$ \\ 
\cline{2-8}
&\multicolumn{3}{r}{$-\frac{\mathrm{N}_{\rm d}}{2} \log(2\pi)$\tnote{$\dagger$}} & -3145.53 & -3145.53 & -3145.53 & -3145.53  \\ 
&\multicolumn{3}{r}{$\frac{\mathrm{N}_{\rm s}}{2} \log(\lambda_{\rm s})$} &  1182.25 & -1507.27 &  1520.64 &  1325.70  \\ 
&\multicolumn{3}{r}{$-\frac{1}{2} \log(\det C_{\rm d})$\tnote{$\dagger$}} & 25983.49 & 25983.49 & 25983.49 & 25983.49  \\ 
&\multicolumn{3}{r}{$-\frac{1}{2} \log(\det C_{\rm s})$} &     0 &  3179.82 &   223.29 &   328.69  \\ 
&\multicolumn{3}{r}{$-\frac{1}{2}\chi^2$} & -1705.31 & -1846.70 & -1735.70 & -1762.97  \\ 
&\multicolumn{3}{r}{$-\frac{1}{2}\lambda_{\rm s} \boldsymbol{s}^T C_{\rm s}^{-1} \boldsymbol{s}$} &  -213.42 &  -162.16 &  -169.34 &  -175.53  \\ 
&\multicolumn{3}{r}{$-\frac{1}{2} \log(\det H)$} & -2063.98 & -2292.50 & -2309.65 & -2250.27  \\ 
&\multicolumn{3}{r}{$\log P$} & 20037.50 & 20209.16 & 20367.20 & 20303.59  \\ 

		\end{tabular}
		\begin{tablenotes}\footnotesize
			\item [$\dagger$] constant
		\end{tablenotes}
	\end{threeparttable}
\end{table*}

\renewcommand{\arraystretch}{1.4}
\begin{table*}
	\centering
	\caption{Mean values and 68 per cent confidence intervals for the lens potential ($\boldsymbol{\eta}$) and source regularization parameters ($\lambda_{\rm s},\boldsymbol{g}_{\rm s}$), and corresponding evidence values. The smooth source (top part of the table) is described in Section \ref{sec:model_smooth}, while NGC3982 and NGC2623 (middle and bottom parts) are described in Section \ref{sec:smooth_complex}. The lens center appears shifted by about half a pixel in the x and y directions due to a corresponding shift in the PSF. The full probability densities for the Gaussian regularization model of the smooth source (top part of the table) are shown in Fig. \ref{fig:results_smooth_corner}.}
	\label{tab:table_mean}
	\begin{tabular}{rrrrrrrr}
		&name & units & Truth & Identity & Curvature & Exponential & Gaussian \\ 
\hline
\multirow{10}{*}{\begin{sideways}Smooth source\end{sideways}}
&$b$ & arcsec & 0.9 & $   0.894_{-   0.001}^{+   0.001}$ & $   0.897_{-   0.003}^{+   0.002}$ & $   0.898_{-   0.003}^{+   0.002}$ & $   0.897_{-   0.002}^{+   0.002}$ \\ 
&$q$ & - & 0.8 & $   0.790_{-   0.002}^{+   0.002}$ & $   0.795_{-   0.005}^{+   0.004}$ & $   0.796_{-   0.005}^{+   0.003}$ & $   0.795_{-   0.004}^{+   0.003}$ \\ 
&$\theta$ & $^\circ$ & -135 & $-136.208_{-   0.750}^{+   0.484}$ & $-135.031_{-   0.504}^{+   0.514}$ & $-135.176_{-   0.598}^{+   0.591}$ & $-135.135_{-   0.633}^{+   0.556}$ \\ 
&$x_0$ & arcsec & 0 & $  -0.022_{-   0.001}^{+   0.001}$ & $  -0.021_{-   0.001}^{+   0.001}$ & $  -0.021_{-   0.001}^{+   0.001}$ & $  -0.021_{-   0.001}^{+   0.001}$ \\ 
&$y_0$ & arcsec & 0 & $   0.022_{-   0.001}^{+   0.001}$ & $   0.022_{-   0.001}^{+   0.001}$ & $   0.022_{-   0.001}^{+   0.001}$ & $   0.022_{-   0.001}^{+   0.001}$ \\ 
&$\gamma$ & - & 0.03 & $   0.028_{-   0.001}^{+   0.001}$ & $   0.029_{-   0.001}^{+   0.001}$ & $   0.029_{-   0.001}^{+   0.001}$ & $   0.029_{-   0.001}^{+   0.001}$ \\ 
&$\phi$ & $^\circ$ & -40 & $ -37.278_{-   0.995}^{+   1.480}$ & $ -39.436_{-   0.993}^{+   1.101}$ & $ -39.352_{-   1.238}^{+   1.171}$ & $ -39.340_{-   1.184}^{+   1.183}$ \\ 
\cline{2-8}
&$\lambda_{\rm s}$ & - & - & $   7.958_{-   0.753}^{+   0.801}$ & $   0.122_{-   0.042}^{+   0.005}$ & $  30.355_{-   9.475}^{+   5.243}$ & $  86.992_{-   9.020}^{+   8.063}$ \\ 
&$l_{\rm s}$ & arcsec & - & - & - & $   0.695_{-   0.182}^{+   0.169}$ & $   0.129_{-   0.004}^{+   0.004}$ \\ 
\cline{2-8}  
&\multicolumn{3}{r}{log $\mathcal{E}$:} & $17388.46\pm0.58$ & $18190.35\pm0.55$ & $18167.62\pm0.55$ & $18208.74\pm0.57$  \\ 
\hline
\hline

\multirow{10}{*}{\begin{sideways}NGC3982\end{sideways}}
&$b$ & arcsec & 0.9 & $   0.896_{-   0.004}^{+   0.001}$ & $   0.891_{-   0.002}^{+   0.002}$ & $   0.895_{-   0.004}^{+   0.004}$ & $   0.895_{-   0.005}^{+   0.005}$ \\ 
&$q$ & - & 0.8 & $   0.793_{-   0.008}^{+   0.001}$ & $   0.785_{-   0.003}^{+   0.003}$ & $   0.791_{-   0.007}^{+   0.007}$ & $   0.791_{-   0.009}^{+   0.008}$ \\ 
&$\theta$ & $^\circ$ & -135 & $-134.743_{-   0.445}^{+   0.499}$ & $-133.491_{-   0.443}^{+   0.425}$ & $-133.837_{-   0.758}^{+   0.709}$ & $-134.473_{-   0.653}^{+   0.599}$ \\ 
&$x_0$ & arcsec & 0 & $  -0.023_{-   0.001}^{+   0.001}$ & $  -0.021_{-   0.001}^{+   0.001}$ & $  -0.022_{-   0.001}^{+   0.001}$ & $  -0.022_{-   0.001}^{+   0.001}$ \\ 
&$y_0$ & arcsec & 0 & $   0.023_{-   0.001}^{+   0.001}$ & $   0.025_{-   0.001}^{+   0.001}$ & $   0.024_{-   0.001}^{+   0.001}$ & $   0.024_{-   0.001}^{+   0.001}$ \\ 
&$\gamma$ & - & 0.03 & $   0.029_{-   0.002}^{+   0.000}$ & $   0.027_{-   0.001}^{+   0.001}$ & $   0.028_{-   0.002}^{+   0.002}$ & $   0.029_{-   0.002}^{+   0.002}$ \\ 
&$\phi$ & $^\circ$ & -40 & $ -40.203_{-   0.942}^{+   0.935}$ & $ -41.950_{-   1.089}^{+   1.134}$ & $ -41.619_{-   1.256}^{+   1.544}$ & $ -40.533_{-   1.007}^{+   1.435}$ \\
\cline{2-8}
&$\lambda_{\rm s}$ & - & - & $  16.197_{-   0.881}^{+   0.871}$ & $   0.126_{-   0.046}^{+   0.005}$ & $  73.260_{-  23.072}^{+  21.781}$ & $  57.157_{-   5.430}^{+   4.580}$ \\ 
&$l_{\rm s}$ & arcsec & - & - & - & $   0.408_{-   0.167}^{+   0.070}$ & $   0.194_{-   0.007}^{+   0.008}$ \\ 
\cline{2-8}
&\multicolumn{3}{r}{log $\mathcal{E}$:} & $17635.43\pm0.57$ & $18296.66\pm0.56$ & $18382.97\pm0.56$ & $18407.46\pm0.58$  \\
\hline
\hline

\multirow{10}{*}{\begin{sideways}NGC2923\end{sideways}}
&$b$ & arcsec & 0.9 & $   0.900_{-   0.001}^{+   0.001}$ & $   0.903_{-   0.002}^{+   0.002}$ & $   0.901_{-   0.001}^{+   0.001}$ & $   0.901_{-   0.001}^{+   0.001}$ \\ 
&$q$ & - & 0.8 & $   0.802_{-   0.002}^{+   0.002}$ & $   0.808_{-   0.004}^{+   0.004}$ & $   0.804_{-   0.002}^{+   0.002}$ & $   0.803_{-   0.002}^{+   0.002}$ \\ 
&$\theta$ & $^\circ$ & -135 & $-135.151_{-   0.376}^{+   0.302}$ & $-132.769_{-   0.521}^{+   0.439}$ & $-134.165_{-   0.747}^{+   0.516}$ & $-134.464_{-   0.614}^{+   0.452}$ \\ 
&$x_0$ & arcsec & 0 & $  -0.020_{-   0.001}^{+   0.001}$ & $  -0.018_{-   0.000}^{+   0.001}$ & $  -0.020_{-   0.001}^{+   0.001}$ & $  -0.020_{-   0.001}^{+   0.001}$ \\ 
&$y_0$ & arcsec & 0 & $   0.020_{-   0.001}^{+   0.001}$ & $   0.019_{-   0.001}^{+   0.001}$ & $   0.021_{-   0.001}^{+   0.001}$ & $   0.021_{-   0.001}^{+   0.001}$ \\ 
&$\gamma$ & - & 0.03 & $   0.033_{-   0.001}^{+   0.001}$ & $   0.033_{-   0.001}^{+   0.001}$ & $   0.033_{-   0.001}^{+   0.001}$ & $   0.033_{-   0.001}^{+   0.001}$ \\ 
&$\phi$ & $^\circ$ & -40 & $ -39.932_{-   0.532}^{+   0.747}$ & $ -44.236_{-   0.771}^{+   0.953}$ & $ -41.881_{-   0.906}^{+   1.327}$ & $ -41.180_{-   0.804}^{+   1.096}$ \\ 
\cline{2-8}
&$\lambda_{\rm s}$ & - & - & $  25.678_{-   1.717}^{+   1.626}$ & $   0.016_{-   0.005}^{+   0.001}$ & $  65.296_{-   7.873}^{+   7.285}$ & $  38.388_{-   5.801}^{+   5.278}$ \\ 
&$l_{\rm s}$ & arcsec & - & - & - & $   0.079_{-   0.013}^{+   0.009}$ & $   0.065_{-   0.006}^{+   0.005}$ \\ 
\cline{2-8}
&\multicolumn{3}{r}{log $\mathcal{E}$:} & $20028.44\pm0.59$ & $20221.07\pm0.58$ & $20330.95\pm0.59$ & $20296.56\pm0.59$  \\ 

	\end{tabular}
\end{table*}

\begin{figure}
	\centering
	\includegraphics[width=0.45\textwidth]{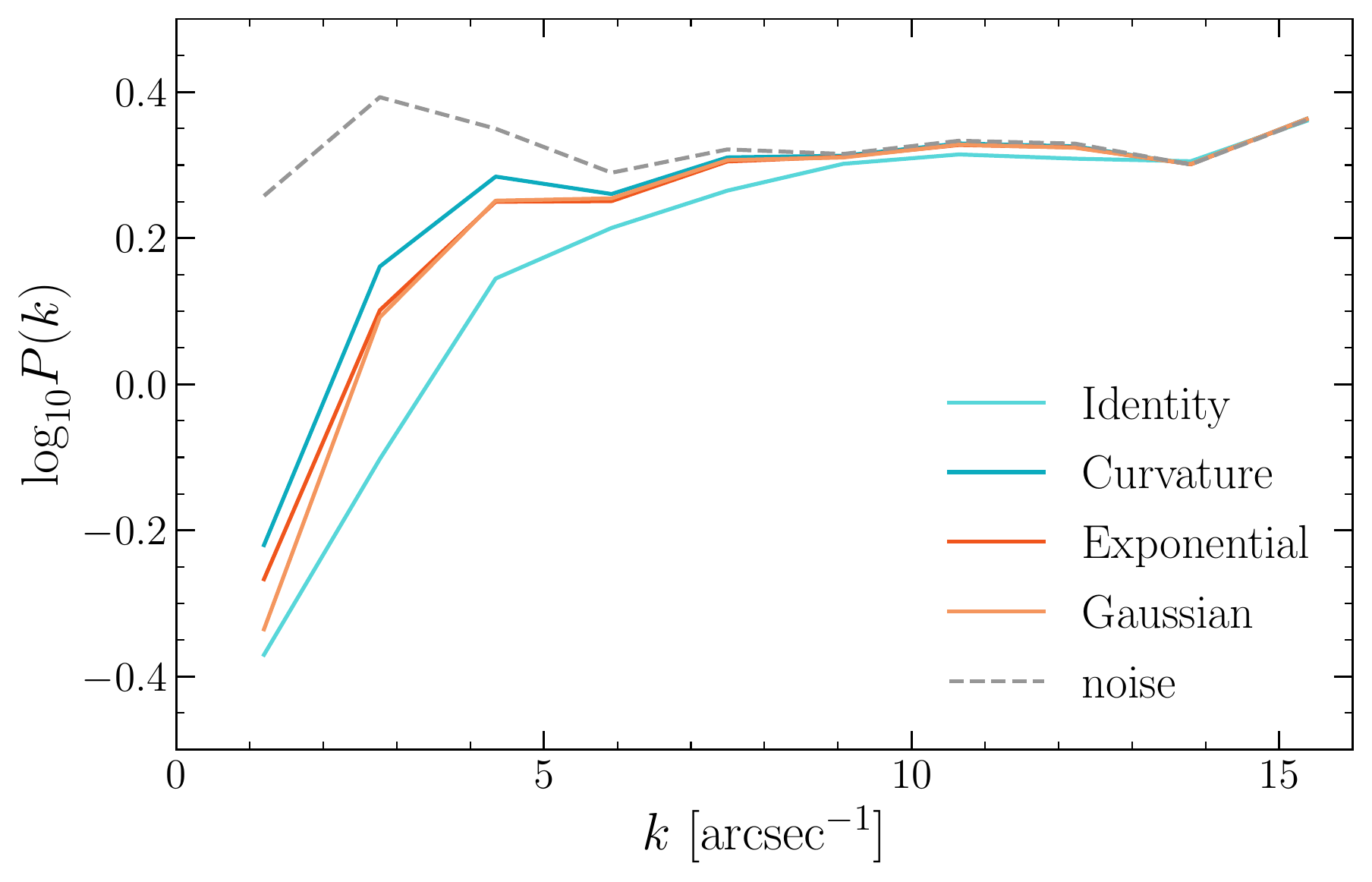}
	\caption{Fourier power spectrum of the model residuals shown in the bottom row of Fig. \ref{fig:results_smooth_images}.}
	\label{fig:results_smooth_res_ps}
\end{figure}

\begin{figure*}
	\centering
	\includegraphics[width=\textwidth]{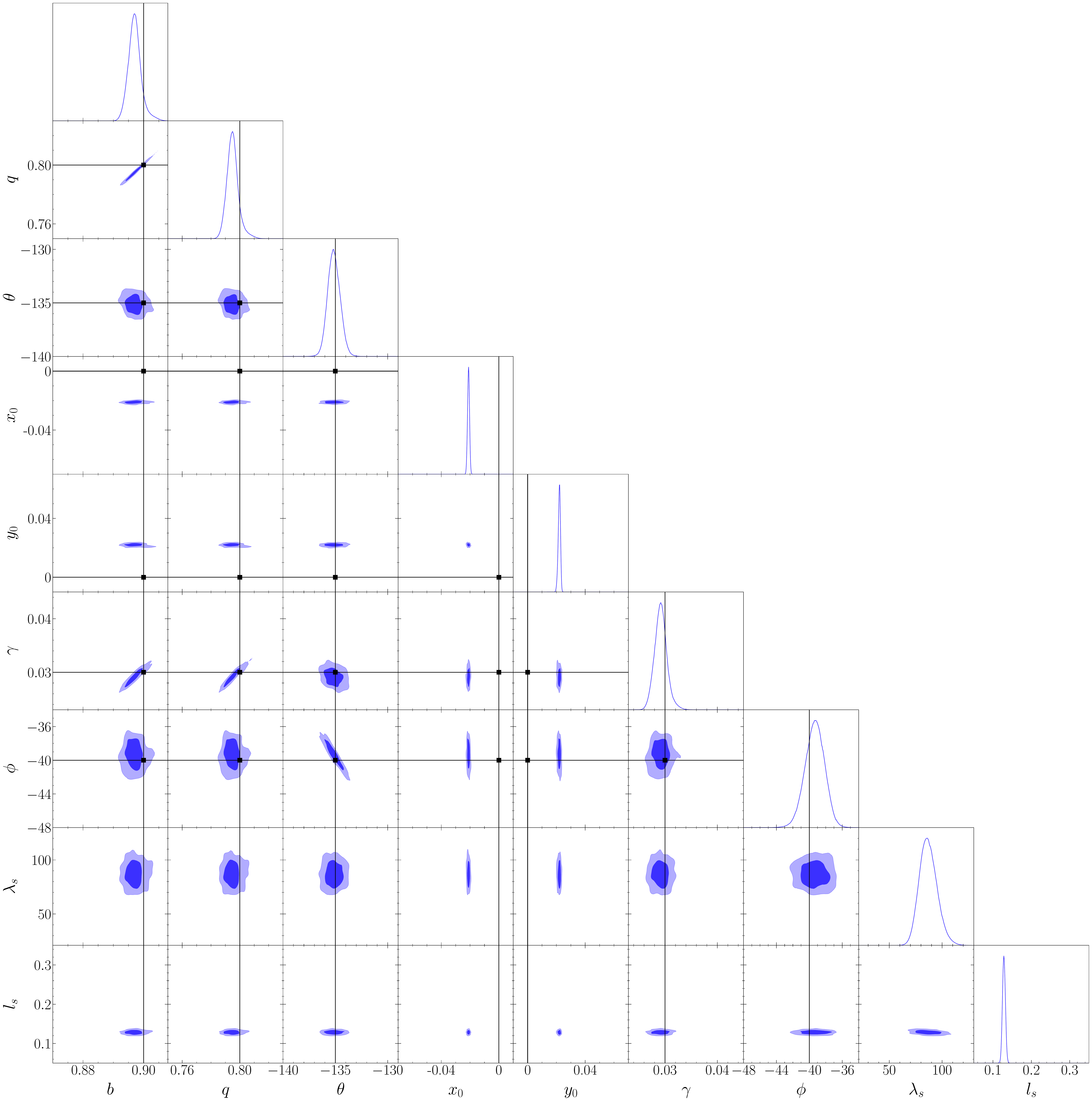}
	\caption{Marginalized probability densities and histograms for the lens potential ($\boldsymbol{\eta}$) and regularization ($\lambda_{\rm s},\boldsymbol{g}_{\rm s}$) parameters for the Gaussian kernel reconstruction of the smooth source described in Section \ref{sec:model_smooth}. The parameter ranges are set to match Fig. \ref{fig:results_combined_corner} and facilitate comparisons with the results described in Sections \ref{sec:model_dpsi} and \ref{sec:both} - a zoomed-in version of this plot that shows the shape of the two-dimensional distributions better is shown in Fig. \ref{fig:app_smooth_corner}. The true values of the smooth potential parameters ($\boldsymbol{\eta}$) are indicated by the vertical and horizontal black lines and the points (squares). Contours are drawn at the 68 and 95 per cent confidence intervals. The corresponding mean values and 68 per cent confidence intervals are listed in Table \ref{tab:table_mean}.}
	\label{fig:results_smooth_corner}
\end{figure*}

\subsection{Smooth lens and smooth source}
\label{sec:model_smooth}
A simulated lens system is created with a single massive lensing galaxy having $(b,q,\theta,x_0,y_0,\gamma,\phi) = (0.9,0.8,-135^\circ,0,0,0.03,-40^\circ)$.
The source brightness distribution consists of two Gaussian components: the first is located at $x,y = (-0.05,0.05)$ arcsec on the source plane, has an axis ratio of $0.6$, position angle of $-70^\circ$, and standard deviation on the $x$ axis of $\sigma_{\rm x} = 0.1$ arcsec, while the second component is at $x,y = (-0.4,0.25)$ arcsec and has $\sigma_{\rm x} = \sigma_{\rm y}= 0.1$ arcsec (circular).
The two components are scaled to have a peak brightness ratio of 0.7, with the first one being the brighter.
The data is simulated on a square 3.5-arcsec 80-pixel field of view, having a pixel size somewhat bigger than 0.04 arcsec.
The corresponding source and resulting lensed images are shown in the left column in Fig.~\ref{fig:results_smooth_images}.

We model the system as a purely parametric smooth lens, without including any grid-based correction to the potential, using $n=3$ for constructing the adaptive source plane grid (selecting 1 out of every $n\times n$ pixels).
In addition to the lens potential parameters, the set of non-linear free parameters includes the regularization of the source, i.e. $\lambda_{\rm s}$ and $\boldsymbol{g}_{\rm s}$.
We use four different source regularization schemes with different associated parameters: identity ($\lambda_{\rm s}$), curvature ($\lambda_{\rm s}$), exponential kernel ($\lambda_{\rm s}$, $l_{\rm s}$), and Gaussian kernel ($\lambda_{\rm s}$, $l_{\rm s}$).
The covariances between source pixels for the latter two schemes are given by equations (\ref{eq:covariance_exp}) and (\ref{eq:covariance_exp_squared}) respectively; we note that the $l_{\rm s}$ are different parameters in these two cases, indicating the length where the correlation drops to roughly half its maximum.
The value of the regularization parameter, $\lambda_{\rm s}$, sets the overall level of regularization and is inversely proportional to the source variance, e.g. smaller values allow for more freedom in the source model.
This parameter is expected to vary between different schemes because of the fundamentally different covariance matrices and cannot be straightforwardly compared.
Instead, one can compare the evidence values to determine which choice of regularization is more justified by the data.
We use the alternative curvature definition for adaptive grids provided in \citet{Vegetti2009a}, which has a fixed regularization pattern/correlation length for a given grid.
In this case, if $H$ is a matrix holding the numerical coefficients for the local curvature of the source then $C_{\rm s} = (H^T H)^{-1}$.

Fig. \ref{fig:results_smooth_images} shows the reconstructed sources, lensed images, and residuals, and Table \ref{tab:table_map} (top) lists the Maximum A Posteriori (MAP) model parameters and the corresponding posterior probability terms from equation (\ref{eq:evidence}), for the four different regularization schemes.
Table \ref{tab:table_mean} lists the mean parameter values, the 68 per cent confidence intervals, and the evidence, $\mathcal{E}$, for each model.
The identity regularization corresponds to a covariance matrix that is the identity matrix, which has a flat power spectrum\footnote{Or a delta function two-point correlation function, which is the inverse Fourier transform of the power spectrum (i.e. the Wiener-Khinchin theorem).} that allows the solution to vary wildly, similarly to white noise, resulting in an unrealistically grainy source.
Despite having the lowest likelihood (i.e. $\chi^2$ term in Table \ref{tab:table_map}), and thus the lowest residuals\footnote{These residuals result from $n=3$ for the adaptive grid and are expected to be reduced by increasing the number of pixels used to describe the source, i.e. $n=2$ or $n=1$.} as shown in Fig. \ref{fig:results_smooth_res_ps}, the identity regularization also has the lowest evidence value.
All other three regularization schemes perform better in recovering the source and the model parameters and give considerably higher evidence values.
However, the Gaussian kernel is decisively preferred over the curvature and exponential kernels, having a Bayes factor of $\log_{\rm 10} K= 7.98$ and $17.85$ respectively \citep[][assuming all models have the same prior probability]{Jeffreys1998}.
Although this is not the best possible kernel, it is still a sufficient approximation to describe the source brightness \citep[see fig. 3 of][]{Vernardos2020}.
As a final note, it can be seen that in all cases there is some overfitting, most prominently for the identity regularization, that suppresses the noise in the large scales ($k<5$ in Fig. \ref{fig:results_smooth_res_ps}).
This can also be seen in the reconstructed sources in Fig. \ref{fig:results_smooth_images}, where the adaptive grid voronoi cells become noisy and don't drop to zero as we move away from the brightest pixels.

\begin{figure*}
	\centering
	\includegraphics[width=\textwidth]{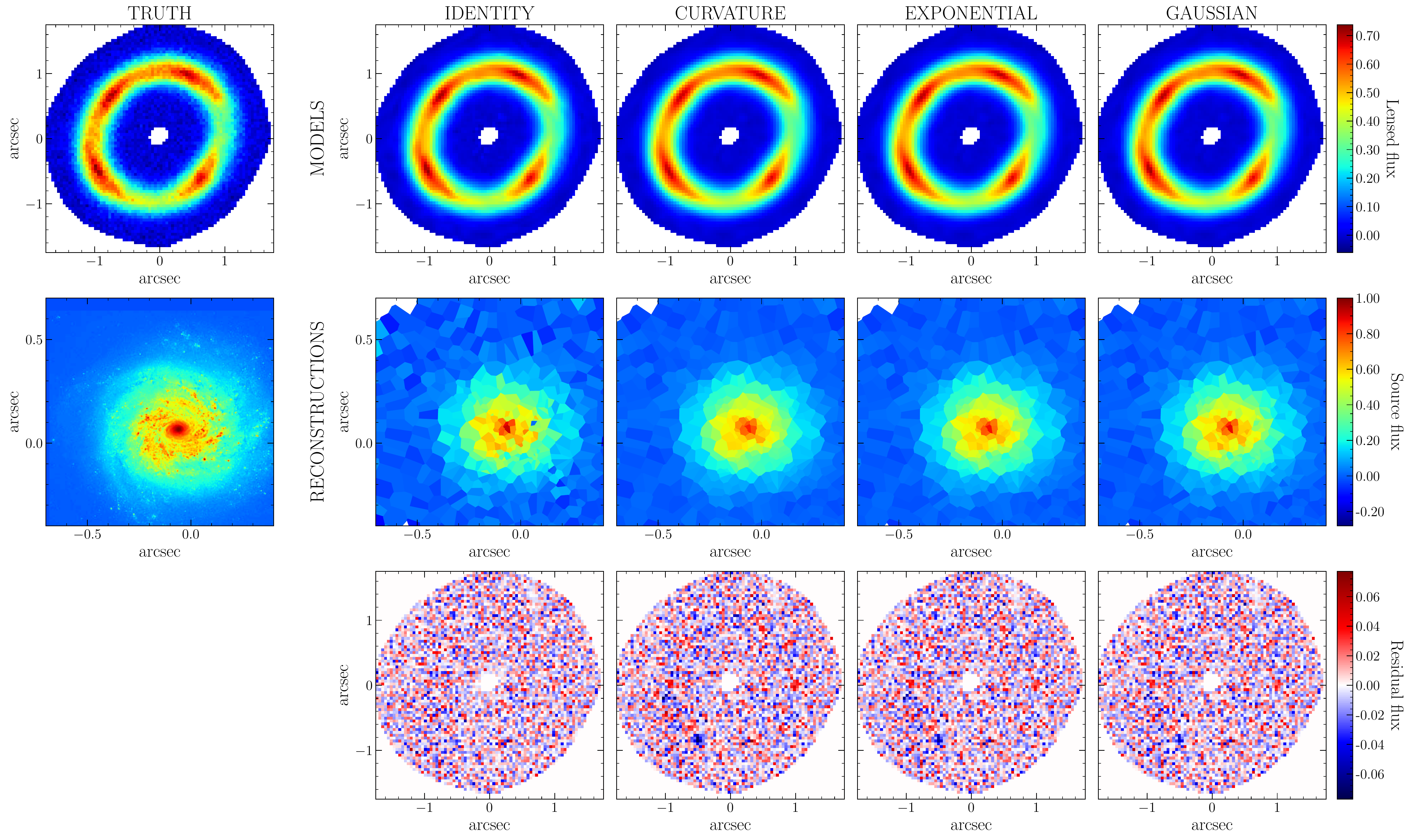}
	\caption{Same as Fig. \ref{fig:results_smooth_images} for NGC3982. \label{fig:results_spiral_images}}
	\vspace{0.5cm}
	\includegraphics[width=\textwidth]{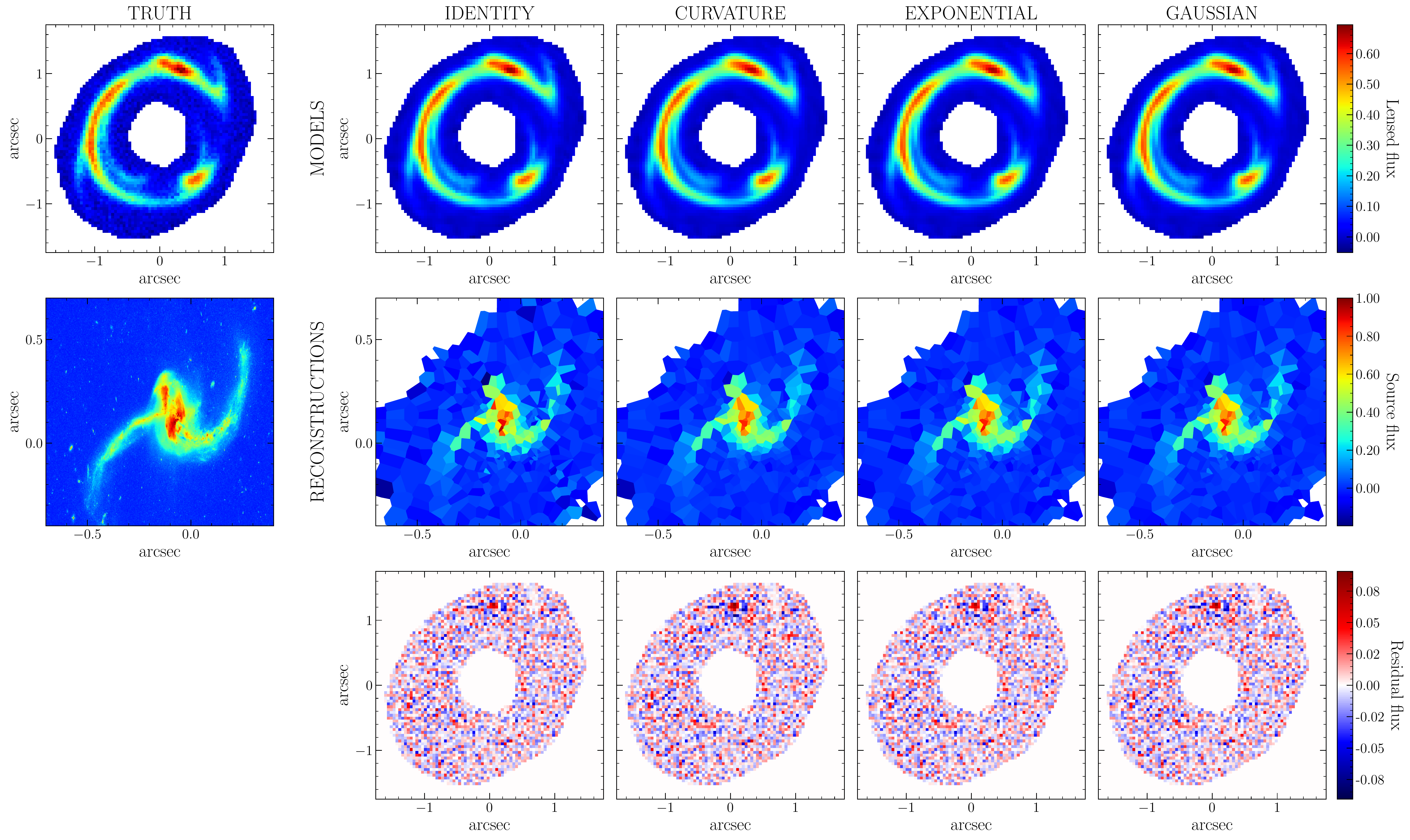}
	\caption{Same as Fig. \ref{fig:results_smooth_images} for NGC2623. \label{fig:results_merger_images}}
\end{figure*}

\begin{figure}
	\centering
	\includegraphics[width=0.45\textwidth]{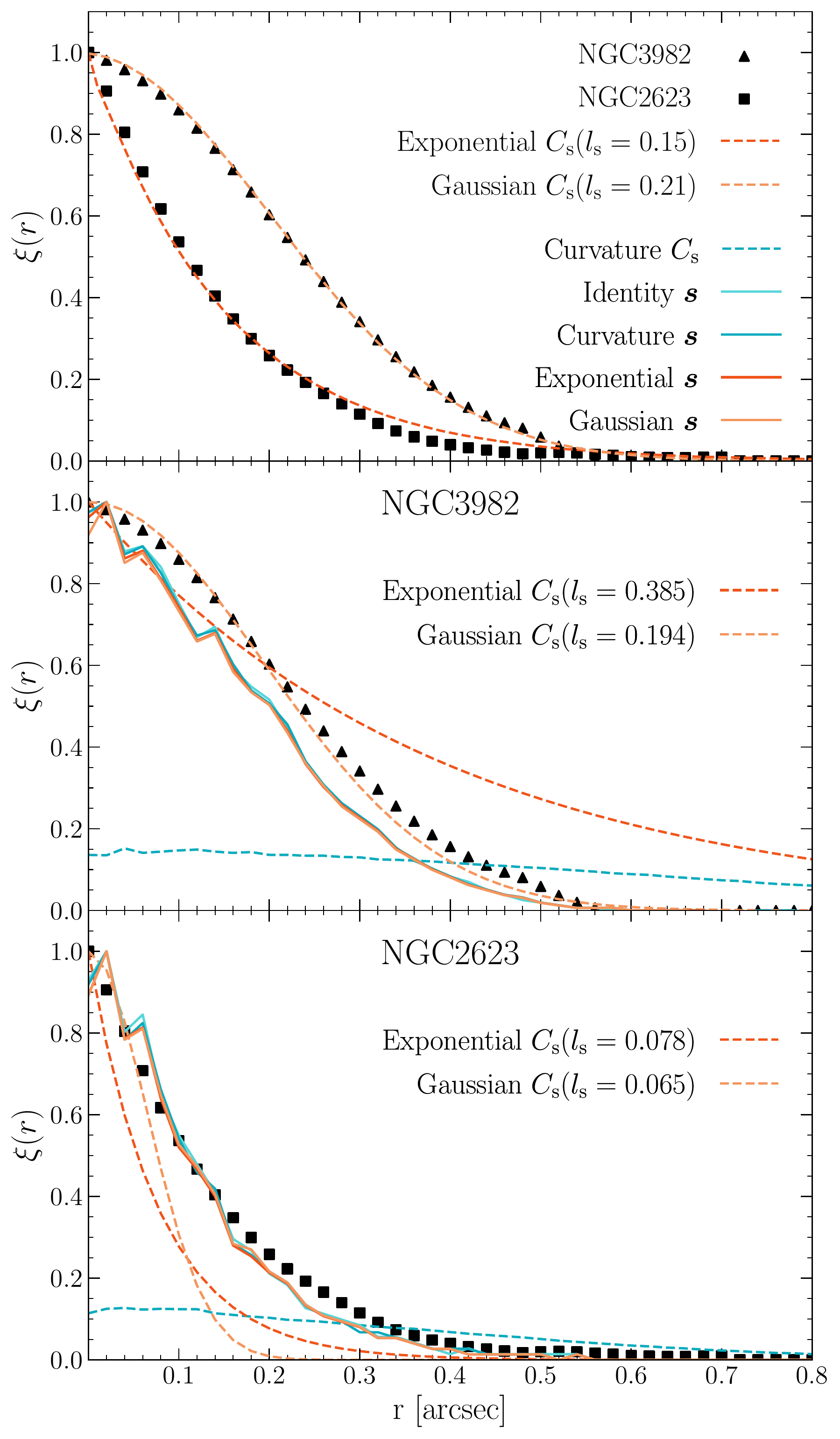}
	\caption{Radially averaged two-point correlation functions for the unlensed (HST-observed) images of NGC3982 and NGC2623 (triangles and squares respectively) and their corresponding reconstructions (solid lines) and priors (dashed lines). The $l_{\mathrm{\delta\psi}}$ parameter for the exponential and Gaussian priors changes in each panel as indicated. We have assumed the angular size of the unlensed sources to be $\approx$1 arcsec, therefore the values on the horizontal axis are scaled accordingly. The functions have been normalized to unity to factor out the effect of $\lambda_{\rm s}$ and the pixel resolution. Top: the exponential and Gaussian theoretical covariance kernels from equations (\ref{eq:covariance_exp}) and (\ref{eq:covariance_exp_squared}) are shown for values of $l_{\rm s}$ selected to visually match the data. Middle and bottom: the $l_{\rm s}$ parameters for the Exponential and Gaussian covariance kernel priors are set to their MAP values (see Table \ref{tab:table_map}). The two-point correlation function of an identity regularization prior would be a delta function centered at zero. \label{fig:spiral_merger_two_point} }
\end{figure}

The full non-linear parameter probability densities for the reconstruction with the Gaussian kernel are shown in Fig. \ref{fig:results_smooth_corner}.
The Nested Sampling method \citep{Skilling2004}, whose MultiNest implementation \citep{Feroz2009} is used here, is designed to compute the Bayesian evidence but can still sample the probability distributions at their peak almost as well as a MCMC algorithm.
However, if such a method is chosen from start it would neither guarantee convergence to the global maximum nor compute the evidence (or be extremely inefficient in doing so).
The recovered probability distributions for the lens model parameters $b,q,\theta,\gamma,$ and $\phi$ contain the true values within confidence intervals of 1 to 2 $\sigma$.
The lens center is systematically offset by approx. half a pixel in the negative x and positive y directions, which is due to a corresponding shift in the PSF's brightest pixel. 
There are no degeneracies observed between the parameters, other than the expected $b-q$ correlation from equation \ref{eq:kappa_sie} and those between $b-\gamma$, $q-\gamma$, and $\theta-\phi$, which reflect the known degeneracy between the strength and orientation of the SIE and the external shear \citep[e.g. see part 2 of][]{Schneider2006}.
The joint probability distribution of $\lambda_{\rm s}$ and $l_{\rm s}$ allows for useful conclusions on the behaviour of the source regularization.
Here, there is a very weak anti-correlation between $\lambda_{\rm s}$ and $l_{\rm s}$, which is somewhat expected: increasing the overall regularization parameter $\lambda_{\rm s}$ smooths out the reconstructed source, as does increasing the correlation length $l_{\rm s}$ in the covariance kernel.
This anti-correlation will become more prominent in the following, but it is worth pointing it out already at this smooth example.
Such information will be increasingly helpful in quantifying the degree of degeneracy between more complex sources and perturbed lens potentials in subsequent examples.

\subsection{Smooth lens and complex source}
\label{sec:smooth_complex}
Setting the lens potential to the same smooth parametric model as before, we now change the source brightness profile to more realistic ones taken from observed galaxies.
We use high resolution HST archival observations of NGC3982 (a spiral galaxy) and NGC2623 (a merger) taken with the ACS instrument, selected to represent a wider range of possible strongly lensed sources.
We scale the source angular size arbitrarily to around 1 arcsec, roughly the same as for the analytic source used in the previous section.
The HST images are scaled down dramatically in size and are significantly oversampled compared to the sampling of the final mock data.
We take this sub-pixel structure into account by heavily oversampling the mock data by a factor of 10, producing very high resolution lensed images, applying an oversampled PSF, and finally averaging to the final pixel scale: the same square 3.5-arcsec 80-pixel field of view as before.
The resulting mock lensed images are shown in Figs. \ref{fig:results_spiral_images} and \ref{fig:results_merger_images}.

Fig. \ref{fig:spiral_merger_two_point} shows the two-point correlation function of the HST observations and indicates that the true underlying covariance properties of these two objects can in principle be captured well by the exponential and Gaussian kernel regularization schemes respectively.
Using these schemes in solving equation (\ref{eq:min_source}) imposes a realistic prior on the reconstructed source that is motivated by real observations, as opposed to, for example, curvature regularization, which implicitly imposes a correlation that is unlikely to match the truth.

\begin{figure}
	\includegraphics[width=0.45\textwidth]{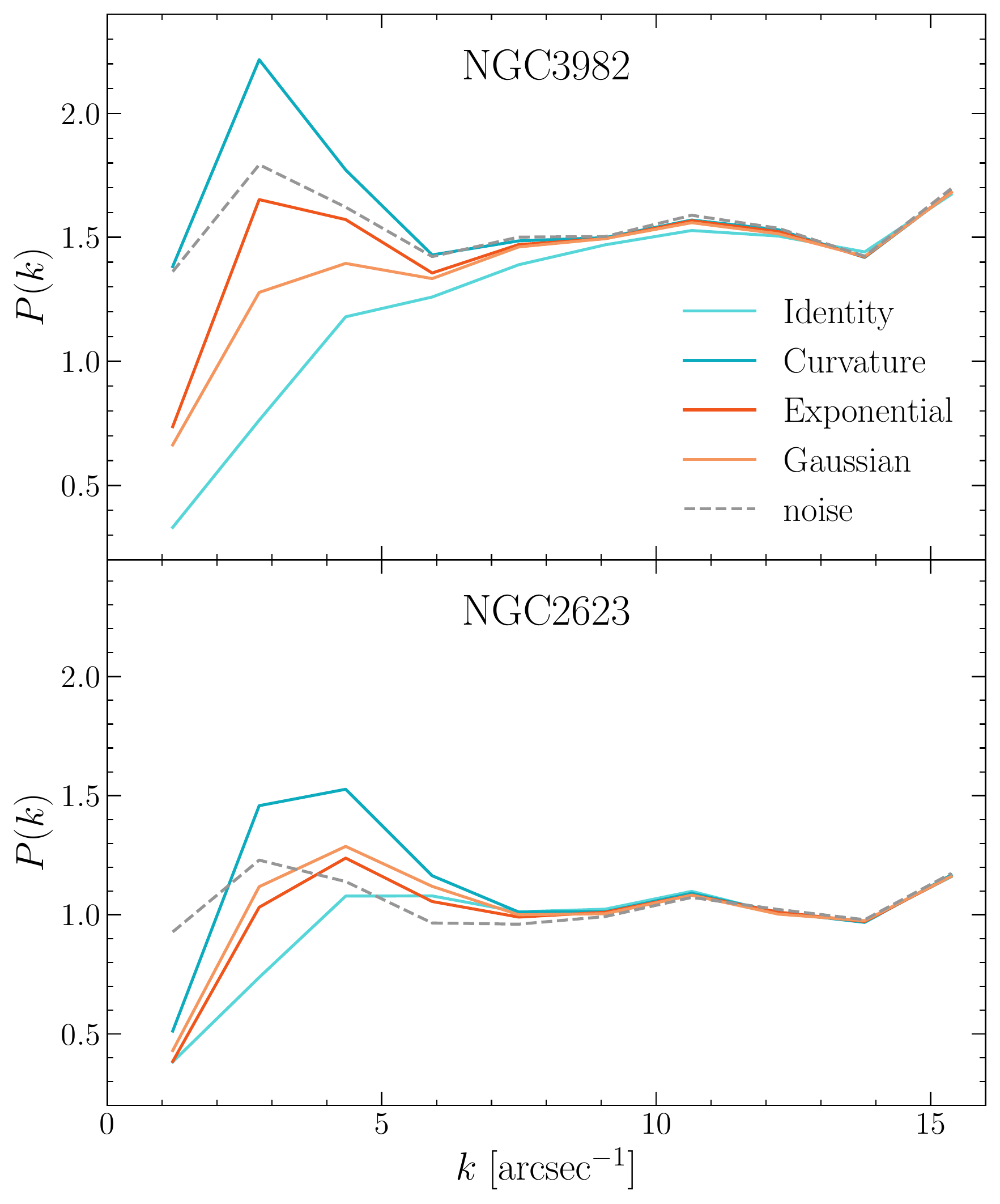}
	\caption{Fourier power spectrum of the model residuals shown in the bottom rows of Figs. \ref{fig:results_spiral_images} (top panel) and \ref{fig:results_merger_images} (bottom panel).}
	\label{fig:spiral_merger_res_ps}
\end{figure}

We model the two systems exactly in the same way as in the previous section, i.e. using $n=3$ and the same four regularization schemes.
The reconstructed sources, lensed images, and residuals are shown in Figs. \ref{fig:results_spiral_images} and \ref{fig:results_merger_images}, while the MAP and mean parameters and evidence terms are listed in Tables \ref{tab:table_map} and \ref{tab:table_mean}.
In Fig. \ref{fig:spiral_merger_two_point} we compare the radially averaged two-point correlation functions of the unlensed (HST-observed) sources and their reconstructions with the priors imposed by the covariance matrix $C_{\rm s}$.
Correlations imposed by curvature regularization have a fixed length (no free parameters) and are quite different from the truth: pixels that are far from each other are much more correlated than, for example, in the case of an exponential kernel, reflecting the implicit smoothness prior.
This is a direct consequence of $C_{\rm s}$ being a quite dense matrix: if $H$ is a matrix holding the numerical coefficients for the local curvature of the source, then $C_{\rm s} = (H^T H)^{-1}$, and although $H$, $H^T$, and $H^T H$ are relatively sparse matrices, $(H^T H)^{-1}$ is not.
However, the quality of the data is high enough to drive the solution close to the truth regardless of the regularization scheme/assumed prior - the two-point correlation functions for all the reconstructions lie on top of each other\footnote{The reconstructions become completely smooth and match almost perfectly the truth and the recovered covariance matrix if the reconstructed sources are interpolated from the adaptive Delaunay grid onto a regular grid with similar resolution.} in Fig. \ref{fig:spiral_merger_two_point}.
Even the reconstruction using the least physically motivated identity regularization manages to recover the correct correlations of the source, suggesting that the solution is driven by the data and not the prior and therefore is not very degenerate.
Nevertheless, the evidence values (see Table \ref{tab:table_mean}) are maximized by the correct regularization scheme in each case, viz. Gaussian for NGC3982 and exponential for NGC2623.
Comparing the mean values and confidence intervals of the correlation length parameter, $l_{\rm s}$, to the truth, i.e. those obtained from the observed images (see Fig. \ref{fig:spiral_merger_two_point}), we find a good agreement for both cases, despite the MAP value for NGC2623 being quite low (see Fig. \ref{fig:spiral_merger_two_point}).

\begin{figure*}
	\includegraphics[width=\textwidth]{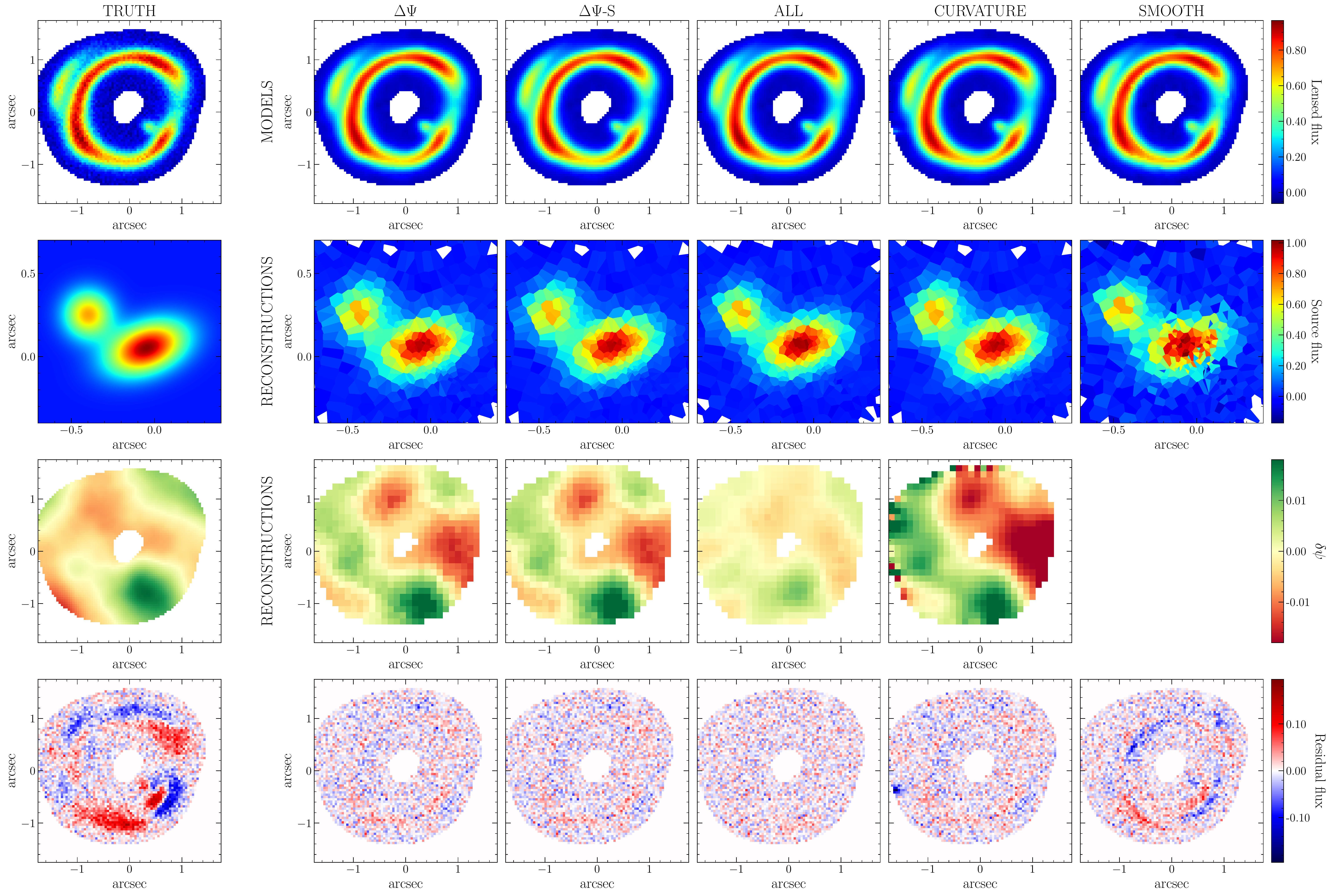}
	\caption{Same as Fig. \ref{fig:results_smooth_images}, with the addition of the true and reconstructed perturbations $\delta\boldsymbol{\psi}$ as described in Section \ref{sec:model_dpsi}. The bottom left panel shows the difference between the perturbed (top left panel) and unperturbed systems (top left panel of Fig. \ref{fig:results_smooth_images}).}
	\label{fig:results_perts_images}
\end{figure*}

Comparing the power spectra shown in Fig. \ref{fig:spiral_merger_res_ps}, we see that the identity regularization performs best, as is the case for the smooth source examined in Section \ref{sec:model_smooth}, which, however, is the result of overfitting.
Curvature regularization produces residuals on the large scales (small wavenumber $k$), while the more physically motivated exponential and Gaussian regularizations result in the smallest residuals and at the same time avoid overfitting.
Despite the successful modelling of the smooth lens potential and finding the correct source prior, there is still some unmodelled flux in the residuals (at SSE in the residuals in Fig. \ref{fig:results_spiral_images} and N in Fig. \ref{fig:results_merger_images}), which results from using $n=3$ to construct the adaptive grid, a value too high to account for the complex small scale source structure.
Such residuals could erroneously be interpreted as spurious lens potential perturbations when modelling real data - this is examined more closely in Section \ref{sec:both}.

\subsection{Modelling potential perturbations}
\label{sec:model_dpsi}
A lens potential fully described by a parametrized smooth lens model, as examined so far, might be an idealized real-world scenario.
Therefore, in this section we introduce and model potential perturbations.
We adopt the same smooth lens potential used in Sections \ref{sec:model_smooth} and \ref{sec:smooth_complex}, which we perturb using a Gaussian Random Field (GRF) of perturbations $\delta\boldsymbol{\psi}$.
GRF perturbations are defined by their power spectrum, which, in this case, we assume to be a power law:
\begin{equation}
\label{eq:power_spectrum}
P(k) = A \; k^{\beta},
\end{equation}
where $A$ is the amplitude, associated to the variance of the zero-mean $\delta\boldsymbol{\psi}$ field \citep[for more details see][]{Chatterjee2018,Bayer2018,Chatterjee2019}, $\beta$ is the slope, and $k$ is the wavenumber of the Fourier harmonics.
%The radially averaged two-point correlation function - the inverse Fourier transform of the power spectrum - is also a Gaussian (see Fig. \ref{fig:results_perts_two_point}).
Regardless of our particular choice of GRF perturbations, the generality of the analysis presented here is not affected - in fact, any form of potential perturbations could be used and modelled.

\begin{figure}
	\includegraphics[width=0.45\textwidth]{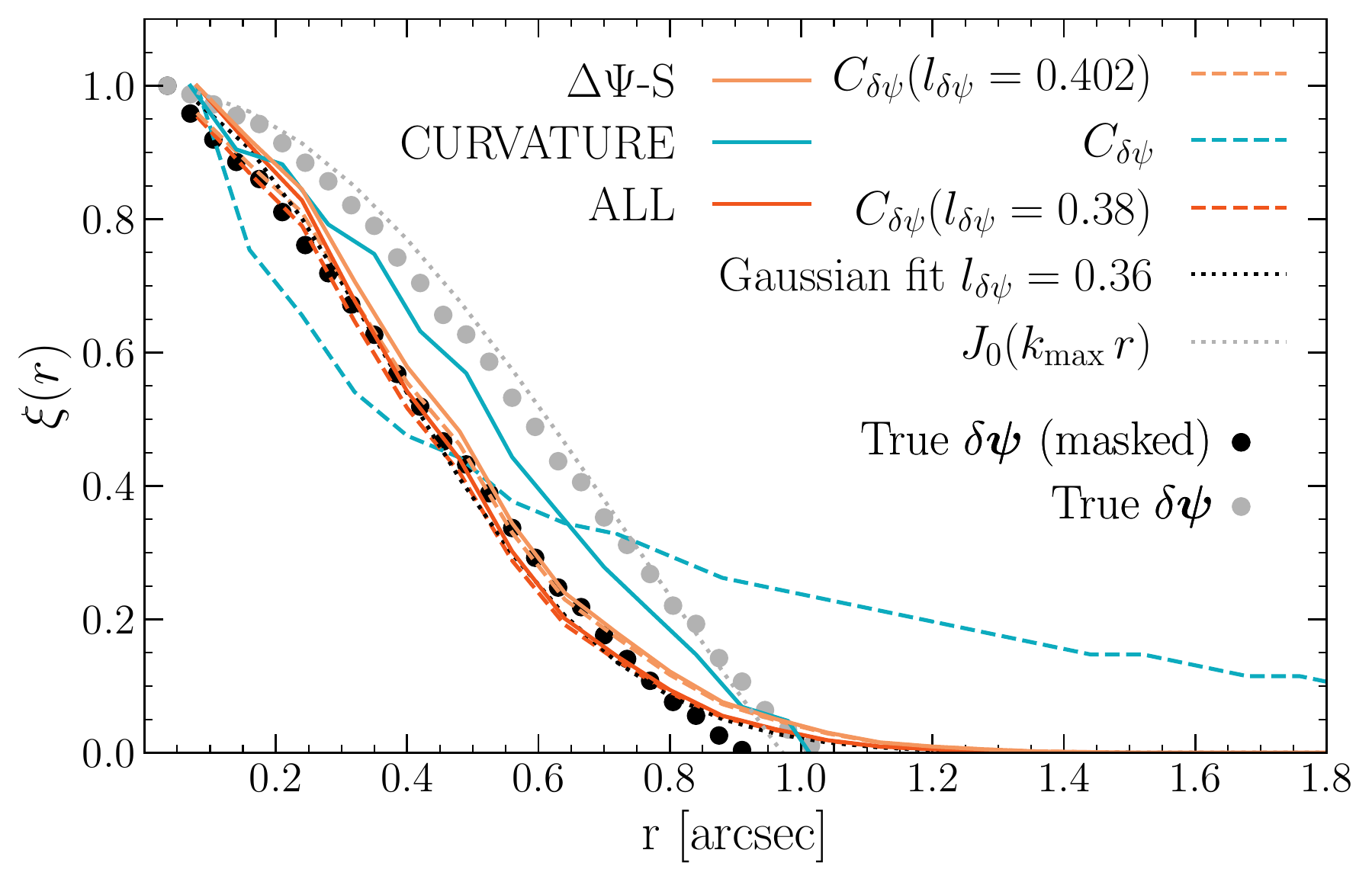}
	\caption{Radially averaged two-point correlation functions of the true $\delta\boldsymbol{\psi}$ field (circles), the reconstructions from the models shown in Fig. \ref{fig:results_perts_images} (solid lines, see Section \ref{sec:model_dpsi} for details), and different $C_{\mathrm{\delta\psi}}$ priors (dashed lines). The priors for the $\Delta\Psi$-S and ALL models are Gaussian with the $l_{\rm \delta\psi}$ parameter MAP value indicated in the parentheses (see Table \ref{tab:table_perts_smooth_map}). The black dotted line is a Gaussian fit to the correlation function of the masked $\delta\boldsymbol{\psi}$ with $l_{\rm \delta\psi} = 0.36$ (using equation \ref{eq:covariance_exp_squared}). The grey dotted line is directly plotted from equation (\ref{eq:two_point_power_law}), i.e. not a fit, with $k_{\mathrm{max}}$ set to the diagonal of the 3.5 arcsec field of view.}
	\label{fig:results_perts_two_point}
\end{figure}

\begin{figure}
	\includegraphics[width=0.45\textwidth]{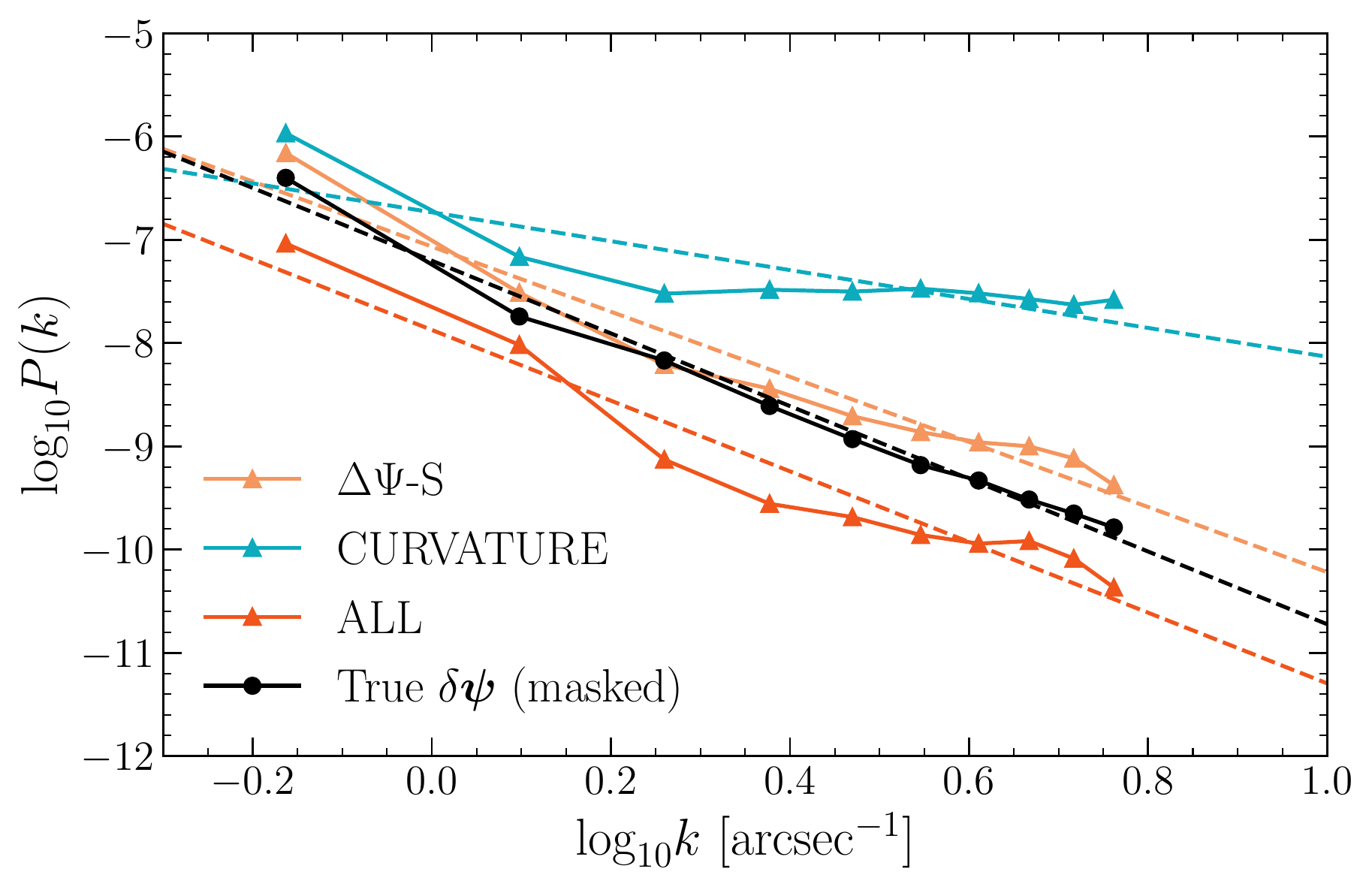}
	\caption{Fourier power spectrum of the perturbations shown in the third row of Fig. \ref{fig:results_perts_images}. The dashed lines are fits using equation (\ref{eq:power_spectrum}) with the corresponding parameters listed in Table \ref{tab:perts_GRF_parameters}. The power spectra are computed within the mask.}
	\label{fig:results_perts_dpsi_ps}
\end{figure}

\begin{figure}
	\includegraphics[width=0.45\textwidth]{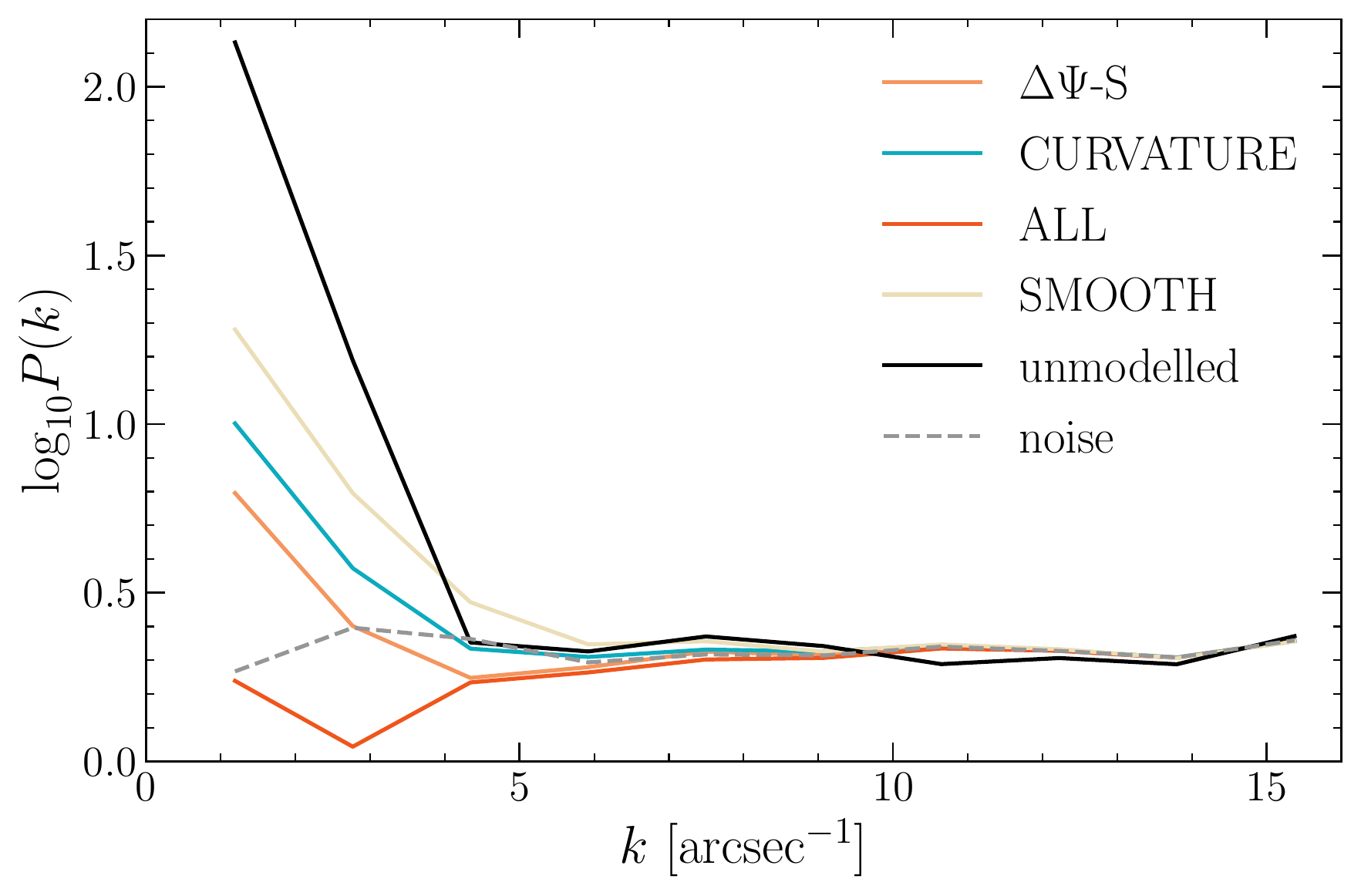}
	\caption{Fourier power spectrum of the model residuals shown at the bottom row of Fig. \ref{fig:results_perts_images}. The ``unmodelled'' residuals correspond to the bottom left panel of Fig. \ref{fig:results_perts_images} and quickly drop to the noise level for $k>4$.}
	\label{fig:results_perts_res_ps}
\end{figure}

\renewcommand{\arraystretch}{1.2}
\begin{table*}
	\caption{MAP parameter values and corresponding probability terms (from equation \ref{eq:evidence}, same as Table \ref{tab:table_map}). Models $\Delta\Psi$, $\Delta\Psi$-S, CURVATURE, and ALL are described in Section \ref{sec:model_dpsi} and model FFF in \ref{sec:both}. Notice that the dimensions of the parameter space are not the same between the models.}
	\label{tab:table_perts_smooth_map}
	\begin{threeparttable}
		\begin{tabular}{rrrrrrrr}
			name & units & Truth & $\Delta\Psi$ & $\Delta\Psi$-S & CURVATURE & ALL & FFF \\ 
\hline
$b$ & arcsec & 0.9 & - & - & - & $   0.895$ & $   0.881$ \\ 
$q$ & - & 0.8 & - & - & - & $   0.799$ & $   0.772$ \\ 
$\theta$ & $^\circ$ & -135 & - & - & - & $-134.351$ & $-133.769$ \\ 
$x_0$ & arcsec & 0 & - & - & - & $  -0.044$ & $  -0.054$ \\ 
$y_0$ & arcsec & 0 & - & - & - & $   0.017$ & $   0.026$ \\ 
$\gamma$ & - & 0.03 & - & - & - & $   0.032$ & $   0.036$ \\ 
$\phi$ & $^\circ$ & -40 & - & - & - & $ -40.767$ & $ -42.472$ \\ 
\hline
 $\lambda_{\rm s}$ & - & -  & ($  88.068$ fixed) & $ 110.271$ & $ 110.983$ & $  83.306$ & $  75.358$ \\ 
 $l_{\rm s}$ & arcsec & -  & ($   0.128$ fixed) & $   0.130$ & $   0.130$ & $   0.129$ & $   0.154$ \\ 
\hline
 $\lambda_{\rm \delta\psi}$ & - & - & $30796.077$ & $32806.464$ & $  18.418$ & $117961.620$ & $20345.966$ \\ 
 $l_{\rm \delta\psi}$ & arcsec & -  & $   0.427$ & $   0.402$ & - & $   0.380$ & $   0.285$ \\ 
\hline
 \multicolumn{3}{r}{$-\frac{\mathrm{N}_{\rm d}}{2} \log(2\pi)$\tnote{$\dagger$}} & -3571.00 & -3571.00 & -3571.00 & -3571.00 & -3145.53 \\ 
 \multicolumn{3}{r}{$\frac{\mathrm{N}_{\rm s}}{2} \log(\lambda_{\rm s})$} &  1534.75 &  1714.22 &  1716.57 &  1612.01 &  1668.14 \\ 
 \multicolumn{3}{r}{$\frac{\mathrm{N}_{\rm \delta\psi}}{2} \log(\lambda_{\rm \delta\psi})$} &  4650.81 &  4679.27 &  1311.00 &  5255.15 &  4579.85 \\ 
 \multicolumn{3}{r}{$-\frac{1}{2} \log(\det C_{\rm d})$\tnote{$\dagger$}} & 24021.55 & 24021.55 & 24021.55 & 24021.55 & 26011.99 \\ 
 \multicolumn{3}{r}{$-\frac{1}{2} \log(\det C_{\rm s})$} &   591.56 &   586.57 &   587.30 &   584.15 &   394.75 \\ 
 \multicolumn{3}{r}{$-\frac{1}{2} \log(\det C_{\rm \delta\psi})$} &   884.23 &   872.31 &  4297.36 &   859.87 &   710.49 \\ 
 \multicolumn{3}{r}{$-\frac{1}{2}\chi^2$} & -1764.76 & -1777.64 & -1804.76 & -1726.69 & -1616.15 \\ 
 \multicolumn{3}{r}{$-\frac{1}{2}\lambda_{\rm s} \boldsymbol{s}^T C_{\rm s}^{-1} \boldsymbol{s} - \frac{1}{2}\lambda_{\rm \delta\psi} \delta\boldsymbol{\psi}^T C_{\rm \delta\psi}^{-1} \delta\boldsymbol{\psi}$} &  -185.53 &  -215.73 &  -208.68 &  -170.42 &  -246.74 \\ 
 \multicolumn{3}{r}{$-\frac{1}{2} \log(\det H)$} & -8199.20 & -8332.19 & -8424.01 & -8749.70 & -7981.98 \\ 
\hline
 \multicolumn{3}{r}{$\log P$} & 17962.42 & 17977.36 & 17925.32 & 18114.93 & 20374.82 \\ 

		\end{tabular}
		\begin{tablenotes}\footnotesize
			\item [$\dagger$] constant
		\end{tablenotes}
	\end{threeparttable}
\end{table*}

\renewcommand{\arraystretch}{1.4}
\begin{table*}
	\centering
	\caption{Mean parameter values, 68 per cent confidence intervals, and evidence terms (same as Table \ref{tab:table_mean}). Models $\Delta\Psi$, $\Delta\Psi$-S, CURVATURE, and ALL are described in Section \ref{sec:model_dpsi} and model FFF in \ref{sec:both}. The full probability densities for models ALL and FFF are shown in Fig. \ref{fig:results_combined_corner}. Notice that although the dimensions of the parameter space differ between the models, this is taken into account while integrating to calculate the evidence. We do not compare model FFF to any other model, hence its evidence value is omitted.}
	\label{tab:table_perts_smooth_mean}
	\begin{tabular}{rrrrrrrr}
		name & units & Truth & $\Delta\Psi$ & $\Delta\Psi$-S & CURVATURE & ALL & FFF \\ 
\hline
$b$ & arcsec & 0.9 & - & - & - & $   0.895_{-   0.005}^{+   0.005}$ & $   0.881_{-   0.002}^{+   0.001}$ \\ 
$q$ & - & 0.8 & - & - & - & $   0.799_{-   0.010}^{+   0.011}$ & $   0.772_{-   0.005}^{+   0.002}$ \\ 
$\theta$ & $^\circ$ & -135 & - & - & - & $-134.349_{-   2.171}^{+   2.085}$ & $-133.769_{-   0.460}^{+   0.643}$ \\ 
$x_0$ & arcsec & 0 & - & - & - & $  -0.044_{-   0.004}^{+   0.003}$ & $  -0.054_{-   0.000}^{+   0.001}$ \\ 
$y_0$ & arcsec & 0 & - & - & - & $   0.017_{-   0.003}^{+   0.003}$ & $   0.026_{-   0.001}^{+   0.001}$ \\ 
$\gamma$ & - & 0.03 & - & - & - & $   0.032_{-   0.003}^{+   0.003}$ & $   0.036_{-   0.002}^{+   0.001}$ \\ 
$\phi$ & $^\circ$ & -40 & - & - & - & $ -40.768_{-   3.640}^{+   4.199}$ & $ -42.472_{-   1.063}^{+   1.076}$ \\ 
\hline
 $\lambda_{\rm s}$ & - & - & -  & $ 110.703_{-  10.443}^{+   9.289}$ & $ 111.384_{-  10.931}^{+   9.815}$ & $  83.742_{-   9.443}^{+   8.318}$ & $  77.163_{-  16.099}^{+  15.816}$ \\ 
 $l_{\rm s}$ & arcsec & - & -  & $   0.130_{-   0.004}^{+   0.004}$ & $   0.130_{-   0.004}^{+   0.004}$ & $   0.129_{-   0.004}^{+   0.004}$ & $   0.159_{-   0.047}^{+   0.022}$ \\ 
\hline
 $\lambda_{\rm \delta\psi}$ & - & -  & $31208.361_{-6465.456}^{+4786.931}$ & $33295.353_{-6464.454}^{+4828.483}$ & $  18.628_{-   3.068}^{+   2.487}$ & $120906.122_{-30053.728}^{+20792.873}$ & $21494.286_{-7949.972}^{+5691.486}$ \\ 
 $l_{\rm \delta\psi}$ & arcsec & -  & $   0.431_{-   0.047}^{+   0.053}$ & $   0.405_{-   0.054}^{+   0.043}$ & - & $   0.381_{-   0.038}^{+   0.038}$ & $   0.292_{-   0.071}^{+   0.035}$ \\ 
\hline
 \multicolumn{3}{r}{log $\mathcal{E}$:} & $17956.22\pm0.23$ & $17963.48\pm0.35$ & $17913.27\pm0.33$ & $18082.77\pm0.56$ & \\ 

	\end{tabular}
\end{table*}

\begin{figure*}
	\includegraphics[width=\textwidth]{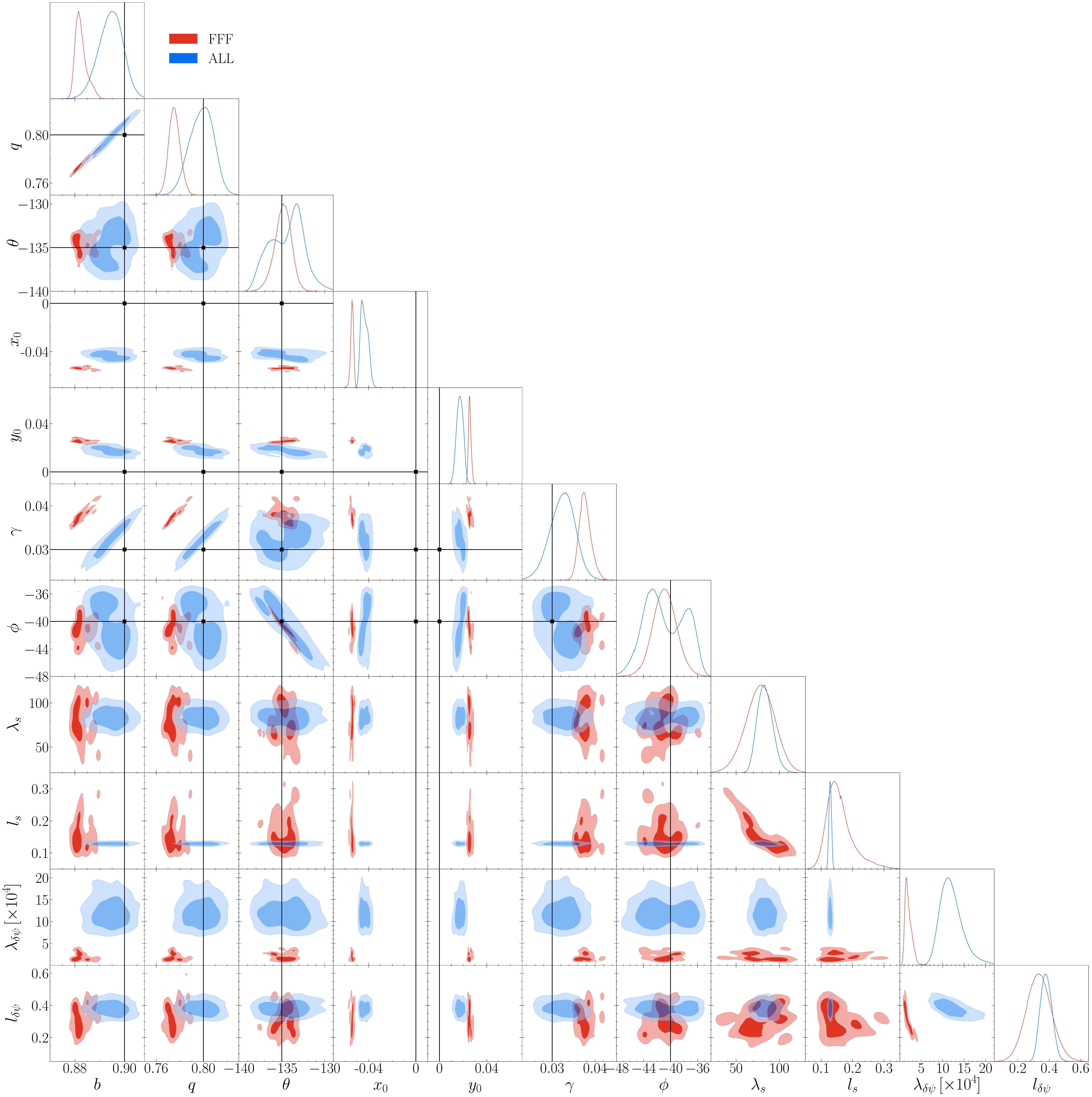}
	\caption{Same as Fig. \ref{fig:results_smooth_corner}, including the perturbation parameters $\lambda_{\rm \delta\psi}$,$\boldsymbol{g}_{\rm \delta\psi}$, for the ALL (blue) and FFF (red) models, described in Sections \ref{sec:model_dpsi} and \ref{sec:both} respectively. The two models are actually the same and have the same free parameters, i.e. the smooth potential and regularization parameters for the source and the potential perturbations, but applied to mock data with different source light profiles. The corresponding mean values and 68 per cent confidence intervals are given in Table \ref{tab:table_perts_smooth_mean}.}
	\label{fig:results_combined_corner}
\end{figure*}

We generate a single realization of $\delta\boldsymbol{\psi}$ from a GRF having $\log_{\rm 10} (A),\beta = (-7.8,-5.5)$, in the same $80\times80$ pixel grid as the mock image.
Within the masked region of the field of view, the GRF field has slightly different $A$ and $\beta$ parameters (see Table \ref{tab:perts_GRF_parameters} and Fig. \ref{fig:results_perts_two_point}).
The resulting perturbations vary in magnitude between roughly $\pm13$ per cent of the average smooth lens potential (within the mask).
The source (the same as the one used in Section \ref{sec:model_smooth}), the perturbations, and the corresponding lensed image are shown on the left column in Fig. \ref{fig:results_perts_images}.
The difference\footnote{We first subtract the perturbed and unperturbed mock lens images without any noise, and then add an artificial white noise realization with the same signal-to-noise ratio as the unperturbed case.} between the mock data with the purely smooth underlying lens model used in Section \ref{sec:model_smooth} (top left panel in Fig. \ref{fig:results_smooth_images}) and its perturbed version used here (top left panel in Fig. \ref{fig:results_perts_images}) is shown in Fig. \ref{fig:results_perts_images}, bottom left panel.

An important and basic observation we need to make here is that in order to be able to reconstruct any perturbing $\delta\boldsymbol{\psi}$ there needs to be some lensed light locally around it. 
This can be understood by examining matrix $M_{\rm r}$ (equation \ref{eq:combined_M}), which extends the smooth lens modelling framework presented in Section \ref{sec:method} to include potential perturbations: if there is no source light (strictly speaking, if the source light is constant, i.e. its derivative is zero) then the terms $D_{\mathrm{s}}(\boldsymbol{s_{\rm p}})$ introduced in equation (\ref{eq:dpsi_residuals}), and consequently the entire perturbing part of $M_{\rm r}$, vanish.
The $\delta\boldsymbol{\psi}$ are then reconstructed based mainly on the regularization prior.
As a result, in general, the further a reconstructed $\delta\psi$ value is from pixels with some lensed light in them the less accurate its estimate based on the data becomes.
In the following, we do not attempt to mitigate this and our reconstructed $\delta\boldsymbol{\psi}$ away from pixels with brightness should be viewed as an extrapolation regularized by the prior.
A similar argument holds for the smooth potential as well.

The covariance matrix of a GRF field is derived from its two-point correlation function, which is simply the inverse Fourier transform of its power spectrum.
For a GRF with a power law power spectrum, like the one given in equation (\ref{eq:power_spectrum}), the two-point correlation function is:
\begin{equation}
\label{eq:two_point_power_law}
\xi(r) = 2 \pi A J_{\mathrm{0}}(k_{\mathrm{max}}\, r) \, k_{\mathrm{max}}^{\beta+2},
\end{equation}
where $J_{\mathrm{0}}$ is the zeroth order Bessel function of the first kind, and $k_{\mathrm{max}}$ the maximum wavenumber.
However, the mask truncates the GRF and changes its covariance properties so that the above relation cannot be used to construct a regularization kernel anymore.
In this case, the Gaussian kernel provides a sufficiently good approximation for the two-point correlation function, as shown in Fig. \ref{fig:results_perts_two_point}.

To model the perturbed system, we use a Gaussian regularization kernel for both $\boldsymbol{s}$ and $\delta\boldsymbol{\psi}$, and $n=3$ for reconstructing the adaptive source grid.
The size of the pixel grid to reconstruct the perturbations $\delta\boldsymbol{\psi}$ on and $n$ set the number of free parameters of any model and can be selected by maximizing the Bayesian evidence \citep{Vegetti2012}.
However, this is outside the scope of this work - and a computationally very demanding task.
We use a $30\times30$ pixel grid for $\delta\boldsymbol{\psi}$, which has enough resolution to capture the details of the true underlying GRF perturbations while still leading to tractable computations \citep[such a grid has been also used in the case of a single perturbing substructure, e.g.][]{Koopmans2005,Vegetti2012}.
We model the perturbed lens in three different set-ups: i) we fix the smooth lens model to the truth and the source regularization parameters to the mean values of the Gaussian kernel model obtained in Section \ref{sec:model_smooth} (see Table \ref{tab:table_mean}) and we sample only $\lambda_{\rm \delta\psi}$,$\boldsymbol{g}_{\rm \delta\psi}$ (model $\Delta\Psi$), ii) we fix the smooth lens model to the truth and sample both $\lambda_{\rm s}$,$\boldsymbol{g}_{\rm s}$ and $\lambda_{\rm \delta\psi}$,$\boldsymbol{g}_{\rm \delta\psi}$ (model $\Delta\Psi$-S), and iii) we sample $\boldsymbol{\eta}$, $\lambda_{\rm s}$, $\boldsymbol{g}_{\rm s}$, $\lambda_{\rm \delta\psi}$, and $\boldsymbol{g}_{\rm \delta\psi}$ simultaneously (model ALL).
Fig. \ref{fig:results_perts_images} shows the resulting lensed images, reconstructed $\boldsymbol{s}$ and $\delta\boldsymbol{\psi}$, and residuals, Table \ref{tab:table_perts_smooth_map} lists the MAP model parameters and the posterior probability terms from equation (\ref{eq:evidence}), and Table \ref{tab:table_perts_smooth_mean} lists the mean parameter values, their 68 per cent confidence intervals, and the evidence for each set-up.
Models $\Delta\Psi$ and $\Delta\Psi$-S give almost identical results.
Models $\Delta\Psi$-S and ALL recover a similar correlation length for the source, in very good agreement with the unperturbed case presented in Section \ref{sec:model_smooth} - this is also true for the parameters $\boldsymbol{\eta}$ recovered by the ALL model.
The correlation length of the perturbations, $l_{\rm \delta\psi}$, has a very similar value for all the models; the values from $\Delta\Psi$ and $\Delta\Psi$-S and the corresponding covariance matrices, $C_{\rm \delta\psi}$, are in fact so close that their determinants differ by very little (see Table \ref{tab:table_perts_smooth_map}).

\begin{figure*}
	\centering
	\includegraphics[width=\textwidth]{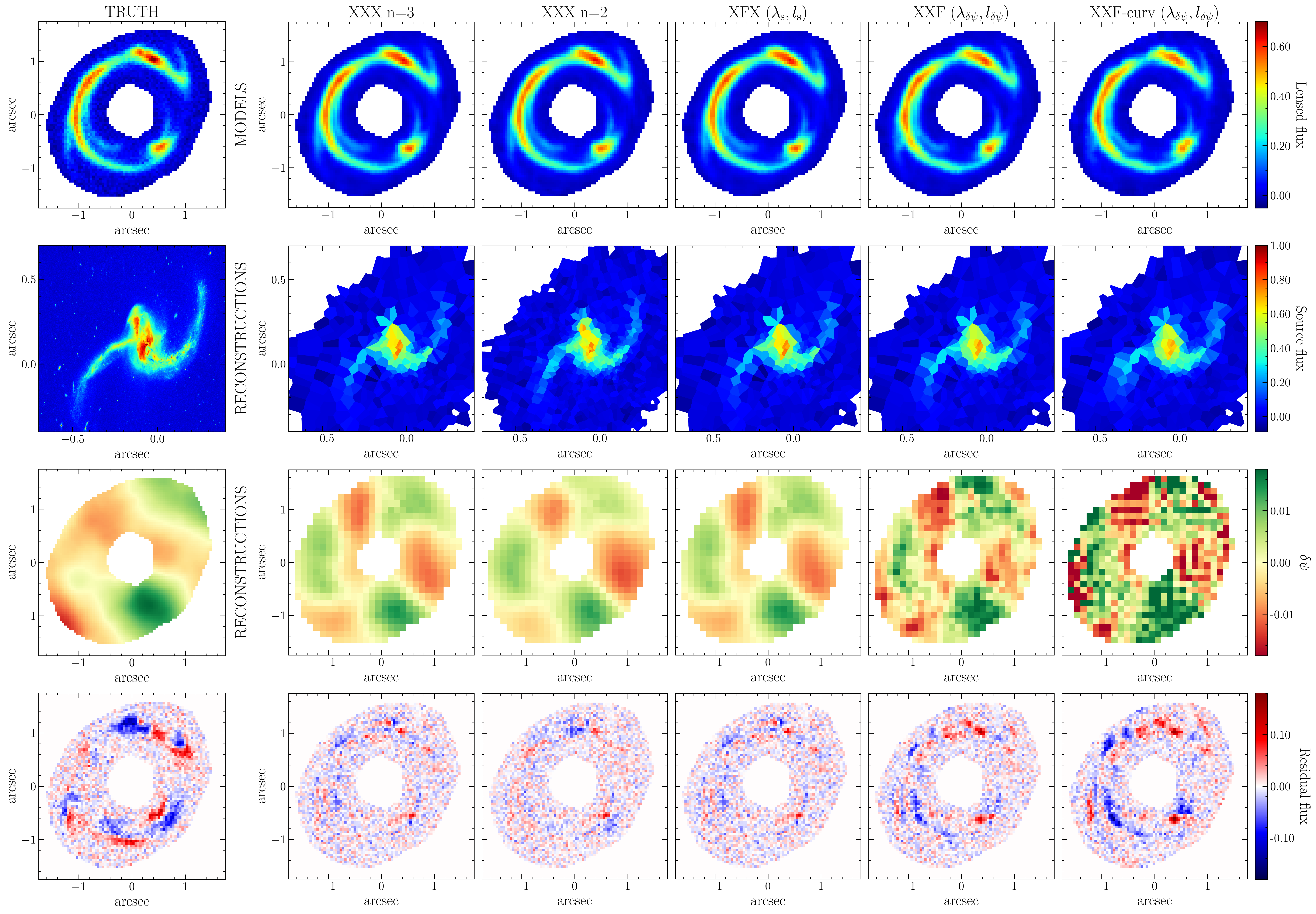}
	\caption{Same as Fig. \ref{fig:results_perts_images} for NGC2623. The bottom left panel shows the difference between the perturbed (top left panel) and unperturbed systems (top left panel of Fig. \ref{fig:results_merger_images}). We list the free parameters of each model in the parenthesis next to its name at the top (see Section \ref{sec:both} for details).}
	\label{fig:results_both_images}
\end{figure*}

\begin{figure*}
	\includegraphics[width=\textwidth]{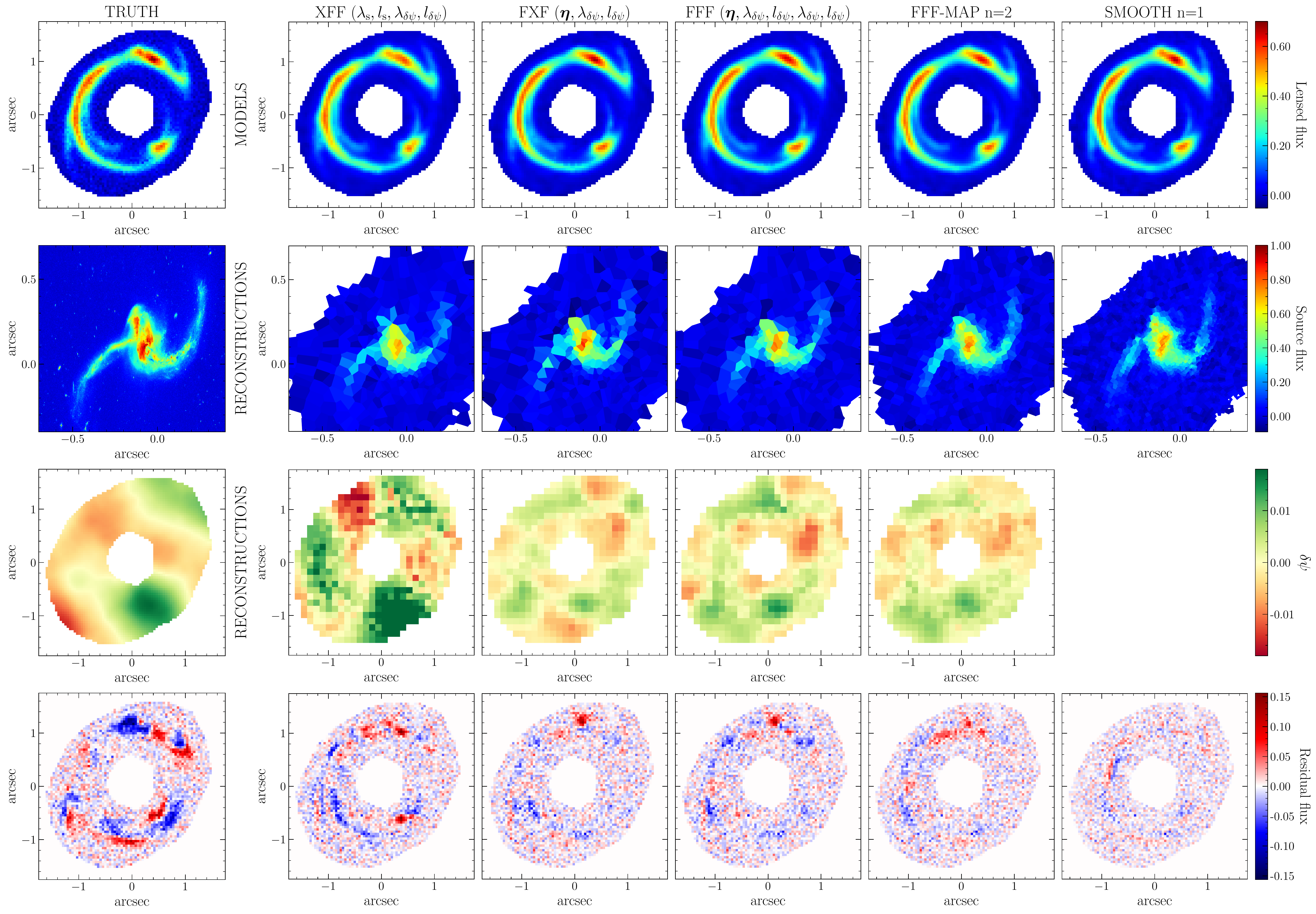}
	\contcaption{}
\end{figure*}

\begin{figure}
	\includegraphics[width=0.45\textwidth]{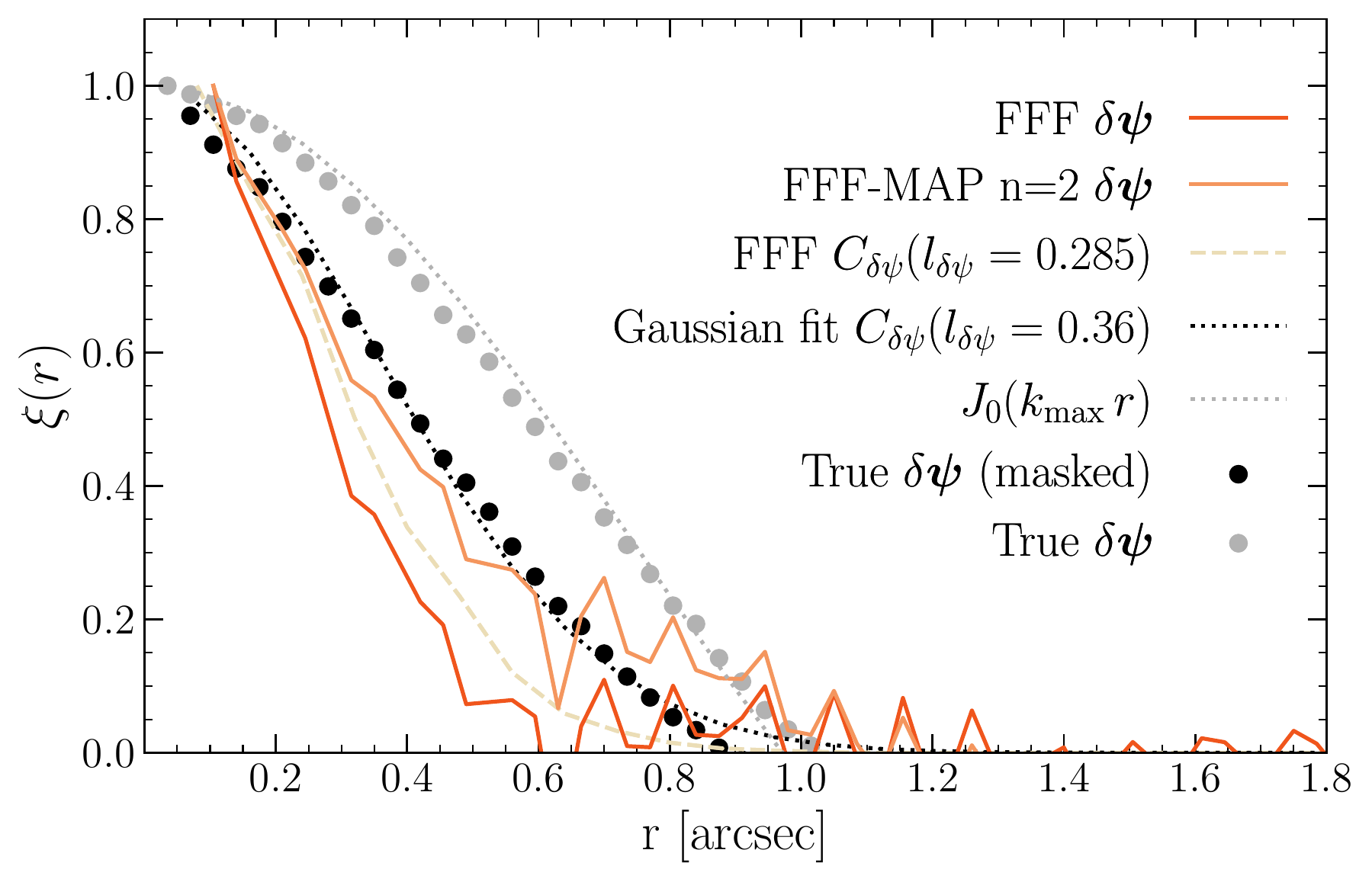}
	\caption{Radially averaged two-point correlation functions of the $\delta\boldsymbol{\psi}$ reconstructions from the FFF and FFF-MAP $n=2$ models (see Section \ref{sec:both}). We include the prior (dashed lines), with the $l_{\rm \delta\psi}$ parameter for the Gaussian covariance kernel set to its MAP value, i.e 0.285 (see Table \ref{tab:table_perts_smooth_map}). The true two-point correlation functions of the full GRF (grey circles) and the one within the mask (black cirles) are shown, together with a Gaussian fit to the latter with $l_{\rm \delta\psi} = 0.36$ (using equation \ref{eq:covariance_exp_squared}, dotted black line) and equation (\ref{eq:two_point_power_law}) with $k_{\mathrm{max}}$ set to the diagonal of the 3.5 arcsec -wide image (dotted grey line, not a fit).}
	\label{fig:results_both_two_point}
\end{figure}

\begin{figure}
	\includegraphics[width=0.45\textwidth]{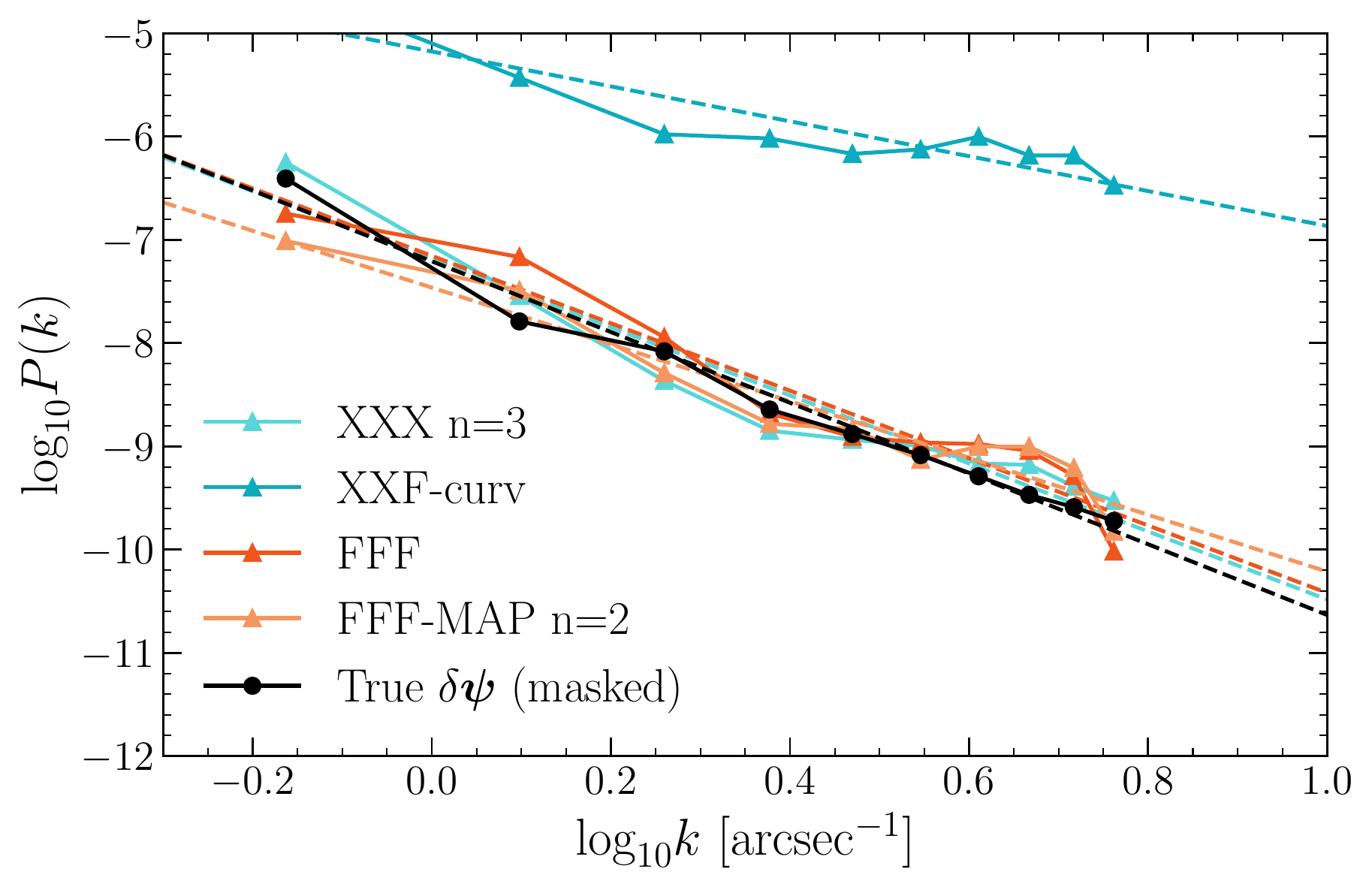}
	\caption{Fourier power spectrum of some of the $\delta\boldsymbol{\psi}$ reconstructions shown in the third row of Fig. \ref{fig:results_both_images}. The dashed lines are fits using equation (\ref{eq:power_spectrum}) with the corresponding parameters listed in Table \ref{tab:perts_GRF_parameters}. The power spectra are computed within the mask.}
	\label{fig:results_both_dpsi_ps}
\end{figure}

\begin{figure}
	\includegraphics[width=0.45\textwidth]{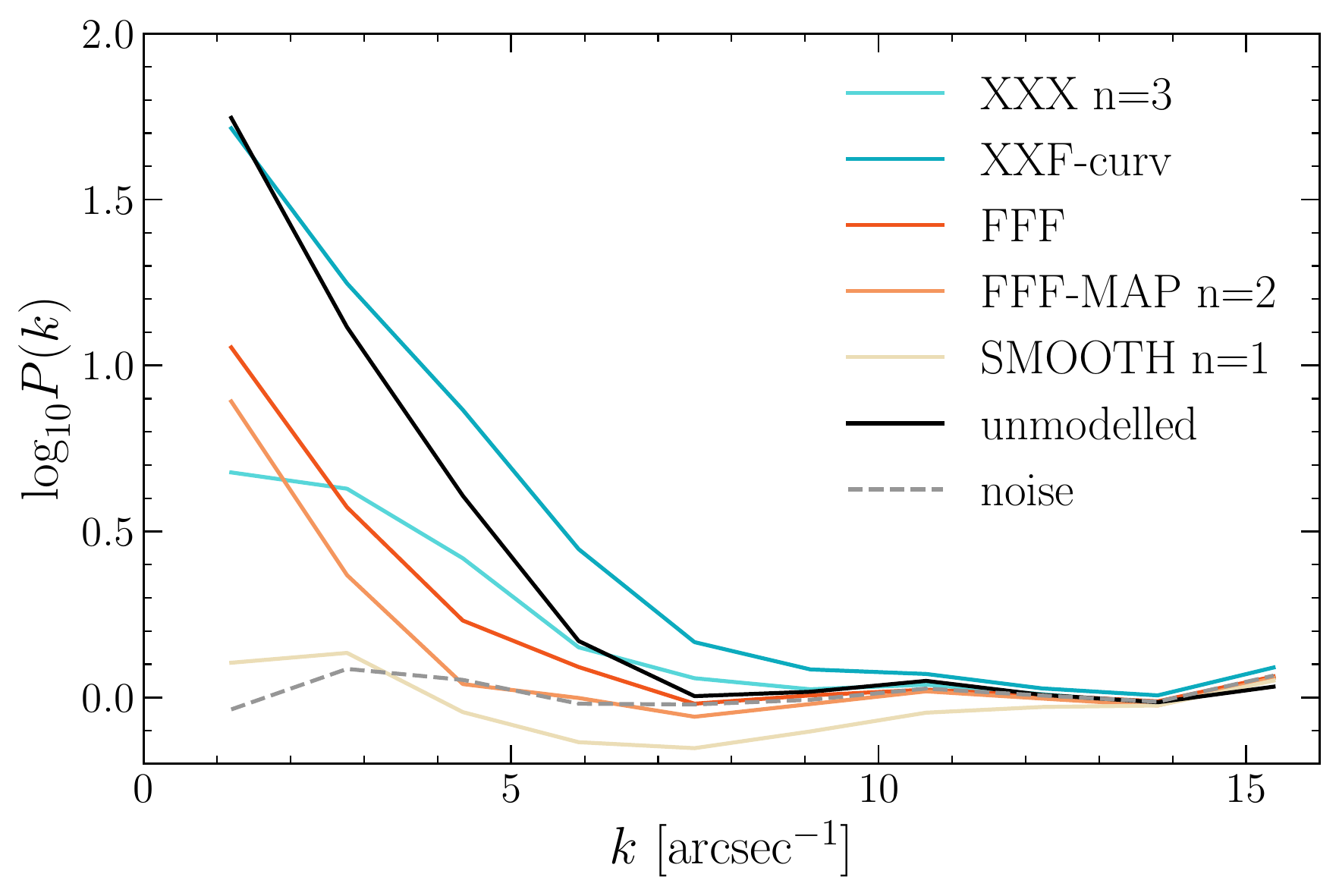}
	\caption{Fourier power spectrum of the residuals shown at the bottom row of Fig. \ref{fig:results_both_images} for the models used in Fig. \ref{fig:results_both_dpsi_ps}. The ``unmodelled'' residuals correspond to the bottom left panel of Fig. \ref{fig:results_both_images} and drop to the noise level for $k>7$.}
	\label{fig:results_both_res_ps}
\end{figure}

To further investigate the effect of the prior on the lens potential perturbations, we evaluate a model using curvature regularization for $\delta\boldsymbol{\psi}$.
To do this, we fix $\boldsymbol{\eta}$ to their true values and sample $\lambda_{\rm s}$, $\boldsymbol{g}_{\rm s}$, and $\lambda_{\rm \delta\psi}$ (there are no $\boldsymbol{g}_{\rm \delta\psi}$ parameters in this case).
First, we notice that the values of $\lambda_{\rm s}$ and $l_{\rm s}$ are almost identical with the $\Delta\Psi$-S model, however, the evidence has a much lower value, despite the latter model having an additional free parameter.
In Fig. \ref{fig:results_perts_two_point}, we show the two-point correlation function from this model and compare it with the one from the true underlying $\delta\boldsymbol{\psi}$ field and the reconstructions from the $\Delta\Psi$-S and ALL models.
It is evident that in this case the data and not the prior is driving the $\delta\boldsymbol{\psi}$ reconstruction.
In Fig. \ref{fig:results_perts_dpsi_ps} we show the power spectra of the reconstructions and in Table \ref{tab:perts_GRF_parameters} list the coefficients of the corresponding fits using equation (\ref{eq:power_spectrum}).
The connection between the slope of the power spectrum and stronger large scale correlations is evident: the flattest power spectrum belongs to the model with curvature regularization, while the slope decreases as the correlation function becomes narrower (or $l_{\delta\psi}$ becomes smaller), first for the $\Delta\Psi$-S and then for the ALL model.
We note, however, that although $\Delta\Psi$-S gives the value for the amplitude closest to the truth, its parameters $\boldsymbol{\eta}$ are fixed to the true underlying smooth model (a quite unrealistic scenario), which means that the dimensions of the parameter space to explore are significantly fewer compared to the ALL model.

The ALL model has the smooth potential parameters $\boldsymbol{\eta}$ free, which in principle could absorb part of the perturbations.
However, as discussed in Appendix \ref{app:A}, this is not the case.
The fitted smooth potential model is very close to the truth, meaning that any differences between the true total and reconstructed potentials is mostly due to the $\delta\boldsymbol{\psi}$.

A parametric-only, purely smooth model is also evaluated, which is obviously insufficient to correctly model the lens, leading to biased values of $\boldsymbol{\eta}$ and reconstructed $\boldsymbol{s}$, and prominent residuals above the noise level (bottom right panel in Fig. \ref{fig:results_perts_images}).
These residuals are lower in amplitude and different from the (unmodelled) residuals between the smooth and perturbed data (bottom left panel in Fig. \ref{fig:results_perts_images}), having a correlation coefficient of $0.26$.
This means that the perturbations are absorbed into the smooth model parameters and the source to some extent, but not fully \citep[see][for a thorough exploration of this effect]{Bayer2021}.
This can be seen in the residual power spectrum, shown in Fig. \ref{fig:results_perts_res_ps}, where the ``unmodelled'' residuals that appear on the large scales have significant power (above the noise) for $k<4$ and the smooth model residuals have 2 to 7 times less power in the same range, yet still also 2 to 7 times more than the noise.

\renewcommand{\arraystretch}{1.0}
\begin{table}
	\centering
	\caption{Power law fits from equation (\ref{eq:power_spectrum}) to the power spectra of the true and reconstructed $\delta\boldsymbol{\psi}$ shown in Figs. \ref{fig:results_perts_dpsi_ps} (top part, Section \ref{sec:model_dpsi}) and \ref{fig:results_both_dpsi_ps} (bottom part, Section \ref{sec:both}). The $\Delta\Psi$ and $\Delta\Psi$-S models give identical fits.}
	\label{tab:perts_GRF_parameters}
	\begin{tabular}{rrr}
									&	$\log_{\rm 10} A$	&	$\beta$ 		\\
		\hline
%		theoretical 				& 	$-7.8$				&	$-5.5$			\\
		True $\delta\boldsymbol{\psi}$ (masked)				&	$-7.20\pm0.01$		&	$-3.52\pm0.02$	\\
		CURVATURE					&	$-6.73\pm0.03$		&	$-1.40\pm0.03$	\\
		$\Delta\Psi$/$\Delta\Psi$-S &	$-7.07\pm0.02$		&	$-3.15\pm0.07$	\\
		ALL							&	$-7.87\pm0.03$		&	$-3.42\pm0.10$	\\
		\hline
		True $\delta\boldsymbol{\psi}$ (masked)	&$ -7.21  \pm  0.01 $ & $ -3.43  \pm  0.03 $ \\
		XXX n=3									&$ -7.19  \pm  0.02 $ & $ -3.29  \pm  0.09 $ \\
		XXF-curv								&$ -5.18  \pm  0.02 $ & $ -1.69  \pm  0.07 $ \\
		FFF										&$ -7.16  \pm  0.02 $ & $ -3.26  \pm  0.09 $ \\
		FFF-MAP n=2								&$ -7.46  \pm  0.02 $ & $ -2.75  \pm  0.07 $ \\
		\end{tabular}
\end{table}

In Fig. \ref{fig:results_combined_corner}, we show the full non-linear parameter probability densities for the ALL model.
In general, the parameters $\boldsymbol{\eta}$ are distributed similarly to Fig. \ref{fig:results_smooth_corner} but with larger statistical uncertainty.
A systematic bias is introduced in $b$, whose lower values become more probable, because the inclusion of perturbations $\delta\boldsymbol{\psi}$ can now absorb some of the overall strength of the lens potential.
Similarly, the presence of the perturbations causes $x_0$ to be offset by one pixel instead of half, which was the case in Section \ref{sec:model_smooth}.
The same degeneracies are observed as in Fig. \ref{fig:results_smooth_corner} between the parameters $b-q$, $b-\gamma$, $q-\gamma$, and $\theta-\phi$.
The latter two have a bi-modal distribution with an extent of roughly $\pm5\deg$.
Such small angular offsets between the SIE and the external shear can be understood in terms of the smoothness of the source, which allows for the perturbing field $\delta\boldsymbol{\psi}$ to make up for the difference and still provide solutions with high probability (low residuals).
There are no correlations between $\boldsymbol{\eta}$ and the regularization parameters for the source or the potential perturbations, neither between the latter two.
However, we observe again the expected anti-correlation between $\lambda_{\rm s}$ and $l_{\rm s}$ and a similar one between $\lambda_{\rm \delta\psi}$ and $l_{\rm \delta\psi}$ (better shown in Fig. \ref{fig:app_all_corner}), i.e. increasing the overall regularization parameters $\lambda$ smooths out the reconstructed fields, as does increasing the correlation length $l$ in the covariance kernels.

\subsection{Perturbed lenses and complex sources}
\label{sec:both}
In reality, we expect complex sources to be lensed by non-smooth lens potentials.
Here we combine the perturbed lens potential from the previous section with the complex brightness profile of NGC2623 (a merger) used as source in Section \ref{sec:smooth_complex}.
The resulting lensed images are shown in the left column of Fig. \ref{fig:results_both_images}.
Although such a lensing scenario could be unrealistically complex - lensing of merging galaxies is not very probable - it serves as an extreme scenario for degeneracies to emerge as a result of the non-linear behaviour approximated by matrix $M_{\rm r}$; from equation (\ref{eq:expanded_dsdpsi}), perturbed deflection angles are associated with incoming rays from a highly structured source, and this information can be lost within the finite resolution of the mock data considered in our examples.

We model the system fixing the regularization kernels to the best-performing ones, i.e. an exponential kernel for the source (see Section \ref{sec:smooth_complex}) and a Gaussian for the perturbations (see Section \ref{sec:model_dpsi}).
We reconstruct $\delta\boldsymbol{\psi}$ in the same $30\times30$-pixel grid as before, and use $n=3$ for the adaptive source grid, unless otherwise stated.
For each of the models presented in Fig. \ref{fig:results_both_images} we either fix (X) or set free (F) each of the three parameter sets $\boldsymbol{\eta},(\lambda_{\rm s},l_{\rm s}),(\lambda_{\rm \delta\psi},l_{\rm \delta\psi})$ and name it accordingly, e.g. model XFX has only ($\lambda_{\rm s},l_{\rm s}$) free to vary.
For the fixed values of the parameters we have: the true values for $\boldsymbol{\eta}$ (e.g. see Section \ref{sec:model_smooth} or Table \ref{tab:table_map}), $l_{\mathrm{s}}=0.15$ and $l_{\rm \delta\psi}=0.36$, which are the values fitted to the true source and perturbations as shown in Figs. \ref{fig:spiral_merger_two_point} and \ref{fig:results_perts_two_point}, $\lambda_{\rm s}=44.031$, the mean value from Section \ref{sec:smooth_complex} (see Table \ref{tab:table_mean}), and $\lambda_{\rm \delta\psi}=86780.1$, the mean value from the ALL model presented in Section \ref{sec:model_dpsi} (see Table \ref{tab:table_perts_smooth_mean}).

In the first part of Fig. \ref{fig:results_both_images}, we show two models with all the parameters fixed to the truth, one with $n=3$ and one with $n=2$, and three models with only one parameter set allowed to vary.
We first note that there is very little difference in the residuals and the reconstructed $\delta\boldsymbol{\psi}$ between the fixed models (despite the many more source pixels for the case with $n=2$) and the one with the source parameters free (XFX).
However, allowing the $\delta\boldsymbol{\psi}$ regularization parameters to vary leads to a worse reconstruction and the residuals increase.
This is even more prominent if we change the $\delta\boldsymbol{\psi}$ regularization from a Gaussian to a curvature kernel.
These $\delta\boldsymbol{\psi}$ solutions have too much structure (low regularization) because they may be actually overcompensating for a low resolution adaptive source grid.
In the second part of Fig. \ref{fig:results_both_images}, we see that the residuals and the $\delta\boldsymbol{\psi}$ reconstruction do not improve if we set both the perturbation and source regularization parameters free (i.e., compare models XXF and XFF).
As soon as we allow $\boldsymbol{\eta}$ to vary then the residuals do decrease at the cost of a less smooth $\delta\boldsymbol{\psi}$ reconstruction.
This is regardless of fixing the source regularization parameters - models FXF and FFF give very similar results.
However, the adaptive grid resolution affects the residuals: after fixing all parameters to the MAP values from the FFF model, we set $n=2$ and although the $\delta\boldsymbol{\psi}$ reconstruction does not improve too much, the residuals do (see also Fig. \ref{fig:results_both_res_ps}), in particular, the prominent positive residuals due north with respect to the lens in the models FXF and FFF considerably decrease.
This is most likely due to the more degrees of freedom available for the source, which is further supported by a smooth model with $n=1$ that absorbs the perturbations almost down to the noise level.

Looking at the two-point correlation functions of the reconstructed $\delta\boldsymbol{\psi}$ shown in Fig. \ref{fig:results_both_two_point}, we note that the prior and the data lie close to each other, which accordingly drives the FFF model.
Increasing the adaptive grid resolution leads to somewhat stronger correlations on the larger scales and brings the reconstructed $\delta\boldsymbol{\psi}$ even closer to both the prior and the data.
However, in Fig. \ref{fig:results_both_dpsi_ps}, and from the fitted coefficients listed in Table \ref{tab:perts_GRF_parameters}, there is a remarkable agreement between the power spectrum of the FFF model (all the parameters free) and the true $\delta\boldsymbol{\psi}$.
The same holds for the reconstructed $\delta\boldsymbol{\psi}$ of the XXX $n=3$ model that has all the parameters fixed to their true values.
Hence, despite their different appearance (see the third row of panels in Fig. \ref{fig:results_both_images}) the reconstructed $\delta\boldsymbol{\psi}$ of the FFF (and XXX $n=3$) model have an almost identical power spectrum to the truth.
We also note that the residual power spectrum of the FFF and the XXX $n=3$ models, shown in Fig. \ref{fig:results_both_res_ps}, is very similar, with both models being above the noise in the small scales ($k<5$).
Curvature regularization is clearly a bad prior for the GRF $\delta\boldsymbol{\psi}$ as it leads to prominent residuals, even more than the difference between the unmodelled perturbed and unperturbed mock systems (see Fig. \ref{fig:results_both_res_ps}), and more extreme values of the reconstructed $\delta\boldsymbol{\psi}$ (see Fig. \ref{fig:results_both_dpsi_ps} and Table \ref{tab:perts_GRF_parameters}).
Completely ignoring the existence of any perturbations and modelling the system with a purely smooth model with $n=1$ can reach the noise level (see Fig. \ref{fig:results_both_res_ps}).
This is clearly a biased solution that could model away substructure or deviations from the smooth potential.

In Fig. \ref{fig:results_combined_corner} we compare the full non-linear parameter probability densities of the FFF model presented here to the ALL model presented in Section \ref{sec:model_dpsi}).
Its MAP and mean parameter values, and the 68 per cent confidence intervals are listed in Tables \ref{tab:table_perts_smooth_map} and \ref{tab:table_perts_smooth_mean}.
The two models are actually the same but applied to different data, i.e. with a difference source light profile.
We can observe three main characteristics of the distributions: i) smaller statistical uncertainties, ii) larger systematic biases, and iii) fragmentation of the probability surfaces, with various local maxima separated by valleys and saddles, given rise to a complex parameter space configuration.
The latter reflects the complex and degenerate underlying lens potential perturbations and source brightness profile.
The smooth lens potential parameters $\boldsymbol{\eta}$ are correlated in the same way as before but the biases are more significant.
The SIE potential strength $b$ is pushed to even lower values as the $\delta\boldsymbol{\psi}$ are now stronger (e.g. compare the reconstructed MAP perturbations between the ALL and the FFF models in Figs. \ref{fig:results_perts_images} and \ref{fig:results_both_images} respectively), $x_0$ and $y_0$ are offset by approx. 1 pixel, and $q$ and $\gamma$ lie several $\sigma$ further than their true values.
Only the angles $\theta$ and $\phi$ are not biased and are in fact less degenerate than the ALL model, i.e. their distributions are not bi-modal anymore.
This is because of the more detailed structure in the source that cannot be accounted for well by the perturbing field $\delta\boldsymbol{\psi}$ for tilted smooth potentials.
All of the regularization parameters have broader distributions except $\lambda_{\rm \delta\psi}$ that is more narrowly distributed around values 3-4 times smaller than the ALL model.
This means that more structured and larger in amplitude $\delta\boldsymbol{\psi}$ reconstructions are expected, which is indeed the case as shown in Fig. \ref{fig:results_both_images}.
A very strong anti-correlation is observed between the regularization strengths, $\lambda$, and correlation lengths, $l$, in the covariance kernels for both the source and the perturbations.
Finally, the complex probability surfaces between the source and potential perturbation regularization parameters (see also Fig. \ref{fig:app_fff_corner}) mean that the two are quite degenerate.
The smaller values of $\lambda_{\rm \delta\psi}$ in combination with the broader $l_{\rm s}$ distribution towards higher values indicate that the complexity of the source brightness is absorbed by the potential perturbations.

%%%%%%%%%%%%%%%%%%%%%%%%%%%%%%%%%%%%%%%%%%%%%%%%%%%%%%%%%%%%%%%%%%%%%%%%%%%%%%%%%%%%%%%%%%%%%%%%%%%%%%%%%%%%%%%%%%%%%%%%%%%%%%%%%%%%%%%%%%%%%%%%%%%%%%
\section{Discussion}
\label{sec:discussion}
Higher order statistical properties of the brightness profiles of gravitationally lensed galaxies can be incorporated in the semi-linear inversion technique through regularization priors based on physically motivated covariance kernels.
In this work, we created mock gravitational lenses using NGC3982 (a spiral) and NGC2623 (a merger) as sources, whose covariance is well-described by a Gaussian and exponential covariance kernel respectively.
We found that these physically motivated priors outperform other traditionally used regularization schemes, such as identity and curvature, and we can model each system down to the noise level in almost all cases while simultaneously avoiding overfitting (some residuals remain in the case of perturbed potentials).

Using generic covariance priors comes at the cost of introducing additional non-linear parameters (in this case, the correlation length $l_{\rm s}$; see equations \ref{eq:covariance_exp} and \ref{eq:covariance_exp_squared}).
Our modelling framework can handle these new parameters and determine their full probability distribution jointly with the other non-linear parameters (e.g. the smooth mass model parameters, $\boldsymbol{\eta}$) at the cost of a now denser source covariance matrix, $C_{\rm s}$, that needs to be inverted (e.g. see equation \ref{eq:min_r}), and slower convergence due to increasing the dimensions of the non-linear parameter space that needs to be explored.
However, here we used logarithmic priors on a wide range of $l_{\rm s}$, which might be a conservative choice.
One could use observationally driven estimates of $l_{\rm s}$ (or other covariance kernel parameters) derived from populations of putative lensed sources, e.g. constructed from samples of observed lenses, in order to narrow-down the parameter space and speed up the modelling process.
In fact, we performed such a test by fixing $l_{\rm s} = 0.21$ for NGC3982, a value well-justified by the observations (see Fig. \ref{fig:spiral_merger_two_point}), and remodelling the corresponding mock lens, achieving a much faster convergence to the same result.

The quality of the data, viz. high signal-to-noise and resolution, plays a major role in finding an acceptable solution for the source, regardless of the choice of prior, observationally motivated or not, on the source brightness profile.
In the cases examined in Section \ref{sec:smooth_complex}, the data are of sufficiently good quality to drive the solution close to the truth for all tested regularization schemes.
For NGC2623, the recovered $l_{\rm s}$ parameter for the case with an exponential covariance kernel - the one matching the true source - lies further than 3$\sigma$ from the truth, despite having the highest evidence.
The reverse statement, viz. whether the use of a (correct) prior becomes more important in the case of degraded/noisy data, is yet to be systematically explored.
This is particularly relevant for upcoming surveys, such as Euclid and LSST, which are expected to have lower angular resolution than what we examined here.
However, our method does prefer the models with the correct priors based on the Bayesian evidence, for the adopted observational setup.

Once perturbations to the lensing potential are introduced, we need to approach the problem in a different way.
We demonstrated that the effect of $\delta\boldsymbol{\psi}$ can be absorbed in the reconstructed source, especially if the adaptive grid resolution is set to the highest ($n=1$, a common choice), and lead to wrong results on the model parameters, $\boldsymbol{\eta}$, and the source, $\boldsymbol{s}$.
This, in turn, leads to spurious structures in the model residuals, unrelated to the original $\delta\boldsymbol{\psi}$, which can be misinterpreted as the effect of a perturbing field of mass substructure \citep[see also][for another study on this]{Chatterjee2019}.
Hence, a two-step approach of first running a parametric smooth model to constrain $\boldsymbol{\eta}$ and then modelling the perturbations $\delta\boldsymbol{\psi}$ would be unreliable \citep[unless lower choices for $n$ are used, e.g. see][]{Bayer2021}.
The extent of the above statement for perturbed lenses with varying $\delta\boldsymbol{\psi}$ properties, as well as concentrated massive substructures, remains to be explored.
Nevertheless, we showed that simultaneously solving for $\boldsymbol{\eta}$, $\boldsymbol{s}$, and $\delta\boldsymbol{\psi}$ gives accurate results in a self-consistent manner.

Attempting to reconstruct the perturbing $\delta\boldsymbol{\psi}$ requires a regularizing term (prior) in addition to the one for the source.
In contrast to the case of smooth potentials, where the data quality is good enough to drive the source reconstructions to solutions with the desired statistical properties regardless of which regularization scheme is used (see Fig. \ref{fig:spiral_merger_two_point}), the data alone are not sufficient and the form of regularization/prior seems to play a major role in reconstructing $\delta\boldsymbol{\psi}$.
Here we examined specifically the curvature and Gaussian covariance kernels, in connection to our choice of a GRF as the true underlying $\delta\boldsymbol{\psi}$.
The traditionally used curvature regularization is less flexible as it imposes fixed, long range correlations (see Fig. \ref{fig:results_perts_two_point}), which are in fact stronger than they should and irrecoverably lead to unphysically smooth solutions, seemingly regardless of the quality of the data.
The covariance of our assumed GRF, however, can be well approximated by a Gaussian kernel (see Fig. \ref{fig:results_perts_two_point}), but in real galaxies the true covariance of potential perturbations is unknown.
More flexibility could be achieved by assuming a covariance kernel described by a number of free parameters, e.g. a Mat\'{e}rn kernel \citep[e.g.][]{Mertens2017,Vernardos2020}, or even a free form two-point correlation function.
In addition, theoretically justified $\delta\boldsymbol{\psi}$ priors could be derived based on dark matter models or N-body hydrodynamical simulations.
Our method allows for a thorough and quantitative exploration of how different regularization schemes on the $\delta\boldsymbol{\psi}$, as well as on the source, can affect the quality of the reconstructions, eventually ranking them by their Bayesian factors.

In Sections \ref{sec:model_dpsi} and \ref{sec:both} we fully model the smooth potential, source, and perturbations in two example cases whose only difference is the brightness profile of the source, i.e. the smooth lens potential and the perturbative field of $\delta\boldsymbol{\psi}$ remain the same.
Our optimization strategy (described in Section \ref{sec:optimization}) works quite well, but the extent of statistical uncertainty and systematic biases in the recovered parameters $\boldsymbol{\eta}$, as well as the degeneracy between the regularization parameters for the source and the perturbations, depend on the complexity of the source brightness profile.
In the case of the complex source presented in Section \ref{sec:both}, the entire parameter space becomes more structured and degenerate (see Fig. \ref{fig:app_fff_corner}) and systematic biases increase (see Fig. \ref{fig:results_combined_corner}).
Most importantly, smoother sources become more compatible with the data and the freedom of the perturbing $\delta\boldsymbol{\psi}$ is increased (i.e. its smoothness reduced), which leads to the latter absorbing the structure of the source.
The overall amplitude of $\delta\boldsymbol{\psi}$ is also larger, pushing the strength of the smooth potential (parameter $b$) to lower values.
These observations explain why the reconstructed $\delta\boldsymbol{\psi}$ from the FFF model in Fig. \ref{fig:results_both_images} do not visually match the true GRF very well, but despite this the power spectrum is recovered remarkably well (see Figs. \ref{fig:results_perts_dpsi_ps}, \ref{fig:results_both_dpsi_ps}, and Table \ref{tab:perts_GRF_parameters}).

The visual differences of the reconstructed $\delta\boldsymbol{\psi}$ compared to the truth (see the FFF and ALL reconstructions in Figs. \ref{fig:results_both_images} and \ref{fig:results_perts_images}, respectively), could be understood in terms of the ``light-constrains-mass'' effect, which we explain here.
Within the framework of our method, but also more generally, it is important to clarify how is $\delta\boldsymbol{\psi}$ constrained where the lensed source brightness, and/or, more precisely, the gradient of the source is low or zero.
Obviously, in such areas using equation (\ref{eq:dpsi_residuals}) to model brightness residuals becomes problematic; the $D_{\rm s}$ operator, which holds the derivatives of the source at the source plane (deflected) location of the given image pixel(s), becomes zero.
Hence, in order to obtain a reconstruction across the entire field of view (or even within a mask) it now becomes obvious that the regularization will be important, particularly where there is low/no source flux.
This is analogous - but not exactly - to reconstructing the source brightness on pixels that are not constrained by the data, as could be the case in a fixed grid model.
Taking the realization of the GRF $\delta\boldsymbol{\psi}$ field that we used as an example (third-row panel in the left of Fig. \ref{fig:results_perts_images}), the success of our reconstructions depends on how much of the source flux eventually end ups in those crucial areas of the lens plane that have the largest gradients (largest deflection angles).
This could play a role in the more degenerate results of the FFF model and its $\delta\boldsymbol{\psi}$ power spectrum amplitude difference with the ALL model (see Figs. \ref{fig:results_perts_dpsi_ps}, \ref{fig:results_both_dpsi_ps}, and Table \ref{tab:perts_GRF_parameters}).
This could be mitigated by reconstructing the $\delta\boldsymbol{\psi}$ within a carefully selected region of the lens plane around the lensed source brightness, possibly weighed by the values of the operator $D_{\rm s}$.
However, determining the extent of this ``light-constrains-mass'' area may introduce another possible source of degeneracy: the gradient of $\delta\boldsymbol{\psi}$, which is in fact the deflection angle, also enters equation (\ref{eq:dpsi_residuals}), and for any pixel with some given lensed source brightness, regions having the same gradient, e.g. large density differences that lie further away or smaller density differences being closer, can have the same effect.

%%%%%%%%%%%%%%%%%%%%%%%%%%%%%%%%%%%%%%%%%%%%%%%%%%%%%%%%%%%%%%%%%%%%%%%%%%%%%%%%%%%%%%%%%%%%%%%%%%%%%%%%%%%%%%%%%%%%%%%%%%%%%%%%%%%%%%%%%%%%%%%%%%%%%%
\section{Conclusions}
\label{sec:conclusions}
We explored the effect of regularization while reconstructing both the source and potential perturbations using the semi-linear inversion technique.
Below we summarize the conclusions from this work and outline future directions of application and improvement.
\begin{itemize}
	\item Physically motivated priors for the source galaxies, such as Gaussian and exponential kernels, lead to better results than traditional choices, such as identity and curvature regularization.
	\item Curvature regularization, a traditionally popular choice, is fundamentally unsuitable as a prior for the GRF $\delta\boldsymbol{\psi}$ perturbations that we examined here.
	\item The source alone can absorb the structure created by $\delta\boldsymbol{\psi}$ almost down to the noise, especially if a high resolution adaptive grid is used (low value of $n$). This leads to biased source reconstructions and parameters for the smooth potential \citep[see also][]{Bayer2018,Bayer2021,Chatterjee2019}.
	\item The statistical properties of the $\delta\boldsymbol{\psi}$, particularly the power spectrum, are recovered remarkably well, both for smooth and more complex sources.
\end{itemize}
Our study constitutes an initial exploration and test of our new code implementation, and as such we restricted ourselves to the four distinct and incrementally more complex examples presented in Section \ref{sec:results}.
The successful outcome of this study enables further and more in depth investigations of potential perturbation reconstructions in lensed systems.
We propose, but not limit ourselves to, the following directions of future research:
\begin{enumerate}
	\item Here we used a specific GRF as the perturbing field, with specific amplitude ($\approx13$ per cent of the smooth potential) and slope, which we believe is an extreme case, pushing the validity of the approximation of equation (\ref{eq:dpsi_residuals}) to its limit. The type (GRF or other), as well as the associated parameter space of the perturbing field can be now explored more in depth, for different smooth potentials and sources.
	\item One such case of particular interest would be using isolated massive perturbers as the perturbing $\delta\boldsymbol{\psi}$, and determining how the conclusions of this work apply to it, e.g. comparing to the work of \citet{Vegetti2009a}.
	\item We have identified an interplay between data quality and priors in determining the best model, which needs to be explored in both directions: at which level of resolution and/or signal to noise ratio the data are driving the solution and the prior begins to play a secondary role, and inversely.
	\item Our $\delta\boldsymbol{\psi}$ reconstructions away from pixels that contain most of the lensed source flux are constrained mostly by the prior - what we described as the ``light-constrains-mass' effect. A weighed scheme - similar to adaptive regularization - could be devised to suppress terms in the $D_{\rm s}$ appearing in equation (\ref{eq:dpsi_residuals}) that are very low or zero.
\end{enumerate}
Finally, our new implementation of the method, the Very Knotty Lenser code, is made publicly available\footnote{\url{https://github.com/gvernard/verykool}}.

%%%%%%%%%%%%%%%%%%%%%%%%%%%%%%%%%%%%%%%%%%%%%%%%%%%%%%%%%%%%%%%%%%%%%%%%%%%%%%%%%%%%%%%%%%%%%%%%%%%%%%%%%%%%%%%%%%%%%%%%%%%%%%%%%%%%%%%%%%%%%%%%%%%%%%
\section*{Data availability}
The data that support the findings of this study are openly available in github at \url{https://github.com/gvernard/verykool}.

%%%%%%%%%%%%%%%%%%%%%%%%%%%%%%%%%%%%%%%%%%%%%%%%%%%%%%%%%%%%%%%%%%%%%%%%%%%%%%%%%%%%%%%%%%%%%%%%%%%%%%%%%%%%%%%%%%%%%%%%%%%%%%%%%%%%%%%%%%%%%%%%%%%%%%
\section*{Acknowledgements}
GV and LVEK were supported through an NWO-VICI grant (project number 639.043.308).
GV has received additional funding from the European Union’s Horizon 2020 research and innovation programme under the Marie Sklodovska-Curie grant agreement No 897124.

\bibliographystyle{mnras}
\bibliography{bibliography} % if your bibtex file is called ml_lc.bib
%%%%%%%%%%%%%%%%%%%%%%%%%%%%%%%%%%%%%%%%%%%%%%%%%%

\appendix
\section{Corner plots}
\label{app:B}
Zoomed-in versions of the probability distributions shown in Figs. \ref{fig:results_smooth_corner} and \ref{fig:results_combined_corner}.

\begin{figure*}
	\includegraphics[width=\textwidth]{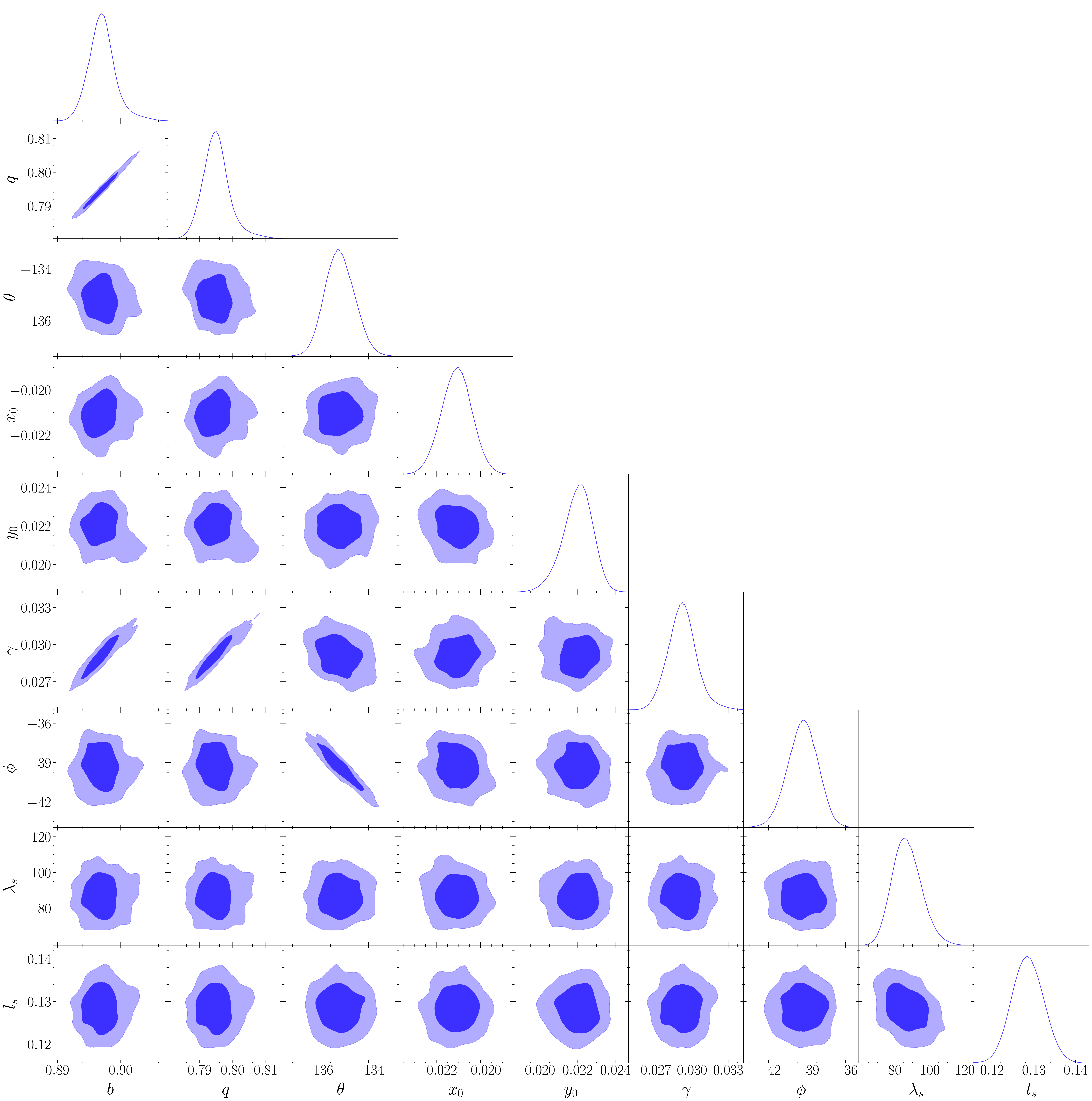}
	\caption{Same as Fig. \ref{fig:results_smooth_corner} with zoomed-in ranges to better show the shape of the two-dimensional distributions.}
	\label{fig:app_smooth_corner}
\end{figure*}

\begin{figure*}
	\includegraphics[width=\textwidth]{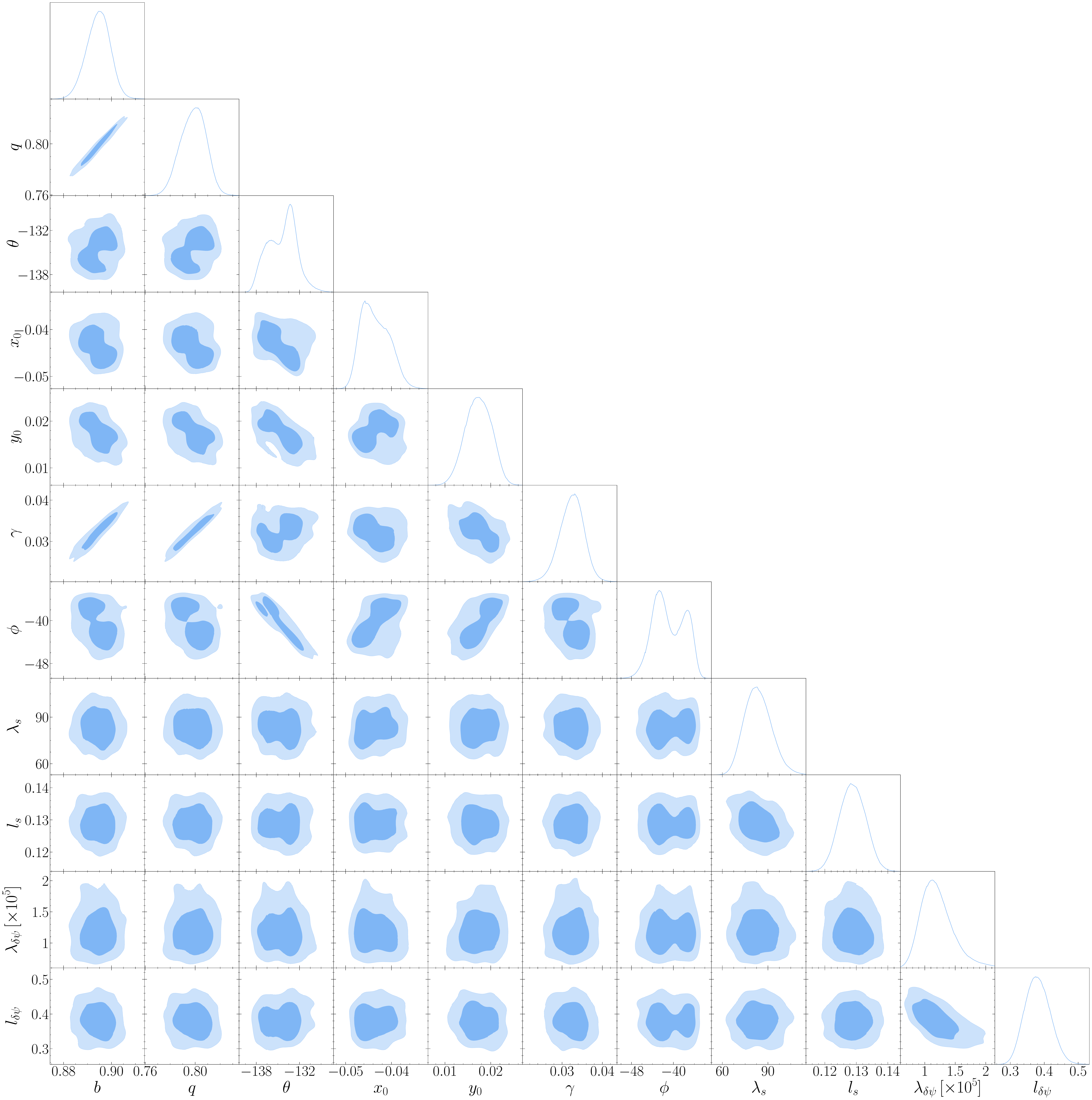}
	\caption{Same as Fig. \ref{fig:results_combined_corner} for the ALL model with zoomed-in ranges to better show the shape of the two-dimensional distributions.}
	\label{fig:app_all_corner}
\end{figure*}

\begin{figure*}
	\includegraphics[width=\textwidth]{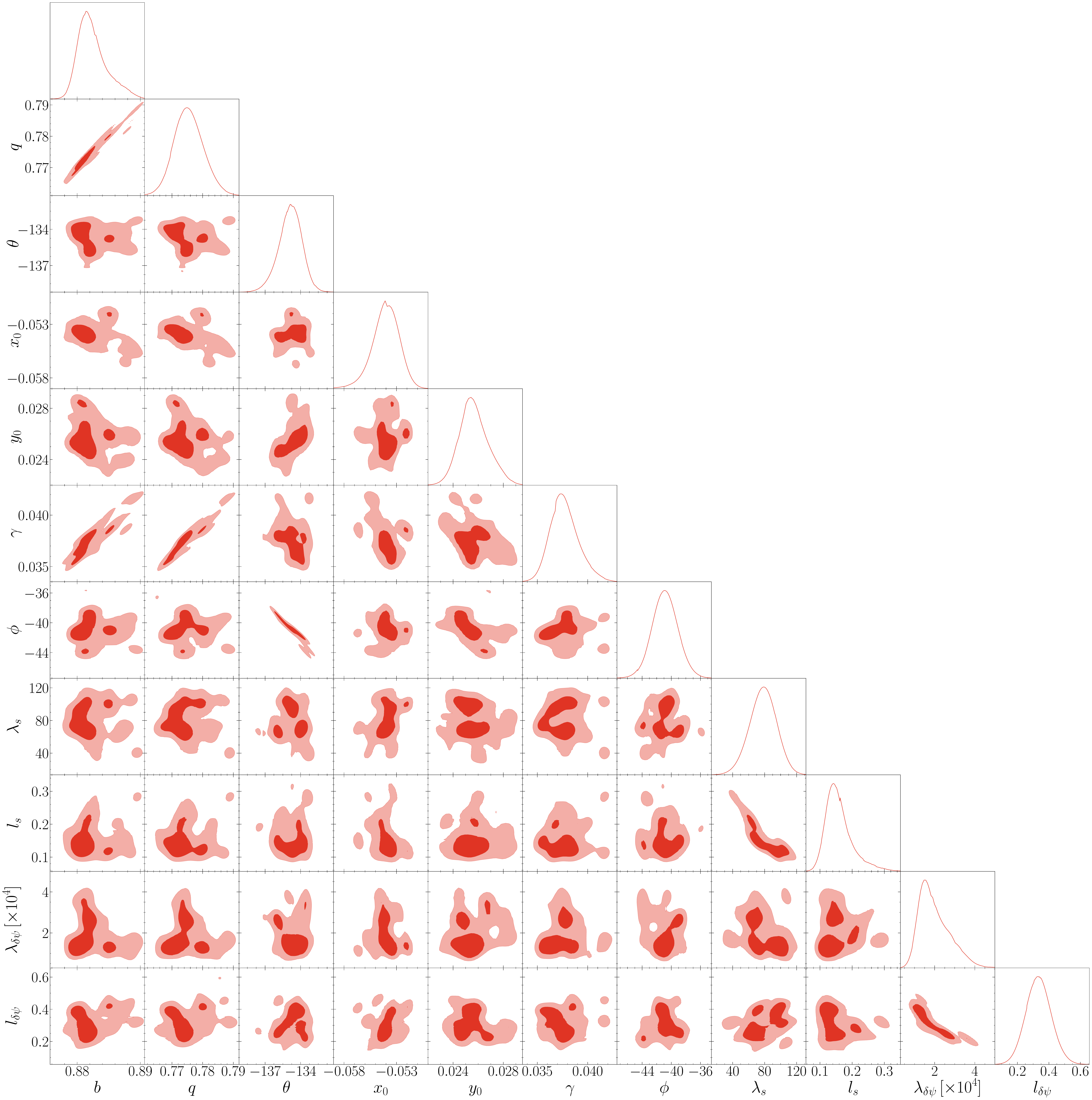}
	\caption{Same as Fig. \ref{fig:results_combined_corner} for the FFF model with zoomed-in ranges to better show the shape of the two-dimensional distributions.}
	\label{fig:app_fff_corner}
\end{figure*}

\section{The smooth potential absorbs the perturbations to a very small extent}
\label{app:A}
\begin{figure}
	\includegraphics[width=0.45\textwidth]{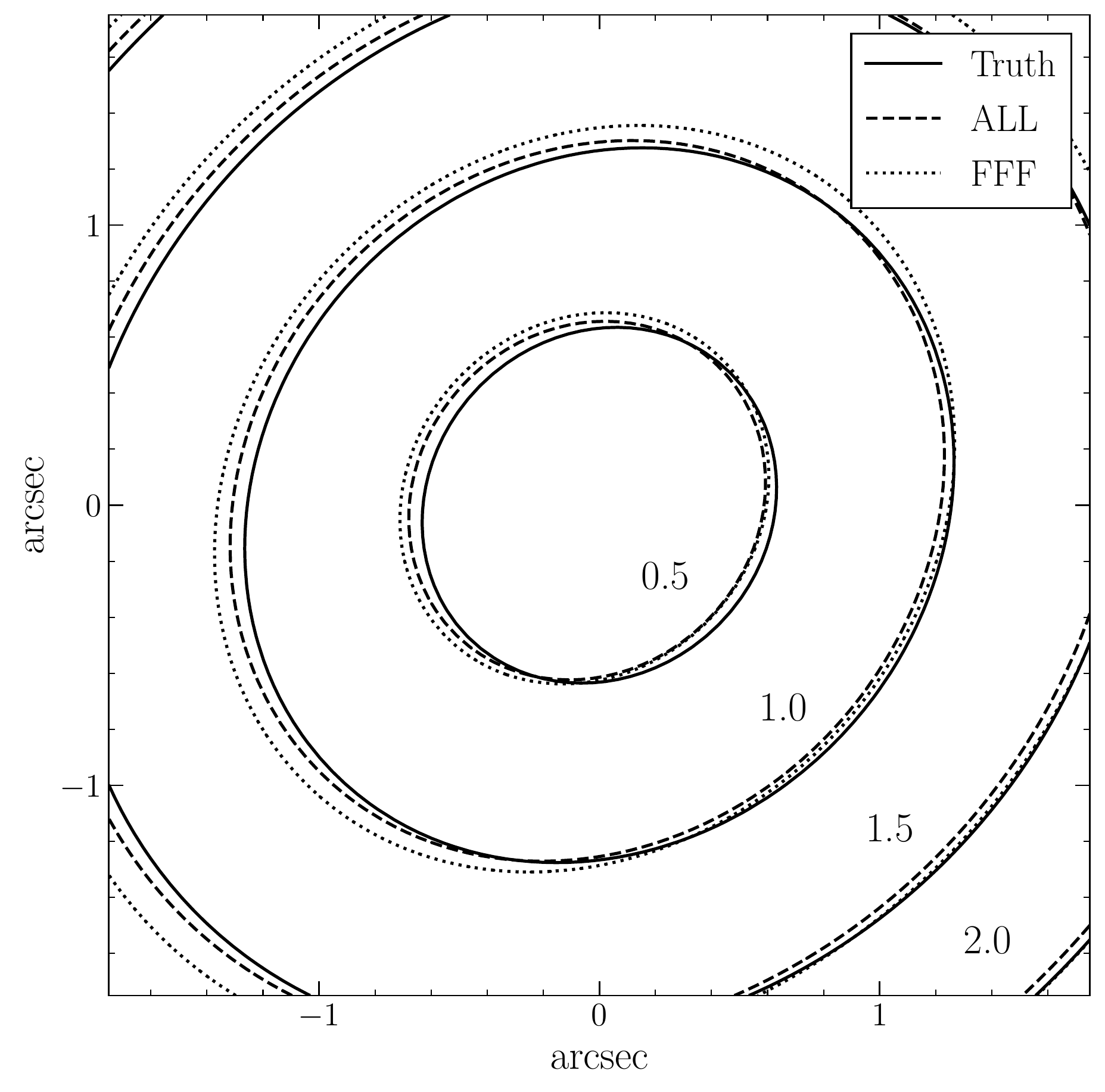}
	\caption{Contours of the true underlying smooth potential described in Section \ref{sec:model_smooth}, as well as its MAP fits by the ALL and FFF models (see Table \ref{tab:table_perts_smooth_map}). These models simultaneously fit the smooth potential and reconstruct its perturbations $\delta\boldsymbol{\psi}$.}
	\label{fig:smooth_psi_contours}
\end{figure}

In Fig. \ref{fig:smooth_psi_contours} we show the true smooth potential, which is a SIE with external shear described by the parameters $\boldsymbol{\eta}$ introduced in Section \ref{sec:model_smooth} (see also Table \ref{tab:table_map}), as well as its MAP fits by the ALL and FFF models described in Sections \ref{sec:model_dpsi} and \ref{sec:both} respectively.
These two models simultaneously fit the smooth potential and reconstruct its perturbations.
Both models recover accurately the smooth potential parameters as listed in Tables \ref{tab:table_perts_smooth_map} and \ref{tab:table_perts_smooth_mean} and shown in Fig. \ref{fig:smooth_psi_contours}.
Therefore, the observed differences in the reconstructions of $\delta\boldsymbol{\psi}$ and $\boldsymbol{s}$ are purely due to their fundamental connection through equation \ref{eq:dpsi_residuals} and the choice of regularization.

\begin{figure}
	\includegraphics[width=0.45\textwidth]{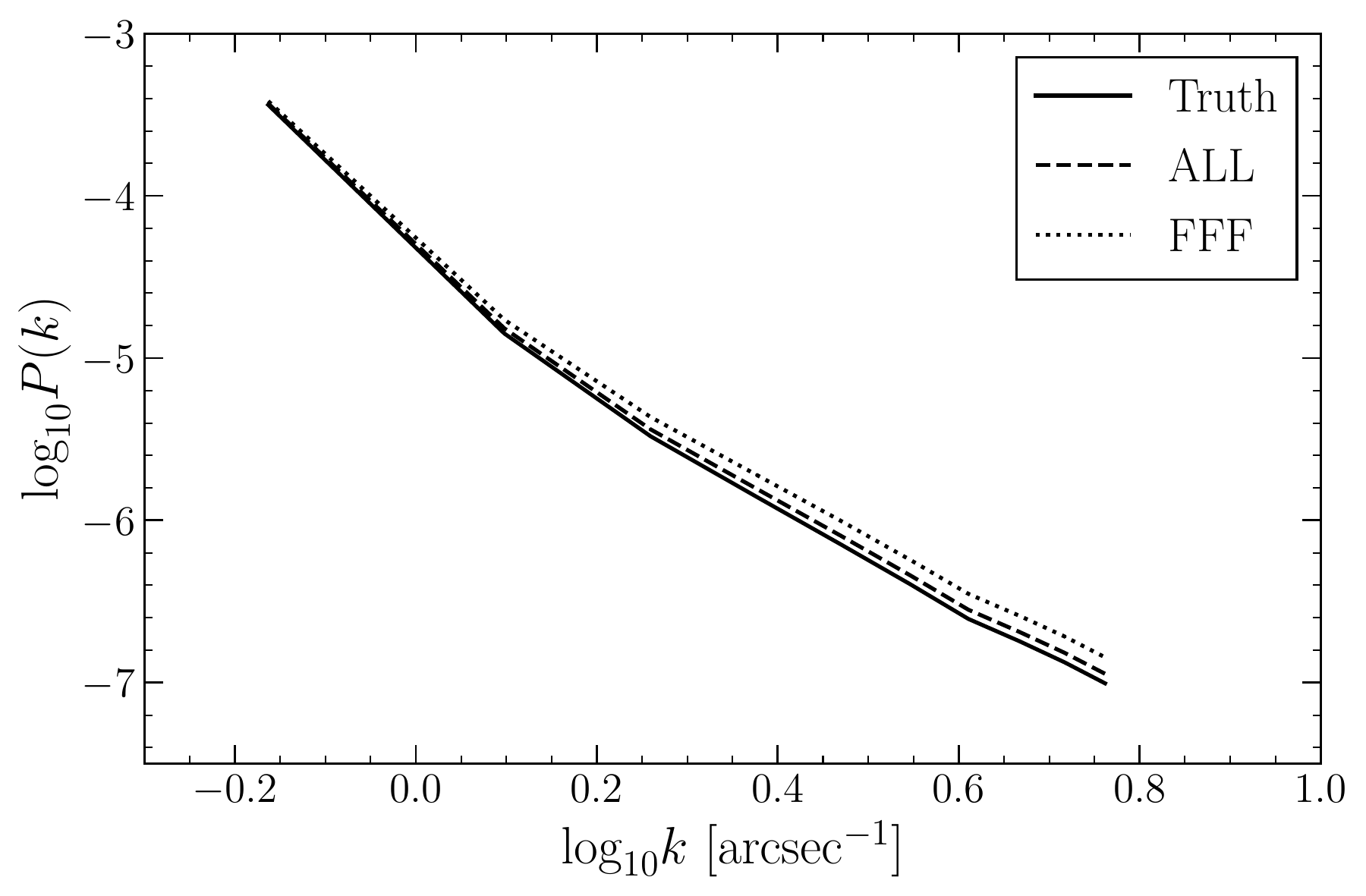}
	\caption{Fourier power spectrum of the smooth potentials shown in Fig. \ref{fig:smooth_psi_contours}.}
	\label{fig:ps_smooth_psi}
\end{figure}

Finally, Fig. \ref{fig:ps_smooth_psi} shows the corresponding power spectra of the smooth potentials shown in Fig. \ref{fig:smooth_psi_contours}.
The power spectra are almost identical and drop smoothly with wavenumber $k$.
Hence, the fitted smooth potentials can neither absorb nor introduce any spurious $\delta\boldsymbol{\psi}$.

% Don't change these lines
\bsp	% typesetting comment
\label{lastpage}
\end{document}